\documentclass[twocolumn,prb,aps,scriptaddress]{revtex4-2}
\usepackage[utf8]{inputenc}
\usepackage{color}
\usepackage{booktabs}
\usepackage{amsmath}
\usepackage{graphicx}
\usepackage{subscript}

\usepackage[ citecolor=blue, linkcolor=blue, urlcolor=black]{hyperref}

\makeatletter

\providecommand{\tabularnewline}{\\}

\makeatother

\begin{document}
\title{Defect engineering in two-dimensional pentagonal PdTe$_{2}$: Tuning electronic, optical, and magnetic properties}

\author{Poonam Sharma, Vaishali Roondhe, Alok Shukla}
\email{shukla@iitb.ac.in}

\affiliation{Department of Physics, Indian Institute of Technology Bombay, Mumbai
400076, India}
\begin{abstract}
Recently, the successful synthesis of the pentagonal form of PdTe$_{2}$
monolayer (\emph{p}-PdTe$_{2}$) was reported [Liu~\emph{et al.}, Nature Materials \textbf{23}, 1339 (2024)]. In this paper, we present an extensive first-principles density-functional theory based computational study of vacancies in this material. Our paper covers the evolution of the electronic, optical, and magnetic properties of various defect configurations and compares those to the pristine monolayer (\emph{p}-PdTe$_{2}$). We find that V$_{Pd}$ (V$_{Te}$) is the most stable defect in the~\emph{p}-PdTe$_{2}$ monolayer with 0.95 (1.60) eV of formation energy in the Te-rich (Pd-rich) limit. The defects alter the electronic properties of the monolayer significantly, leading to changes in their magnetic and optical properties due to the emergence of midgap impurity states. The defect complex V$_{Pd+4Te}$ is found to induce spin polarization in the system with a total magnetic moment of 1.87 $\mu_{B}$. The obtained low diffusion energy barrier of 1.13 eV (in-plane) and 0.063 eV (top-bottom) corresponding to V$_{Te}$ indicates its facile migration probability is higher in the top-bottom direction in comparison to the in-plane direction at room temperature, as revealed by \emph{ab initio} molecular dynamics simulations as well. In order to guide the experimentalists, we also simulated the scanning-tunneling microscope images corresponding to all the defect configurations. Moreover, we also computed the electron-beam energies required for creating mono-vacancies. In the optical absorption spectra of the defective configurations, finite peaks appear below the band edge that are unique to the respective defective configuration. We have also computed the excess polarizability of the defective configurations with respect to the pristine one and found that maximum changes occur in the infrared and visible regions, providing insights into the change in their optical response as compared to the pristine monolayer. Our paper will open prospects of defect engineering in this and related materials with the aim of tuning their electronic, optical, and magnetic properties from the point of view of device applications. 
\end{abstract}
\maketitle

\section{Introduction}

Layered van der Waals materials, such as two-dimensional (2D) transition-metal dichalcogenides (TMDs), are one of the most extensively investigated classes of materials by the research community owing to their potential applications in optoelectronics and spintronics devices~\cite{henriques2021calculation,PhysRevB.90.125440, PhysRevB.110.045425}.
These materials provide an engaging area for investigating an extensive range of physical phenomena, including exotic charge density waves, unconventional superconductivity, topological states, excitons, valley-related physics, etc.~\cite{he2021two,PhysRevMaterials.2.094001,rossnagel2011origin,rahimi2017unconventional,lin2022phonon}.
In general, all materials have naturally occurring defects such as vacancies, interstitials, antisites, and grain boundaries, which can influence their properties profoundly. Similarly, it is also possible to introduce defects in the materials in a controlled manner, including TMDs, with the aim of tuning their properties~\cite{maguire2018defect}.
In spite of the fact that creating defects may cost extra energy, there are many experimental techniques that offer different ways to create defects and study their impact on material properties~\cite{PhysRevB.99.115312,PhysRevMaterials.5.044002,PhysRevB.106.075428,PhysRevLett.109.035503,krasheninnikov2002production}.
In order to create defects, many techniques, such as ion bombardment,
electron irradiation, scanning tunneling microscope (STM) voltage
pulsing, and vacuum annealing, have been extensively used because
they enable controlled modification of the properties of materials
after growth~\cite{PhysRevB.81.153401,zheng2013manipulation,dagdeviren2016surface}.
Therefore, it is important also to study theoretically the evolution
of properties of materials in the presence of defects.

In the TMD systems, the majority of the materials exhibit the honeycomb
(hexagonal) type of crystal structure (i.e., graphene-like) and possess
MX$_{2}$ stoichiometry, where the metal atom (M) is sandwiched between
the two layers of chalcogen atoms (X). However, in a handful of cases,
one observes a preference for the pentagonal phase as well, i.e.,
those with buckled pentagons as their fundamental building units.
Although theoreticians, based on first-principles calculations, have
predicted a wide array of 2D pentagonal materials such as PdS$_{2}$,
PdSe$_{2}$, and SnS$_{2}$, many of them exist in metastable states,
making their practical synthesis challenging~\cite{shen2022pentagon}.
Therefore, these pentagonal dichalcogenides remain relatively unexplored
due to the challenges associated with their synthesis, which sets
them apart from the other TMDs, such as MoX$_{2}$ and WX$_{2}$ (X
= S, Se, Te), which have been extensively studied both theoretically
and experimentally. However, these pentagonal materials show unparalleled
characteristics and distinct potential applications because of their
lower spatial symmetry~\cite{shen2022pentagon,marfoua2019high}.
With the identification of robust structural stability and exceptional
carrier mobility within monoclinic pentagonal PdS$_{2}$, as uncovered
by Wang~\emph{et al.}~\cite{wang2015not}, the interest in the studies
of the penta-MY\textsubscript{2} class of materials (M = Sn, Pd,
Pt, Y = Te, Se, S) has grown tremendously~\cite{figueiredo2022hydrogen,gonzalez2016layer,PhysRevMaterials.2.094001}.
As far as PdTe$_{2}$ is concerned, the primary focus of research has been on studying superconductivity and topological states \cite{PhysRevMaterials.2.094001,pdte2-superconductivity-PhysRevB.100.224501},
thermoelectric properties~\cite{marfoua2019high,p-pdx2-thermoelectric-theory-2019}, catalytic properties~\cite{zuo2023defect}, etc.

Recently, Liu~\emph{et al}. have reported successful synthesis of
the pentagonal form of the PdTe$_{2}$ monolayer (\emph{p}-PdTe$_{2}$)
by the symmetry-driven epitaxy method~\cite{liu2024metastable}. Differing
from typical TMDs, such as MoS$_{2}$ and WS$_{2}$, a single layer
of \emph{p}-PdTe$_{2}$ is composed of Pd atoms surrounded by four
Te atoms. This arrangement leads to the creation of four-coordinated
PdTe$_{4}$ units, rendering the monolayer highly responsive to a
range of defects, in general, and vacancies, in particular. Recently,
it was found that the introduction of defects in 2D layered PdTe$_{2}$
significantly increased the catalytic properties of the material~\cite{zuo2023defect}. Additionally, there has been extensive research on defect-induced magnetism in several TMD monolayers, such as PtX$_2$(X=S, Se, Te)~\cite{manchanda2021defect,manchanda2016hydrogen}.

Given that there have been no previous
theoretical or experimental studies of defects in the \emph{p}-PdTe$_{2}$ monolayer,
in this paper, we have undertaken a systematic first-principles density-functional theory (DFT) based
study of vacancies in this compound. By means of total energy calculations,
we compute the formation energies of various defects. Furthermore,
we also analyze the electronic, optical, and magnetic properties of
the \emph{p}-PdTe$_{2}$ monolayer in the presence of various
vacancies. We have also calculated the STM images corresponding to
all defect configurations considered in this paper to guide the
experimentalists and computed the electron-beam energy for creating
such defect configurations. The remainder of this paper is organized
as follows. In the next section (sec. \ref{sec:COMPUTATIONAL-DETAILS}),
we present the computational details of our DFT calculations. In sec.
\ref{sec:RESULTS-AND-DISCUSSION}, we present and discuss our results,
while in the final section \ref{sec:Conclusion} we conclude our findings.

\section{COMPUTATIONAL DETAILS}

\label{sec:COMPUTATIONAL-DETAILS}

All the calculations were carried out within the framework of first-principles
DFT implemented in the Vienna~\emph{Ab-initio} Simulation Package
(VASP) \cite{hohenberg1964inhomogeneous,kohn1965self,kresse1996efficient}.
The generalized gradient approximation (GGA) with the Perdew-Burke-Ernzerhof
(PBE) parametrization was used to describe the electronic exchange-correlation functional~\cite{perdew1996generalized,perdew1997generalized}. $\Gamma$-centred
\textbf{k} meshes of 14$\times$14$\times$1 and 3$\times$3$\times$1
were employed for the unit cell and supercell calculations, respectively,
for the geometry optimization, along with the kinetic energy cutoff
of 500 eV. A given geometry was considered to converge only after
the Hellman-Feynman force on each atom was less than 1$\times$10$^{-2}$
eV~\cite{monkhorst1976special}. A vacuum of 21~{Å} was used in the $z$ direction for the purpose of preventing interactions between adjacent periodic layers. The valence electronic configurations
of Pd and Te atoms for the projector augmented wave pseudopotentials were taken as 4$d^{9}$5$s^{1}$
and 5$s^{2}$5$p^{4}$, respectively. For studying the thermodynamic
stability of considered vacancies, the formation energy ($\Delta$$E_{f}$)
was calculated using the following equation: 
\begin{equation}
\Delta E_{f}=E_{defect}-E_{prisitne}+\sum_{i}\Delta n_{i}\mu_{i},\label{eq:eqform}
\end{equation}

where $E_{defect}$ and $E_{prisitne}$ denote the total
energies of the defective and pristine systems, respectively. The
number of vacant atoms of the $i$th species is denoted by $\Delta n_{i}$,
while $\mu_{i}$ represents the corresponding chemical potential.
For studying the optical properties of the system, the frequency-dependent
real and imaginary parts of the dielectric constant, $\epsilon_{1}(\omega)$ and $\epsilon_{2}$($\omega$), were computed. Moreover, using $\epsilon_{1}(\omega)$
and $\epsilon_{2}$($\omega$), the real and imaginary parts of the
electronic polarizability were computed using \cite{tian2019electronic}:

\begin{equation}
\epsilon_{2D}^{\parallel}=1+\frac{\alpha_{2D}^{\parallel}}{\epsilon_{0}L},
\end{equation}

where $\alpha_{2D}$ and $L$, respectively, represent
the polarizability per unit area and the length along the z direction separating the two consecutive cell images of the 2D material under
consideration. STM images were generated using the Tersoff-Hamann
theory~\cite{tersoff1985theory} as implemented in the HIVE program
of Vanpoucke and Brocks~\cite{vanpoucke2008formation}, which uses
the output data of the DFT calculations for this purpose. The spin-orbit
coupling (SOC), as implemented in VASP, was also included in all the
calculations. To validate our GGA+SOC results, we
repeated some of our calculations using the HSE06 functional. In
light of this, our computational study of vacancies in the \emph{p}-PdTe$_{2}$
monolayer and their influence on other properties of the system is
computationally quite accurate.

\section{RESULTS AND DISCUSSION}

\label{sec:RESULTS-AND-DISCUSSION}

\subsection{Perfect crystal~\emph{p}-PdTe$_{2}$}

\begin{figure}[ht]
\includegraphics[width=1\linewidth]{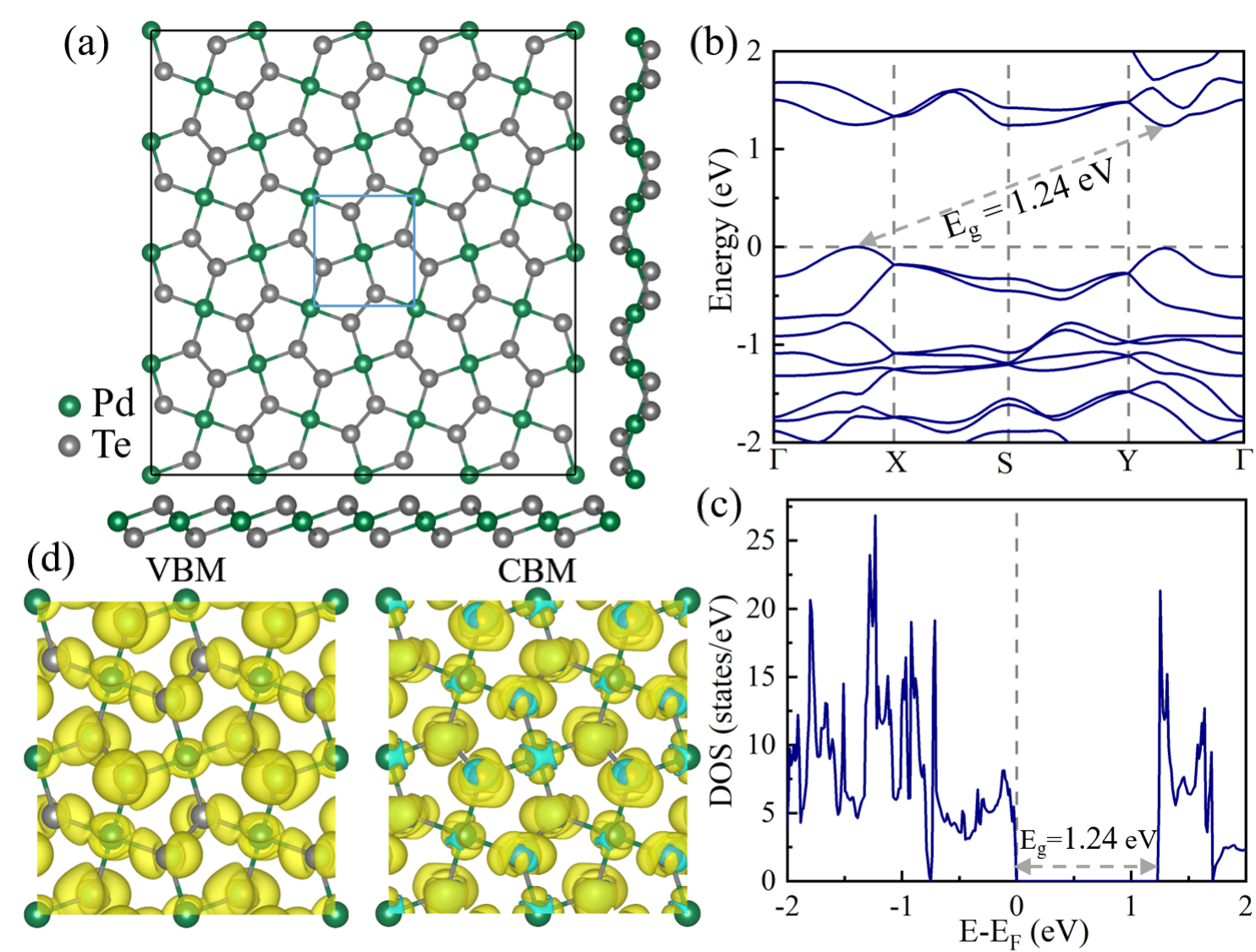}\caption{(a) The optimized 4$\times$4$\times$1 supercell structure of the \emph{p}-PdTe$_{2}$ monolayer with top and side views. (b) Electronic band structure, (c) the total density of states (TDOS) of the pristine unit cell of \emph{p}-PdTe$_{2}$, and (d) band decomposed charge densities for the VBM (left side) and CBM (right side) for the isovalues of 0.0005 and 0.0002 e/Å$^{3}$, respectively. Green and gray colors represent the Pd
and Te atoms, respectively. The Fermi level is set at 0 eV. All our calculations include SOC.}
\label{fig:fig1} 
\end{figure}

Fig.~\ref{fig:fig1}(a) shows the $4\times4\times1$ supercell containing
96 atoms of the pentagonal monolayer of palladium ditelluride (\emph{p}-PdTe$_{2}$)
with top and side views. Each Pd atom in the PdTe$_{2}$ monolayer
links with four Te atoms (two from the top and two from the bottom), while each Te atom is bonded to two Pd atoms
and one Te atom (Te-Te top-bottom pair) in such a way that the anisotropic pentagonal structure
of \emph{p}-PdTe$_{2}$ is formed. After geometry optimization, we find that the buckled \emph{p}-PdTe$_{2}$ monolayer exhibits a vertical height of 1.70~{\AA} (see Fig. \ref{fig:fig1}(a)), in excellent agreement with the value of 1.69 {\AA} reported by Liu~\emph{et al.}~\cite{liu2024metastable}. The unit cell of the \emph{p}-PdTe$_{2}$ monolayer (space group P2$_{1}$/c) consists of six atoms, as shown by the blue rectangle in Fig.~\ref{fig:fig1}(a). Our optimized lattice parameters for the \emph{p}-PdTe$_2$ monolayer are $a = 6.14$ \AA~and $b = 6.43$ \AA, with minimum bond lengths of 2.62 \AA~(Pd-Te) and 2.81 \AA~(Te-Te). These results match well with the previously reported values~\cite{p-pdx2-thermoelectric-theory-2019,marfoua2019high}. Before we start the discussion of the electronic structure of the pristine \emph{p}-PdTe$_{2}$ monolayer, we compute its formation energy with respect to bulk metals Pd and Te as follows:
\begin{equation}
\begin{split} & \Delta E_{\emph{f}}^{PdTe_{2}}=E_{PdTe_{2}}-(\mu_{Pd}^{bulk}+2\mu_{Te}^{bulk})\end{split}.
\label{equ:form-pristine}
\end{equation}
Above $E_{PdTe_{2}}$ is the total energy per formula unit of the pristine \emph{p}-PdTe$_{2}$ monolayer, while $\mu_{Pd}^{bulk}$ and $\mu_{Te}^{bulk}$ are the chemical potentials of Pd and Te in
pure bulk palladium and tellurium metals, respectively. For calculating the $\mu_{Pd}^{bulk}$ and $\mu_{Te}^{bulk}$, the most stable cubic and hexagonal structures of Pd and Te with the space group Fm$\bar{3}$m and P3$_{1}$21, respectively, are considered~\cite{kirklin2015open}. From our calculations, $E_{PdTe_{2}}$= -12.41 eV, $\mu_{Pd}^{bulk}$ = -5.19 eV, and $\mu_{Te}^{bulk}$ = -3.26 eV, which on substitution in Eq. \ref{equ:form-pristine} lead to a negative value of the formation energy $\Delta E_{\emph{f}}^{PdTe_{2}}$ = -0.70 eV, indicating the stability of the pristine \emph{p}-PdTe$_{2}$ monolayer. Experimentally, the \emph{p}-PdTe$_{2}$ monolayer is synthesized by depositing Te atoms on a Pd(100) substrate~\cite{liu2024metastable}. Therefore, we performed calculations of the \emph{p}-PdTe$_{2}$ on the Pd(100) substrate by adding a Te adatoms layer in between the Pd(100) substrate and \emph{p}-PdTe$_{2}$ monolayer (see Fig. S1 of the Supplemental Material (SM)~\cite{SI}). For the Pd(100) substrate, we considered a $\sqrt{5}\times\sqrt{5}\times$ 1 supercell and found the optimized parameters to be $a= b= 6.23$ \AA, which differ somewhat from the corresponding values obtained for the freestanding monolayer (see above). After optimization, the \emph{p}-PdTe$_{2}$ was found to be stable along with the adhesive energy -3.43 J/m$^2$. The negative value of the adhesive energy indicates that the \emph{p}-PdTe$_{2}$ monolayer can be stabilized on the top of the Pd(100) substrate.

Figs. \ref{fig:fig1}(b) and \ref{fig:fig1}(c), respectively, show the calculated electronic band structure along the high-symmetry path $\Gamma$-X-S-Y-$\Gamma$, and the total density of states (TDOS) for the pristine~\emph{p}-PdTe$_{2}$ unit cell. Our results reveal that the \emph{p}-PdTe$_{2}$ monolayer is a non-magnetic semiconductor with an indirect band gap of 1.24 eV connecting the valence band maximum (VBM) close to the X point and the conduction band minimum (CBM) near the Y point, as indicated in Fig.~\ref{fig:fig1}(b-c). The obtained indirect band gap is in agreement with the previous theoretical~\cite{p-pdx2-thermoelectric-theory-2019,marfoua2019high} and experimental results~\cite{liu2024metastable}.
Further, Fig.~\ref{fig:fig1}(d) represents the band decomposed charge densities corresponding to VBM and CBM for the isovalues of 0.0005 and 0.0002 e/Å$^{3}$, respectively. As compared to Pd atoms, the charge density
isosurface is found to be mainly concentrated around the Te atoms for both VBM and CBM, indicating the dominant character of Te atoms, which form the zigzag chains in the \emph{p}-PdTe$_{2}$ monolayer and contribute primarily to the formation of STM images. The calculated partial density of states is in good agreement with the band decomposed charge densities analysis (see Fig. S2 of the SM~\cite{SI}).

\subsection{Defective configurations}

\subsubsection{Structures}

In TMD materials, the vacancies
of chalcogen atoms are the most frequently occurring defects \cite{komsa2015native,nguyen20183d}.
Compared to the hexagonal family of TMDs such as MoS$_{2}$, defect formation in~\emph{p}-PdTe$_{2}$ is much more sensitive due
to its structural characteristics, which further results in distinct
optoelectronic properties~\cite{lin2017novel}. In this paper, we have studied the most common native defects, i.e., vacancies in the \emph{p}-PdTe$_{2}$ monolayer, which include mono-vacancies, bi-vacancies, co-vacancies, and vacancy complexes, i.e., palladium (V$_{Pd}$), tellurium (V$_{Te}$), tellurium+tellurium (V$_{Te+Te}$), palladium+tellurium (V$_{Pd+Te}$), and palladium+4tellurium
(V$_{Pd+4Te}$). The calculations were performed using
a 4$\times$4$\times$1 supercell, which we found to be large enough
to consider all types of defect configurations. Although the inclusion
of SOC made our calculations extremely time consuming and computationally
quite costly, we believe that it is important to include
it because of the heavy atoms constituting the monolayer.

\begin{figure}[ht]
\includegraphics[width=1\linewidth]{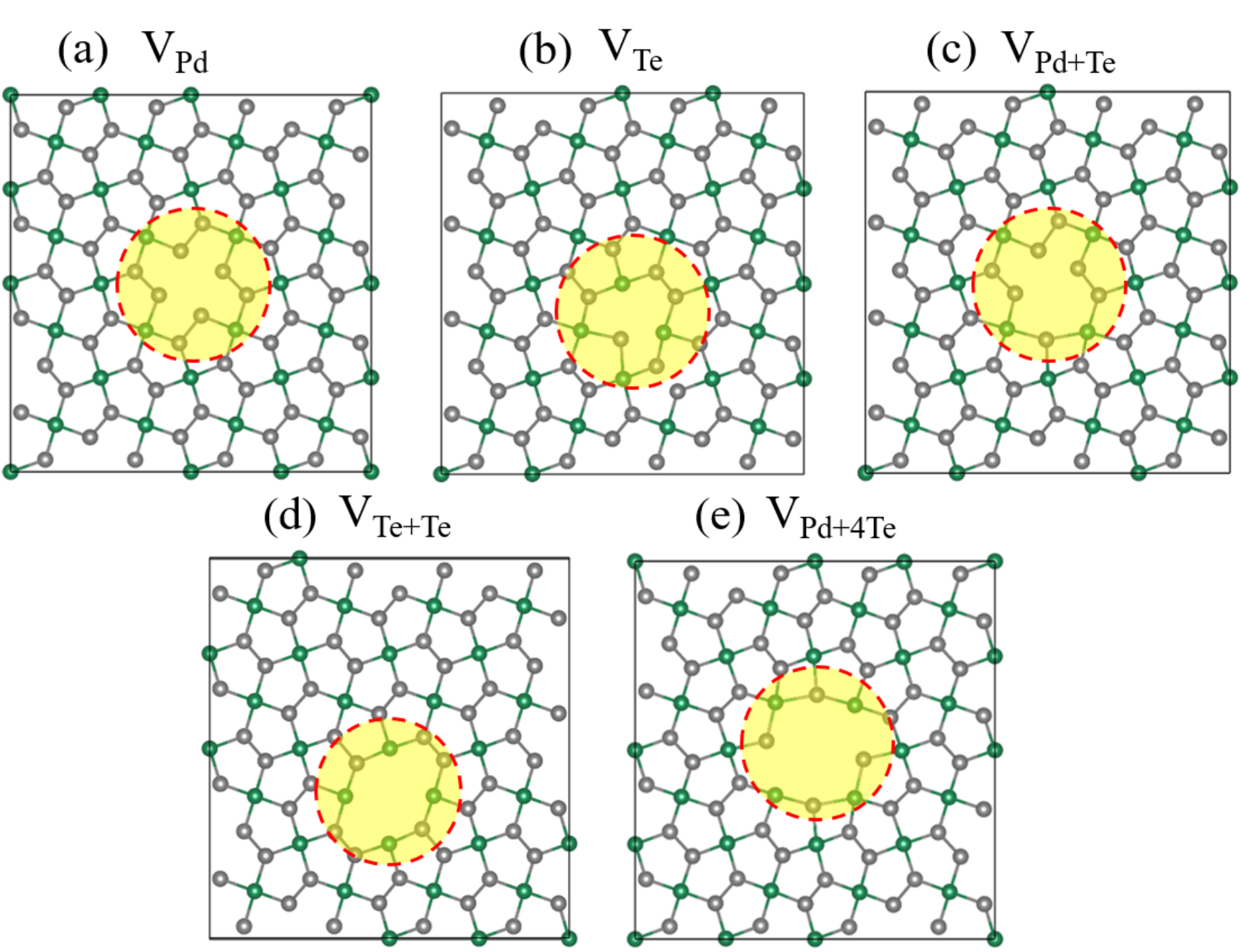}\caption{The optimized 4$\times$4$\times$1 supercells of the \emph{p}-PdTe$_{2}$
monolayer with (a) a single Pd vacancy (V$_{Pd}$), (b) Te vacancy
(V$_{Te}$), (c) Pd+Te vacancy (V$_{Pd+Te}$), (d) Te+Te vacancy (V$_{Te+Te}$),
and (e) Pd+4Te vacancy (V$_{Pd+4Te}$). The highlighted circles indicate
the defect region.}
\label{fig:defect-config} 
\end{figure}

Figs.~\ref{fig:defect-config}(a)-(e) show the optimized structures
corresponding to different considered vacancies. It is evident from
the highlighted circles in Figs.~\ref{fig:defect-config}(a)-(e)
that in the vicinity of the defects, the crystal lattice shows local
distortion.
The different considered defect configurations in
the \emph{p}-PdTe$_{2}$ monolayer
reduce its symmetry from the monoclinic point group ($C_{2h}$) to
the triclinic point group symmetry. The Pd vacancy position can be
chosen such that the center of inversion is preserved in the system.
The symmetry of the system reduces to the $C_{i}$ point group due
to the presence of V$_{Pd}$. However, Te vacancy removes the center
of inversion symmetry, and due to V$_{Te}$, the system acquires the
$C_{1}$ point group. If we remove an even number of Te atoms, we
can preserve the center of inversion symmetry. Therefore, V$_{Te+Te}$
and V$_{Pd+4Te}$ again belong to the $C_{i}$ point group. In V$_{Pd+Te}$,
the single Te vacancy reduces the symmetry to the $C_{1}$ point group.

\subsubsection{Formation Energies of Neutral Vacancies}
Next, we computed the formation energies of the considered vacancies in the contexts of rich in Pd (Te-poor) and
rich in Te (Pd-poor) conditions. In these scenarios, $\mu_{Pd}$ and
$\mu_{Te}$ (i.e., Pd and Te chemical potentials) vary, depending
on the level of abundance of Pd and Te under the experimental conditions.
Chemical potential is a growth-dependent parameter that determines
the system's off stoichiometry and the value of which is never greater than
that of the pure condensed form of the respective element~\cite{martin_2004}.
Using equilibrium thermodynamics, the upper and lower bounds of $\mu_{Pd}^{PdTe_{2}}$
and $\mu_{Te}^{PdTe_{2}}$ can be determined as follows~\cite{zheng2006native,kohan2000first,zhang2001intrinsic}:

\begin{equation}
\mu_{Pd}^{bulk}+\Delta E_{\emph{f}}^{PdTe_{2}}<\mu_{Pd}^{PdTe_{2}}<\mu_{Pd}^{bulk}.\label{equ:equpd}
\end{equation}
\begin{equation}
\mu_{Te}^{bulk}+\frac{1}{2}\Delta E_{\emph{f}}^{PdTe_{2}}<\mu_{Te}^{PdTe_{2}}<\mu_{Te}^{bulk}.\label{equ:equTe}
\end{equation}

\begin{table}[ht]
\centering %\setlength{\tabcolsep}{2pt}
\caption{\label{tab:table1} Formation energies (eV) of various vacancies in
monolayer~\emph{p}-PdTe$_{2}$ calculated under Pd-rich and Te-rich
conditions.}

\begin{ruledtabular}
\begin{tabular}{cccccc}
Vacancy types$\rightarrow$  & V$_{Pd}$  & V$_{Te}$  & V$_{Pd+Te}$  & V$_{Te+Te}$  & V$_{Pd+4Te}$\tabularnewline
\hline 
Pd-rich  & 1.65  & 1.60  & 2.40  & 3.29  & 4.58\tabularnewline
Te-rich  & 0.95  & 1.95  & 2.05  & 3.98  & 5.28\tabularnewline
\end{tabular}
\end{ruledtabular}

\label{Table:table1} 
\end{table}

\begin{figure}[ht]
\includegraphics[width=0.9\linewidth]{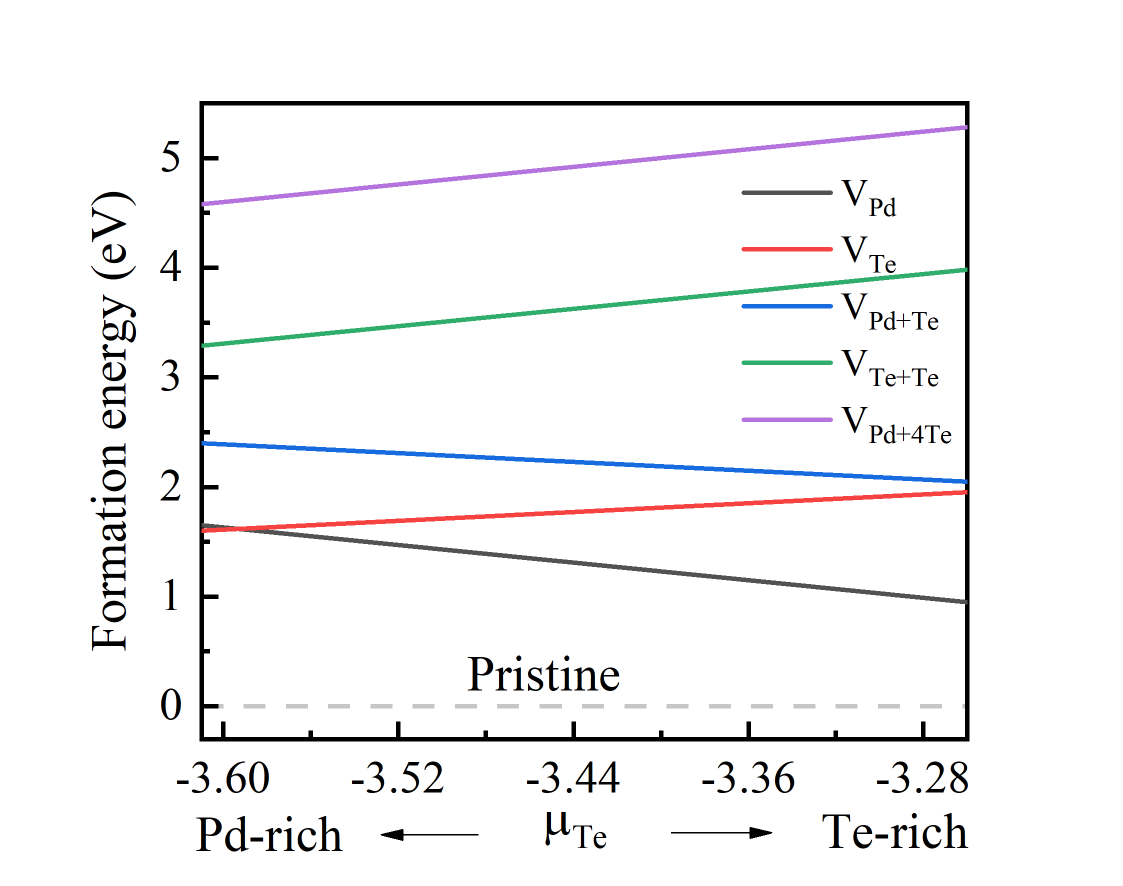}\caption{Formation energy corresponding to different neutral vacancies in the~\emph{p}-PdTe$_{2}$ monolayer with respect to the Te chemical potential ($\mu_{Te}$) in Pd-rich and Te-rich conditions. The SOC is incorporated for computing energies in different vacancies.}
\label{fig:figform} 
\end{figure}

Using Eqs.~\ref{equ:equpd} and \ref{equ:equTe}, $\mu_{Pd}^{PdTe_{2}}$($\mu_{Te}^{PdTe_{2}}$) in the Pd-rich conditions is -5.26 (-3.57) eV, and in the Te-rich conditions, it is -5.89 (-3.26) eV. Fig.~\ref{fig:figform} shows the
formation energy plot (computed using Eq.~\ref{eq:eqform}) for different
vacancies as a function of Te chemical potential in Pd-rich and Te-rich
conditions. The positive values of all the formation energies indicate
that the defective configurations are thermodynamically less stable
as compared to the pristine~\emph{p}-PdTe$_{2}$ monolayer. It is
clear from Fig.~\ref{fig:figform} (Table~\ref{Table:table1}) that
V$_{Te}$ and V$_{Pd}$ are the most dominant defects in the monolayer
of \emph{p}-PdTe$_{2}$ because of their comparatively smaller formation
energies. Under the Pd-rich condition, V$_{Te}$ costs the minimum energy of 1.60 eV; while as we move toward the Te-rich condition, V$_{Pd}$ starts dominating and costs the minimum energy of 0.95 eV.
Previously, Noh~\emph{et al.} reported that in monolayer MoS$_{2}$,
the formation energy of S vacancy in Mo-rich conditions lies in the
range of 1.3-1.5 eV~\cite{noh2014stability}. In the PdSe$_{2}$
monolayer, the formation energy of Se vacancy in Pd-rich conditions
is reported to be $\sim$ 1.5 eV~\cite{nguyen20183d}. Moreover,
the formation energy of O vacancy in MoO$_{3}$ in Mo-rich conditions
was found to be 1.69 eV in our earlier work~\cite{sharma2023influence}.
Thus, in the present paper, the computed formation energy for the Te
vacancy (V$_{Te}$) is found to be within the typical range of formation
energies of anions predicted for other chalcogenides \cite{noh2014stability,nguyen20183d,sharma2023influence,kuklin2021point}.
Further, the defect complex, i.e., V$_{Pd+4Te}$, shows high formation
energy among all considered vacancies for the entire range of Te chemical
potential. On the other, co-vacancy (V$_{Pd+Te}$) and bi-vacancy
(V$_{Te+Te}$) cost less formation energy as compared to the defect
complex (V$_{Pd+4Te}$). Moreover, we find that V$_{Pd+Te}$ needs
less formation energy than the sum of the formation energies of the
two independent mono-vacancies V$_{Te}$ and V$_{Pd}$, indicating
strong interactions between the two, with the ability to fuse and
produce the co-vacancy, whereas tellurium bi-vacancy, i.e., V$_{Te+Te}$,
has a larger formation energy than the sum of the two independent
tellurium vacancies. Consequently, V$_{Te}$ will be more likely to
form and less likely to fuse with another V$_{Te}$. Previously, Horzum
\emph{et al.} also reported that sulfur mono-vacancy (V$_{S}$) distribution
in the ReS$_{2}$ monolayer is more favorable as compared to sulfur
bi-vacancy~\cite{horzum2014formation}. Quite interestingly, this
observed behavior in TMD systems is opposite to that observed in graphene,
in which C vacancies prefer to form vacancy complexes instead of staying
isolated.

\subsubsection{Density of States and Induced Magnetism}
\begin{figure*}
\includegraphics[width=1\linewidth]{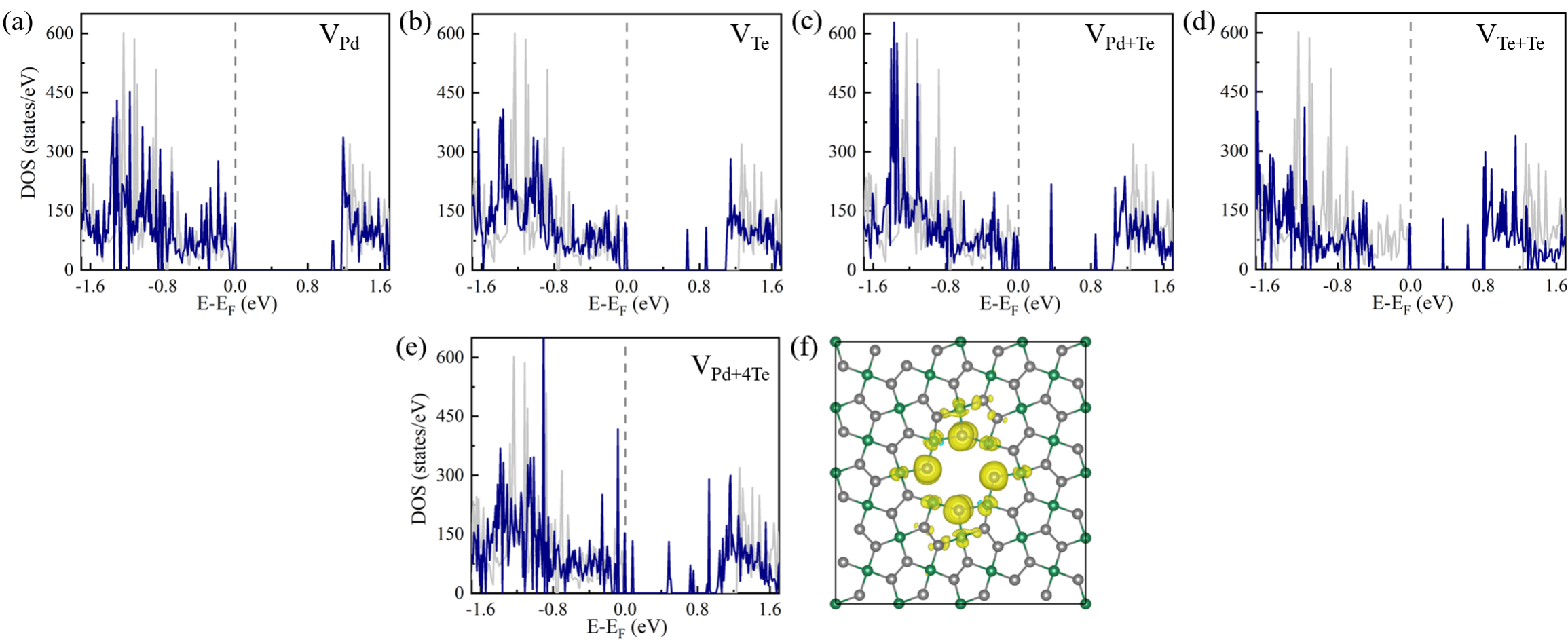}\caption{The total density of states (TDOS) corresponding to (a) V$_{Pd}$,
(b) V$_{Te}$, (c) V$_{Pd+Te}$, (d) V$_{Te+Te}$, and (e) V$_{Pd+4Te}$
vacancies in the~\emph{p}-PdTe$_{2}$ monolayer. The gray color
in the background represents the TDOS corresponding to the pristine
4$\times$4$\times$1 supercell of the~\emph{p}-PdTe$_{2}$ monolayer.
(f) The spin density plot corresponding to V$_{Pd+4Te}$ for the isosurface
value of 0.00042 e/Å$^{3}$. The spin-density isosurface shown in
yellow is mainly concentrated around the Te ions near the vacancy.
The SOC is incorporated in the calculations.}
\label{fig:tdos-defect} 
\end{figure*}
It is a well known fact that vacancies lead to the emergence of additional
states in the band gap region, which can act as electron traps and
thus modify the system’s properties remarkably. One way to understand these midgap states in defective systems is by examining their TDOS, which also allows one to interpret the transmission
electron microscopy and scanning tunneling microscopy based experiments~\cite{borner2016evidence,ugeda2010missing}.
Therefore, we next discuss the calculated TDOS for various defective
configurations of the~\emph{p}-PdTe$_{2}$ monolayer presented in
Figs.~\ref{fig:tdos-defect}(a)-(e). In
the V$_{Pd}$ case, the TDOS is not significantly altered compared
to the TDOS of the pristine system. However, one defect state appears
0.08 eV below the conduction band edge, implying that the system will
behave like an n-type semiconductor. Additionally, the band gap reduces to 1.17 eV in this case. In the V$_{Te}$ case, the band gap slightly reduces to 1.09 eV, with one impurity state 0.21 eV below
the CBM, while another one is 0.42 eV lower than it, indicating that the latter is a deep impurity level. Notably, no impurity states are found
near the VBM, and the system will again behave like an n-type semiconductor.
In the case of V$_{Pd+Te}$, the band gap reduces to 1.03 eV, with
two defect states appearing in the mid-gap region. One of them is
0.17 eV below the CBM, which will act like an electron donor, leading
to n-type doping. The other one is a deep impurity level located almost
in the middle of the gap. For V$_{Te+Te}$, the band gap exhibits
a very small reduction to 1.22 eV as compared to the pristine case; however, the VBM and CBM exhibit significant downward shifts. Moreover, three major impurity states appear in the midgap region, two of which are deep states; one is an occupied state, and the other one is a vacant state. Moreover, the third state is 0.15 eV below the CBM (unoccupied), which will again lead to n-type doping. In the defect complex, i.e., V$_{Pd+4Te}$, the band gap is reduced to 1.00 eV, with the CBM moving downwards and the VBM staying unchanged. Numerous impurity levels appear in the midgap region, out of which several are deep ones. However, one vacant impurity level is just 0.05 eV above the VBM, while towards CBM, one state is found to be 0.07 eV below it. Thus, this system will exhibit mixed doping.

To understand the influence of different functionals on the band edges, in Fig. S3 of SM~\cite{SI}, we present the locations of the VBM and CBM of the pristine and defective configurations of the \emph{p}-PdTe$_{2}$ monolayer that were computed using PBE+SOC and HSE06+SOC approaches. From the obtained results, we note the following trends:\\ 
(a) As far as the position of the VBM is concerned, for all the cases, the VBMs computed using the HSE06+SOC method are lower as compared to those calculated using the PBE+SOC approach.\\ 
(b) Except for V$_{Pd}$ and V$_{Pd+4Te}$,  for all other cases, the CBMs obtained from HSE06+SOC calculations are higher compared to PBE+SOC results. \\
Consequently, and as expected, the band gaps obtained using the HSE06+SOC approach are significantly wider compared to the PBE+SOC ones. Moreover, a comparison of band gaps computed using the two approaches is mentioned in Table S1 of SM~\cite{SI}.

Additionally, the system maintains its non-magnetic behavior in all considered vacancies except for the V$_{Pd+4Te}$, for which a total magnetic moment of 1.87~$\mu_{B}$ is induced. Furthermore, it is found that the nearby Te atoms contribute the most to the magnetic moment as compared to the Pd atoms near V$_{Pd+4Te}$. Out of 1.87~$\mu_{B}$
of magnetic moment, the neighborhood Te atoms contribute 65$\%$, while the remaining 35$\%$ contribution is obtained from the nearest Pd atoms. For the sake of more clarity, we also plotted the TDOS corresponding to all considered vacancies without including SOC and presented in Figs. S4(a)-(e) of the SM~\cite{SI}. Even without SOC, similar behavior is seen; i.e., for all the considered cases, a symmetry between up-spin and down-spin states is found except for the vacancy complex V$_{Pd+4Te}$, for which the induced magnetism is preserved (see Fig. S4(e) of the SM~\cite{SI}), but with a slightly larger magnetic moment of
2.0~$\mu_{B}$.

For a better understanding of the contribution of different atoms
to the magnetic moment, we next performed the spin density analysis
corresponding to the V$_{Pd+4Te}$. The area covered by the spin density
isosurface provides a qualitative picture of the contribution of various
atoms to the magnet moment. Fig.~\ref{fig:tdos-defect}(f) shows
the real-space effective spin density plot of V$_{Pd+4Te}$ obtained
by taking the difference between up-spin ($\rho_{\uparrow}$) and
down-spin ($\rho_{\downarrow}$) channels, i.e., $\Delta\rho=\rho_{\uparrow}-\rho_{\downarrow}$
for the isovalue of 0.00042 e/Å$^{3}$. As one can see, the spin density isosurface is primarily concentrated around the Te atoms, indicating the major contribution of Te atoms to the magnetic moment. Also, the
spin-density isosurface shows a dumbbell-like shape, suggesting that
the magnetic moment in the system is mostly induced by the $p$ orbitals
of Te atoms.
\subsubsection{Formation Energies of Charged Vacancies}

\label{sec:formation-energy-charged-vacancies}

\begin{figure}[ht]
\includegraphics[width=1\linewidth]{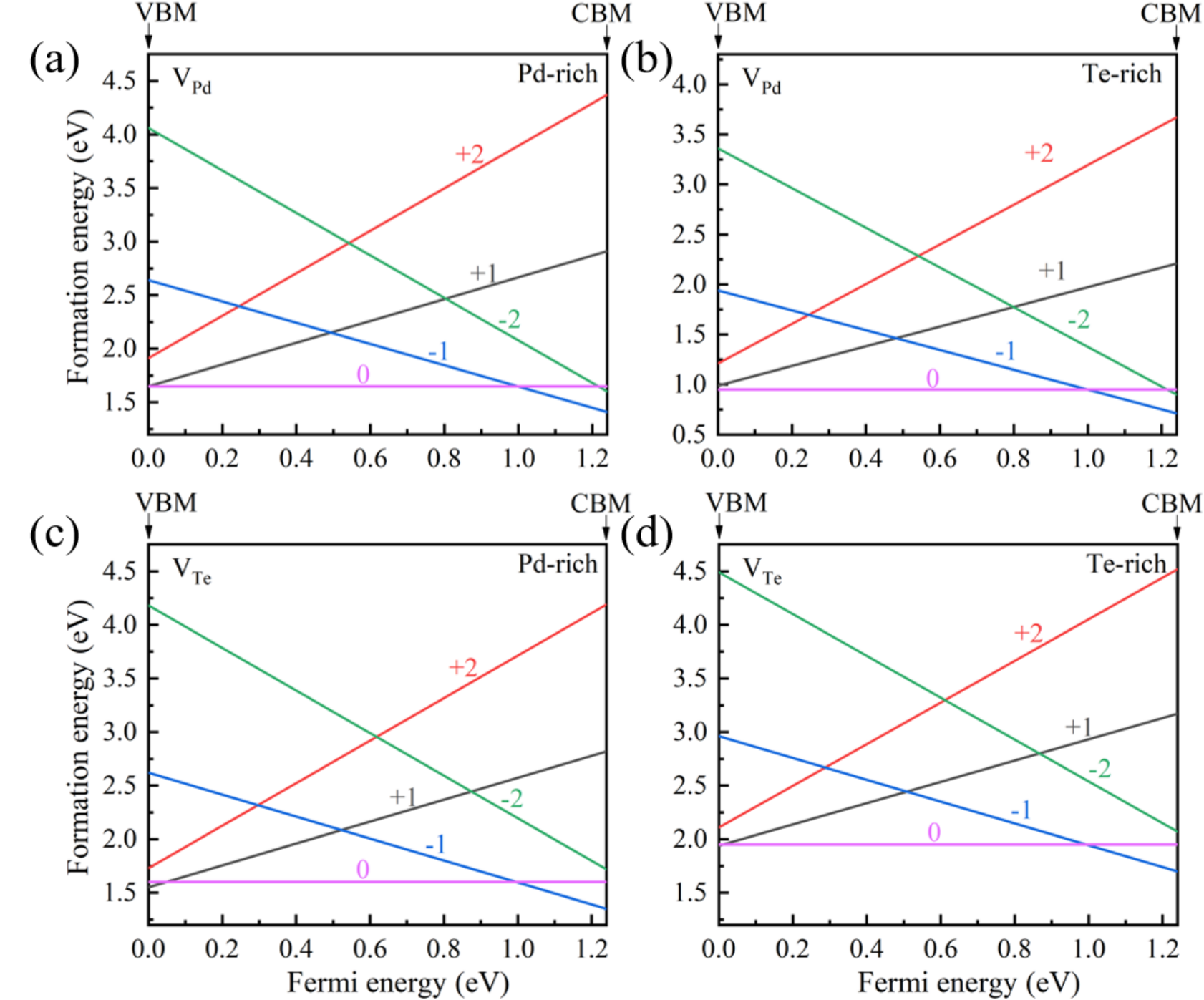}\caption{Formation energy plots of V$_{Pd}$ and V$_{Te}$ in different considered
charged states with respect to Fermi energy in Pd-rich and Te-rich
conditions. The SOC is incorporated in the calculations.}
\label{fig:vpd-vte-form-energy} 
\end{figure}

In addition to the neutral vacancies, charged vacancies are also frequently
found in materials in general~\cite{RevModPhys.86.253}, and TMDs,
in particular~\cite{tan2020stability,komsa2012finite}. Experimentally, the STM technique has been used not only to trigger the migration of defects but also to charge the neutral vacancies and bring them back to the neutral states again~\cite{nguyen20183d,fu2019defects,zheng2013manipulation,hla2000inducing}.
Therefore, in addition to the neutral vacancies, we also considered various charged vacancies in the~\emph{p}-PdTe$_{2}$ monolayer. We mainly considered the different charged states for V$_{Pd}$ and V$_{Te}$ because they are found to be the most prominent defects in the \emph{p}-PdTe${_{2}}$ monolayer, as discussed earlier. We considered four charged states for each considered case, i.e., +1, +2, -1, -2, and further computed the formation energy corresponding to each charged defect configuration. We adopted a supercell approach for calculating charged defect formation energy
in the dilute limit \cite{komsa2013finite,wang2015determination,komsa2018erratum}. In the supercell approach, which has proven to estimate the formation energy accurately, the cell dimensions $a$, $b$, and $c$ are scaled concomitantly in such a way that the dimensions of the supercell in the three directions $L_{a}$, $L_{b}$, and $L_{c}$ are approximately
equal \cite{komsa2012finite,komsa2013finite}. The expression $\Delta$$E_{f}^{(q)}=E_{defect}^{(q)}-E_{pristine}^{(0)}+\sum_{i}\Delta n_{i}\mu_{i}+q(E_{VBM}+E_{F})$ has been used in Pd-rich and Te-rich conditions as a function of the Fermi energy ($E_{F}$) for the charged defect formation energy calculation, which is similar to Eq. \ref{eq:eqform} except for some additional corrections due to consideration of the charged state. We denote the charged state of a vacancy by $q$, where $q>(<)$ 0 implies the electron removal (electron addition). Here, the value of the Fermi energy $E_{F}$ is defined with respect to the energy of the VBM ($E_{VBM}$) of the pristine \emph{p}-PdTe$_{2}$ monolayer, and its value varies between $E_{VBM}$ to $E_{CBM}$, i.e., in the band gap region. From the obtained results, the formation energies are fitted to the formula $c_{0}+c_{1}/L+c_{2}/L^{3}$ \cite{komsa2012finite,komsa2013finite}, where $c_{0}$, $c_{1}$, and $c_{2}$ are constants, and $L$ is the supercell dimension. Next, the limit $L\rightarrow\infty$ is taken (see Fig. S5 of the SM~\cite{SI}), leading the value of the formation energy to be $c_{0}$ in the infinite supercell limit, which is presented in Fig.~\ref{fig:vpd-vte-form-energy} and Table~\ref{Table:table2}). It is clear from Fig.~\ref{fig:vpd-vte-form-energy} that for both V$_{Pd}$ and V$_{Te}$, negatively charged states require less energy and are more stable as compared to both the positively charged vacancies and the neutral ones, towards the conduction band, irrespective of whether the conditions are Pd rich or Te rich. Moreover, with the decrease in the Fermi energy, i.e., towards the valence band, the formation energies of the positively charged states decrease but still are less stable than the neutral states. In particular, in both V$_{Pd}$ and V$_{Te}$, irrespective of Pd-rich and Te-rich conditions, the formation energies of V$_{Pd}^{1-}$ and V$_{Te}^{1-}$ are negative, forming the most stable state when $E_{F}>$ 1.0 eV (towards the CBM). Moreover, in both Pd-rich and Te-rich conditions, V$_{Pd}^{2-}$ shows more stability almost at the CBM edge, i.e., when $E_{F}>$ 1.22 eV, whereas in the case of V$_{Te}^{2-}$, the neutral state remains most stable. As we move toward the lower value of the Fermi energy (toward the VBM), positively charged states show more stability as compared to the negatively charged states but are less stable than the neutral states in the case of V$_{Pd}$, irrespective of Pd-rich and Te-rich conditions. In the V$_{Te}$ case, in the Pd-rich conditions, the formation energy of the +1 charged state turns out to be more stable for a small energy range ($E_{F}<$ 0.05 eV) close to the valence band edge, while in Pd-rich conditions,  again the neutral state forms the most stable state. Therefore, based on these results, we can expect that even for other considered vacancies in the \emph{p}-PdTe$_{2}$ monolayer close to the conduction band, the negatively charged states will be more stable compared to the neutral and positively charged ones.

%\scriptsize

\begin{table}
\centering %\setlength{\tabcolsep}{2pt}
\scriptsize
\caption{Formation energy values (eV) of charged vacancies in~\emph{p}-PdTe$_{2}$
monolayer calculated under Pd-rich and Te-rich conditions.\label{Table:table2}}
\begin{tabular}{cccccccccccc}
\toprule 
Defect &  & \multicolumn{4}{c}{V$_{Pd}$} &  & \multicolumn{4}{c}{V$_{Te}$} \tabularnewline
\midrule 
charge on defect &  & +1 & +2 & -1 & -2 & & & +1 & +2 & -1 & -2 \tabularnewline
\midrule 
Pd-rich & VBM & 1.65 & 1.91 & 2.64 & 4.06 & & & 1.55 & 1.73 & 2.62 & 4.18 \tabularnewline
 & CBM & 2.91 & 4.37 & 1.41 & 1.6 & &  & 2.82 & 4.19 & 1.35 & 1.72 \tabularnewline
Te-rich & VBM & 0.99 & 1.21 & 1.94 & 3.36 & & & 1.94 & 2.11 & 2.96 & 4.49 \tabularnewline
 & CBM & 2.21 & 3.67 & 0.71 & 0.9 & & & 3.17 & 4.52 & 1.7 & 2.07 \tabularnewline
\bottomrule
\end{tabular}
\end{table}

\subsubsection{Diffusion Energy Barriers for Neutral Mono-vacancies}

\begin{figure}[ht]
\includegraphics[width=1\linewidth]{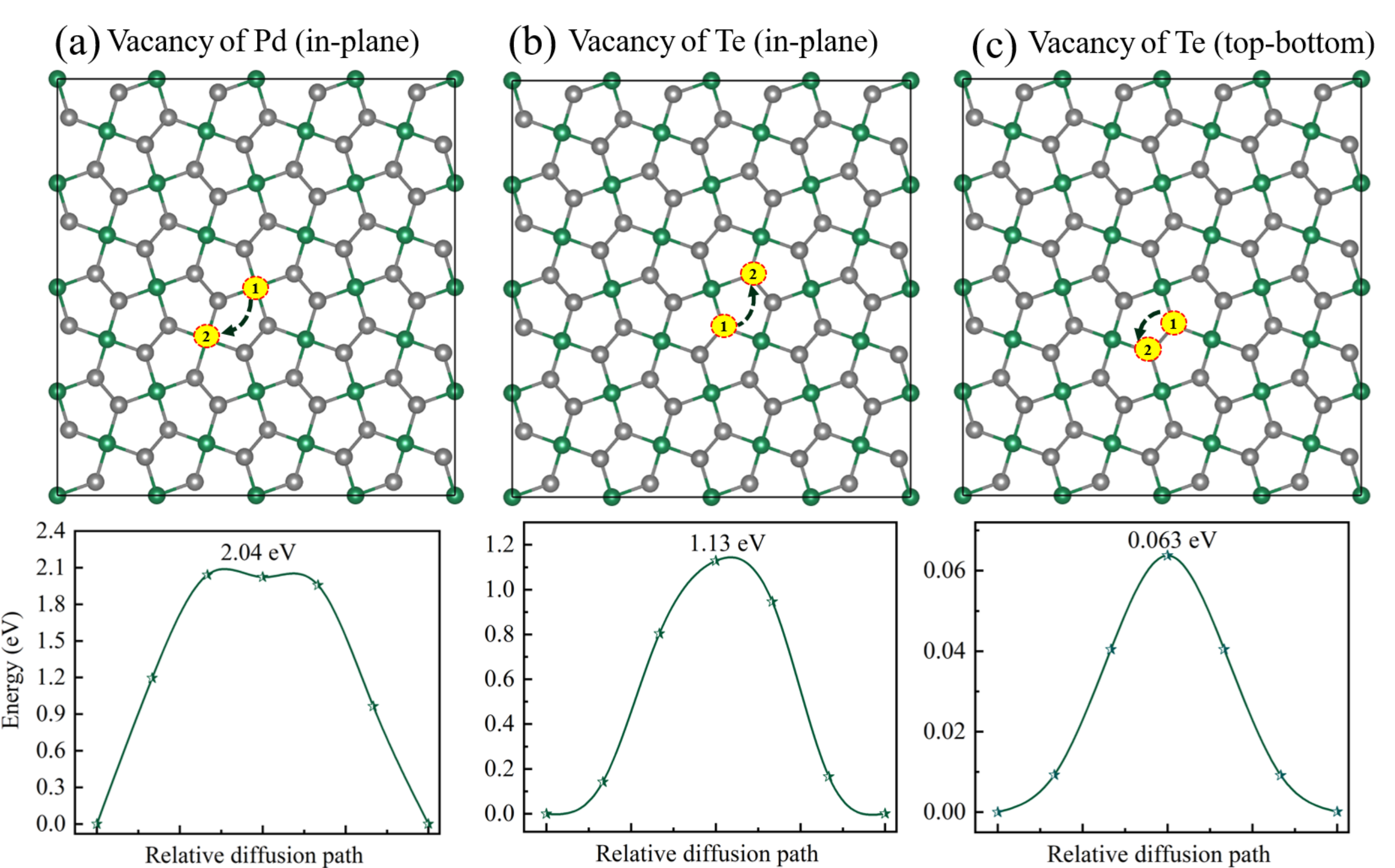}\caption{Top panels show the initial (1) and final (2) positions of vacancy diffusion corresponding to (a) V$_{Pd}$, (b) V$_{Te}$ (in-plane), and (c) V$_{Te}$ (top-bottom). Bottom panels present
the potential energy curves for the corresponding diffusion paths,
along with their diffusion energy barriers. The SOC is included in
the calculations. \label{fig:climage}}
\end{figure}

An important aspect of the manipulation of point defects is our ability to trigger their migration in a controlled manner using, for example, the STM technique. The tendency of migration of defects, including vacancies, is largely controlled by their diffusion energy barriers, which we will discuss next for mono-vacancies V$_{Pd}$
and V$_{Te}$. We have restricted our discussion of diffusion energy barriers only to mono-vacancies because the migration
of vacancy complexes through the lattice is expected to be less favorable. We compute the diffusion energy barriers of the monovacancies using the climbing image nudged elastic band (CI-NEB) approach \cite{henkelman2000climbing}, in which an appropriate number of intermediate images between the initial and final configurations are chosen, thus defining a reaction
path. In our case, initial and final configurations are nothing but the initial and final locations of the vacancy in question, while the reaction path is defined by an appropriately chosen (by us) set of intermediate locations of the vacancy within the supercell, as shown in the top panels of Figs.~\ref{fig:climage}(a-c) for V$_{Pd}$
and V$_{Te}$. Next, using the CI-NEB method, optimization of these intermediate states is performed in order to calculate the saddle points and energy minima along that path for a given specific configuration.
In the case of V$_{Pd}$, the reaction path is shown in Fig.~\ref{fig:climage}(a) top panel, while the corresponding calculated potential energy curve is given in the bottom panel. Our calculations predict a high diffusion energy barrier of 2.04 eV for V$_{Pd}$, thereby making its migration quite improbable at room temperature. For V$_{Te}$, we considered two possible cases for initial and final positions, i.e., in-plane and top-bottom, as shown in the top panels of Figs.~\ref{fig:climage}(b) and ~\ref{fig:climage}(c). In the top-bottom case, the initial position of Te vacancy is considered in the top sublayer, and the final position is in the bottom sublayer, while for the in-plane case, they are in
the same sublayer. Our calculations predict an in-plane energy barrier of 1.13 eV for the V$_{Te}$, which is much lower than the sulfur vacancy energy barrier of 2.30 eV in MoS$_{2}$~\cite{komsa2013point}.
This lower energy barrier in the \emph{p}-PdTe$_{2}$ is due to comparatively weaker covalent bonds between Te atoms, as
compared to S atoms in the hexagonal MoS$_{2}$, as originally argued by Komsa \emph et al.~\cite{komsa2013point}. A similar
study of diffusion energy barriers for $V_{Se}$ was undertaken by Nguyen \emph et al. in the \emph{p}-PdSe$_{2}$
monolayer~\cite{nguyen20183d}. The energy barrier for the same in-plane migration of V$_{Se}$ in \emph{p}-PdSe$_{2}$ was computed to be 1.00 eV \cite{nguyen20183d}, which is in good agreement with our value of the V$_{Te}$ barrier (1.13 eV). In the V$_{Te}$ top-bottom case, a very small barrier of 0.063 eV is obtained, which is roughly double the corresponding barrier of 0.03 eV for V$_{Se}$ in the \emph{p}-PdSe$_{2}$ monolayer \cite{nguyen20183d}.
Therefore, based on the calculated diffusion energy barriers, we conclude the following: \\
(a) The barrier for V$_{Pd}$ migration at 2.04 eV is very high, implying that its controlled migration will be relatively difficult to achieve.\\
(b) The diffusion energy barrier for V$_{Te}$ in-plane migration is 1.13 eV, which is significantly
lower, and therefore, its controlled migration will be much easier by using the STM. \\
(c) The smallest diffusion energy barrier of 0.063 eV is obtained for V$_{Te}$ migration across the top-bottom plane.
The small value of the top-bottom diffusion energy barrier corresponds to twice the room temperature; therefore, its spontaneous thermal activation at room temperature is highly unlikely. However, given its very low value, this across-the-plane migration can be easily achieved using an STM tip.

In order to obtain better clarity, \emph{ab initio} molecular dynamics (AIMD) simulations are performed for V$_{Te}$ for a time period of 5 ps at 400 K, and the corresponding result is presented in Fig. S6 of the SM~\cite{SI}. A difference in the initial and final structures with V$_{Te}$ migration from position 1 to position 2, as indicated in Fig.~\ref{fig:climage}(c), is noticed. Therefore, our AIMD simulations reveal that spontaneous thermally induced diffusion is possible at temperatures higher than the room temperature in the V$_{Te}$ case in the top-bottom (vertical) direction due to a very small diffusion energy barrier in that direction.

Moreover, we further analyzed the small diffusion barrier for V$_\text{Te}$ in the top-bottom direction (0.063 eV) using the Arrhenius equation~\cite{WANG2001111}
\begin{equation}
\nu = \nu_0 e^{-\frac{E_a}{kT}}, 
\end{equation}
where $\nu$ is the hopping rate, $\nu_0$ is the pre-exponential factor (commonly associated with the Debye frequency and typically of the order of $10^{12}$/sec for impurities in crystals), $E_a$ is the diffusion energy barrier, $k$ is the Boltzmann constant, and $T$ is the absolute temperature. At room temperature ($kT \approx 25$ meV), this yields a hopping rate ($\nu$) of approximately $10^{11}$/sec, corresponding to a characteristic timescale of $\sim 10$ ps for individual hopping events. The absence of hopping events in our AIMD simulations is due to the fact that the simulation time spans only 5 ps, which is half of the characteristic timescale (10 ps) calculated above for these hopping events. On experimental timescales, such as STM scanning ($\sim$ 1 sec), the observed configuration would likely appear as a time-averaged superposition of the top and bottom configurations, making static DFT predictions less relevant at room temperature. Distinct top or bottom configurations could only be observed experimentally at significantly lower temperatures, such as those achievable using liquid helium, for which the hopping rate would be sufficiently reduced to trap individual states.

\subsubsection{Diffusion Energy Barriers of Charged Mono-vacancies}

\begin{figure}[ht]
\includegraphics[width=1\linewidth]{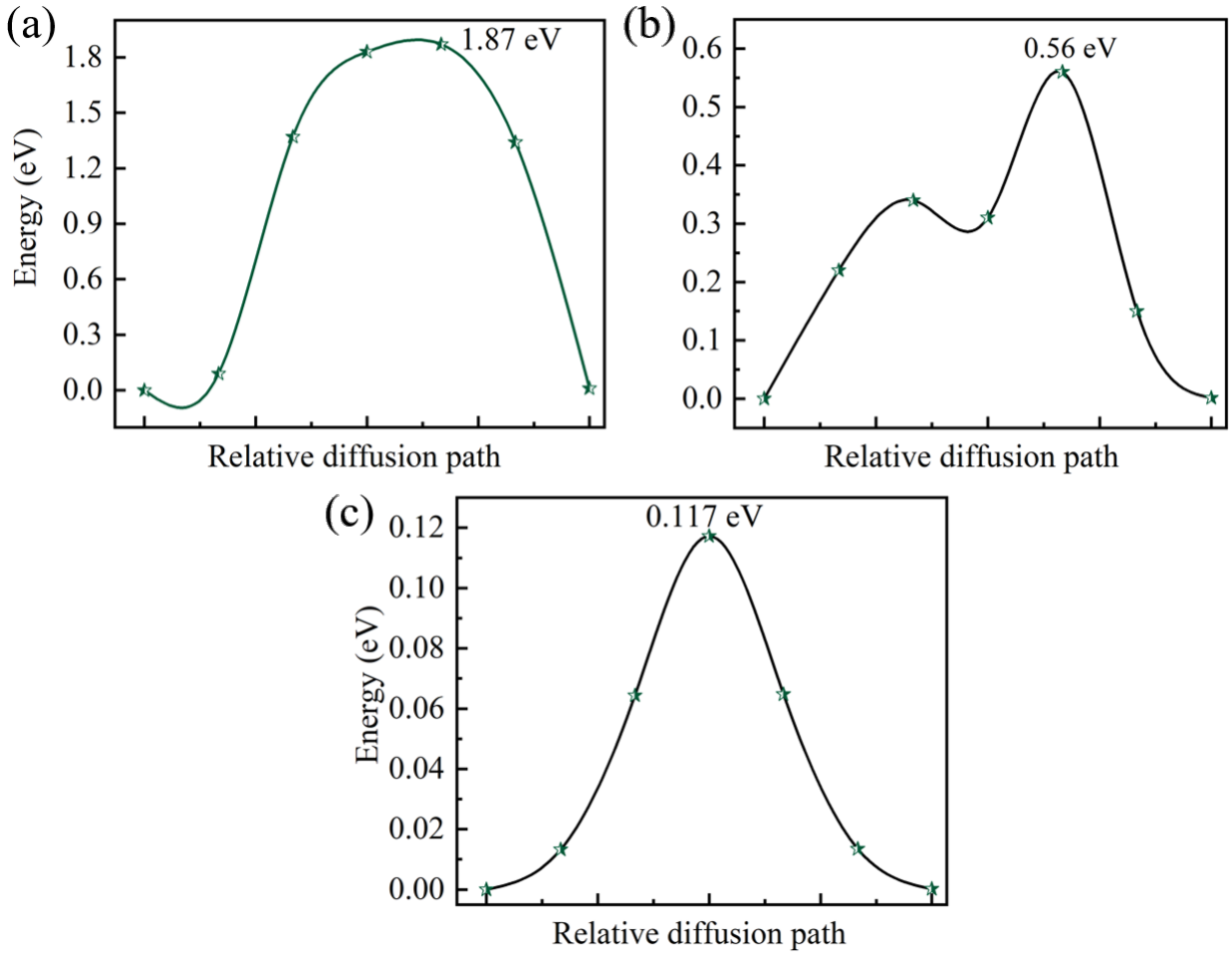}\caption{Diffusion energy barriers of (a) V$_{Pd}^{1-}$, (b) in-plane V$_{Te}^{1-}$, and (c) top-bottom V$_{Te}^{1-}$. The SOC is included in the calculations.}
\label{fig:diff_charged_curve} 
\end{figure}

In Sec.~\ref{sec:formation-energy-charged-vacancies}
we reported that the negatively charged mono-vacancies V$_{Pd}^{1-}$
and V$_{Te}^{1-}$ are more stable as compared to their neutral and
positively charged counterparts towards the CBM for $E_{F}>$ 1.0
eV. Therefore, it is also of interest to study their migration through
the lattice by computing their diffusion-energy barriers. We applied the same methodology of CI-NEB to compute the energy barriers of the charged vacancies, and our results are shown in Fig.~\ref{fig:diff_charged_curve}.
We draw the following conclusions from our results: (a) for V$_{Pd}^{1-}$, a barrier height of 1.87 eV is obtained, which is lower than 2.04 eV, the height of its neutral counterpart V$_{Pd}$, (b) for the in-plane migration of V$_{Te}^{1-}$, the energy barrier is 0.56 eV, again roughly half of the corresponding barrier of V$_{Te}$, and (c) finally for the top-bottom migration of V$_{Te}^{1-}$, the barrier is 0.12 eV, which is roughly twice the corresponding barrier
for the neutral vacancy V$_{Te}$. We note that as far as these negatively charged mono-vacancies are concerned, especially for V$_{Te}^{1-}$ in both directions, their diffusion energies are relatively small and, thus, within the reach of STM-triggered controlled migration.

\begin{figure*}
\includegraphics[width=1\linewidth]{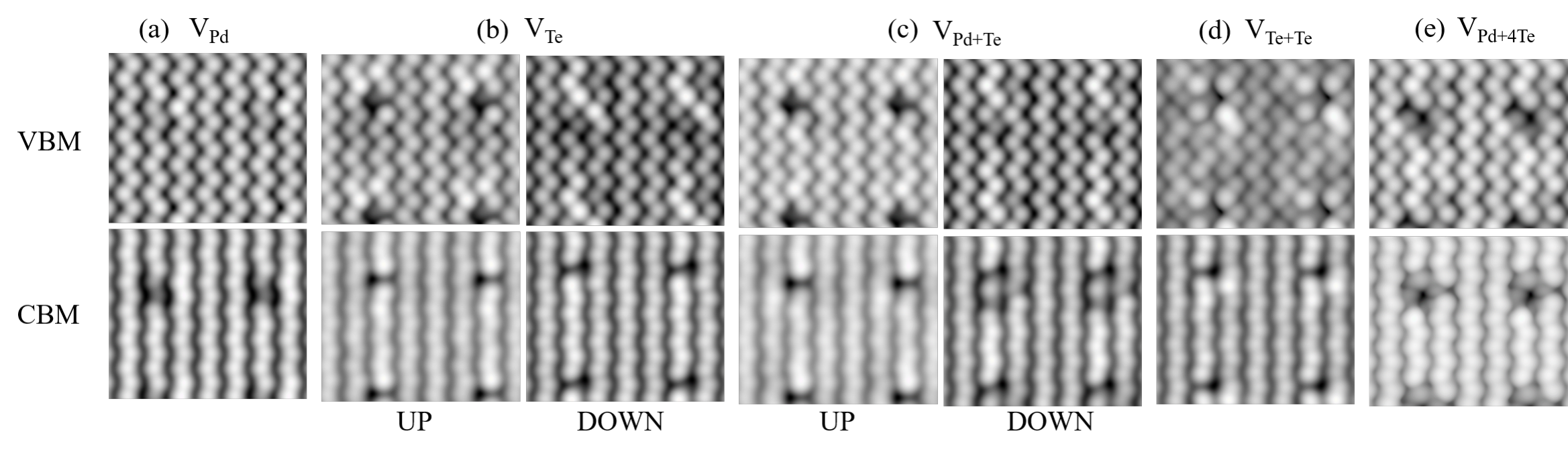}\caption{STM images of VBM and CBM of the~\emph{p}-PdTe$_{2}$ monolayer with
(a) Pd vacancy (V$_{Pd}$) with bias voltage Vs = -0.7 and 1.8 eV, (b) Te vacancy (V$_{Te}$) up and down views with Vs = -0.6 and 1.7 eV, (c) Pd+Te vacancy (V$_{Pd+Te}$) up and down views with Vs = -0.7 and 1.7 eV, (d) Te+Te-vacancy (V$_{Te+Te}$)
with Vs = -0.6 and 1.7 eV, and (e) Pd+4Te vacancy (V$_{Pd+4Te}$) with Vs = -0.5 and 1.7 eV with respect to the Fermi level in the 4$\times$4$\times$1 supercell, respectively. Up and down show the two different views of STM images.}
\label{fig:stm} 
\end{figure*}

\subsubsection{Creation of Vacancies Using Electron Irradiation}

We have studied the response of the~\emph{p}-PdTe$_{2}$ monolayer to electron irradiation with the aim of creating post-synthesis defects in a controlled manner. In this perspective, the knowledge of the displacement threshold energy $T_{d}$, i.e., the minimum energy required to introduce defects by electron irradiation, is crucial for understanding the interaction of the \emph{p}-PdTe$_{2}$ monolayer with the electron beam~\cite{krasheninnikov2010ion}. For calculating T$_{d}$, one can use density-functional tight-binding theory or AIMD simulations by assigning the initial velocity to the respective atom, as reported previously~\cite{PhysRevLett.109.035503,kotakoski2010electron,zobelli2007electron}. Moreover, Komsa~\emph{ et al.} found a close quantitative agreement between $T_{d}$ and the formation energy ($\Delta$$E_{f}$) of the corresponding vacancy, which is calculated using a non-relaxed supercell (also called "frozen-lattice approximation'') for a number of TMD materials~\cite{PhysRevLett.109.035503}. This is because of the fact that electron-nucleus collisions take place almost instantaneously during the electron beam irradiation so that a very small amount of energy gets transferred to the neighboring atoms due to the quick sputtering event and the structure's stiffness. In this scenario, certain atoms can be considered to have been immediately knocked out from the lattice. In the relaxed geometry, atomic relaxation causes a significant reduction in $\Delta$$E_{f}$ for some systems, losing its resemblance to $T_{d}$. 

\begin{table}[b]
\centering %\setlength{\tabcolsep}{2pt}
\caption{\label{tab:table3} Displacement threshold energies, $T_{d}$, in
eV, and the incident electron beam energies, $E_{beam}$, in keV corresponding
to neutral mono-vacancies in the~\emph{p}-PdTe$_{2}$ monolayer.
The SOC is included.}

\begin{ruledtabular}
\begin{tabular}{ccc}
Vacancy types$\rightarrow$  & V$_{Pd}$  & V$_{Te}$ \tabularnewline
\hline 
T$_{d}$  & 1.62  & 2.11 \tabularnewline
E$_{beam}$  & 73  & 111 \tabularnewline
\end{tabular}
\end{ruledtabular}

\label{Table:tebeam} 
\end{table}

Further, defect formation occurs when the maximum energy ($T_{max}$) transferred from the incident beam to the atomic nucleus is equal to or greater than $T_{d}$, i.e., $T_{max}$$\geq$$T_{d}$. Taking relativistic effects into consideration, $T_{max}$ is related to the incident beam energy $E_{beam}$ by the relation~\cite{garcia2014analysis,zobelli2007electron}
\begin{equation}
T_{max}=\frac{E_{beam}(E_{beam}+2m_{e}c^{2})}{E_{beam}+\frac{Mc^{2}}{2}\left(1+\frac{m_{e}}{M}\right)^{2}}.\label{equ:equtmax}
\end{equation}
Above, $M$ denotes the mass of the removed atom, $m_{e}$ is the rest mass of the electron, and $c$ is the speed of light. Eq.~\ref{equ:equtmax} enables the estimation of the $E_{beam}$ at which an atom is knocked out from its position within the lattice by the direct "knock-on" mechanism~\cite{zobelli2007electron}. In our paper, we adopted the
approach of Komsa~\emph{et al.}~\cite{PhysRevLett.109.035503} to calculate $T_{d}$. The calculated displacement threshold energy (T$_d$) and the corresponding incident beam energy ($E_{beam}$) for the considered mono-vacancies are shown in Table~\ref{Table:tebeam}. As far as complex vacancies involving multiple atoms are concerned, Eq. \ref{equ:equtmax} is not valid because it was derived for the removal of a single atom of mass $M$. In order to calculate the beam energies needed for creating multiple-atom vacancies, we need to take into account complex interactions among the involved atoms and vacancies with their surroundings, which is beyond the scope of this paper.

$T_{d}$ is found to be 1.62 eV for V$_{Pd}$, and 2.11 eV for V$_{Te}$. The corresponding values of $E_{beam}$ are computed to be 73 and 111 keV, respectively. We note that our calculated values of $T_{d}$ and $E_{beam}$
for creating chalcogen atom vacancy in~\emph{p}-PdTe$_{2}$ are
smaller than those corresponding to MoTe${_{2}}$~\cite{PhysRevLett.109.035503}.
It is found that the theoretical approach based on DFT generally overestimates
($\sim$10 $\%$) the incident beam energy values, irrespective of
irradiated material due to consideration of only ground-state dynamics
in the calculations~\cite{PhysRevLett.109.035503,zobelli2007electron}.
Further, Kretschmer~\emph{et al.} reported that with the involvement
of electronic excitations, defects can be created with energies lower
than those predicted from the knock-on process \cite{kretschmer2020formation}.
This implies that the vacancies in the~\emph{p}-PdTe$_{2}$ monolayer
can be created even at incident energies substantially lower than
those indicated in Table~\ref{Table:tebeam}, indicating the $p$-PdTe$_{2}$
monolayer to be a good candidate for defect creation using the electron-beam irradiation technique.

\subsubsection{Calculated STM Images}

Next, in Figs.~\ref{fig:stm}(a)-\ref{fig:stm}(e), we present our calculated STM
images for all the considered vacancies. For the vacancies that are
asymmetrically placed with respect to the~\emph{p}-PdTe$_{2}$ monolayer,
we present two different views, i.e., up and down, while for the ones
that are symmetrically placed, only one view is shown. The STM images
serve many purposes, such as phase identification, defect origin identification,
analysis of distortion induced due to lattice imperfection. As
discussed previously, in the pristine~\emph{p}-PdTe$_{2}$ monolayer,
the Te surface charge dominates (see Fig.~\ref{fig:fig1}(d)) both
for VBM and CBM. From our STM images (see Fig.~\ref{fig:stm}), it
is obvious that this trend holds for the defective configurations
as well. For the case of V$_{Pd}$ (see Fig.~\ref{fig:stm}(a)),
it is difficult to identify the Pd vacancy directly in the STM image
due to the dominant Te-Te surface charge. However, the nearby Te atoms
of V$_{Pd}$, which acquire more charge in the vacancy-induced charge
redistribution process, appear bright in the VBM (negative bias image) and the same look dark in the CBM (positive bias image). Further, in the case of V$_{Te}$ (see Fig.~\ref{fig:stm}(b)), dips in both
VBM and CBM in the up view of the STM images are noticed, whereas
some bright meteorite-like features are obtained in the down view
of VBM, while the same appears dark in the CBM. These bright spots
in the VBM are the Te atoms close to V$_{Te}$, which acquire more
charge (0.27 e) as compared to other Te atoms, as revealed by Bader
charge analysis. The STM images of V$_{Pd+Te}$ and V$_{Te+Te}$ (see
Fig.~\ref{fig:stm}(c) and Fig.~\ref{fig:stm}(d)) also reveal similar
features as V$_{Te}$, where bright atoms in the VBM represent nearby
Te atoms of the respective vacancy. For the case of V$_{Pd+4Te}$,
due to four missing Te atoms, tetragonal type of shades in the VBM
and a hole-like feature in the CBM are prominent. In summary, different
defect configurations show different kinds of shades in the STM images,
which will be helpful to the experimentalists in identifying the defect
types in this material.

\subsection{Optical Properties of Pristine and Defective Systems}
It is clear from the previous discussion that defects substantially
alter the electronic properties of the~\emph{p}-PdTe$_{2}$ monolayer.
This change can be seen in the optical properties as well, owing to
the fact that optical properties are closely related to the electronic
properties of the system. Therefore, in this section, the impact of vacancies on the optical properties of the~\emph{p}-PdTe$_{2}$ monolayer is discussed. First, the optical characteristics of the pristine \emph{p}-PdTe$_{2}$ monolayer are evaluated, followed by an investigation of how vacancies alter those characteristics. The frequency-dependent real ($\epsilon_{1}(\omega)$) and imaginary ($\epsilon_{2}(\omega)$) parts of the dielectric constant are extremely useful entities in determining various other optical properties of the system, such as optical absorption coefficient, optical conductivity, electron energy loss function, and extinction coefficient. At the photon energy of 0 eV, the real part of the dielectric constant corresponds to the static dielectric constant ($\epsilon_{1}$(0)).

\begin{figure}[ht]
\includegraphics[width=1\linewidth]{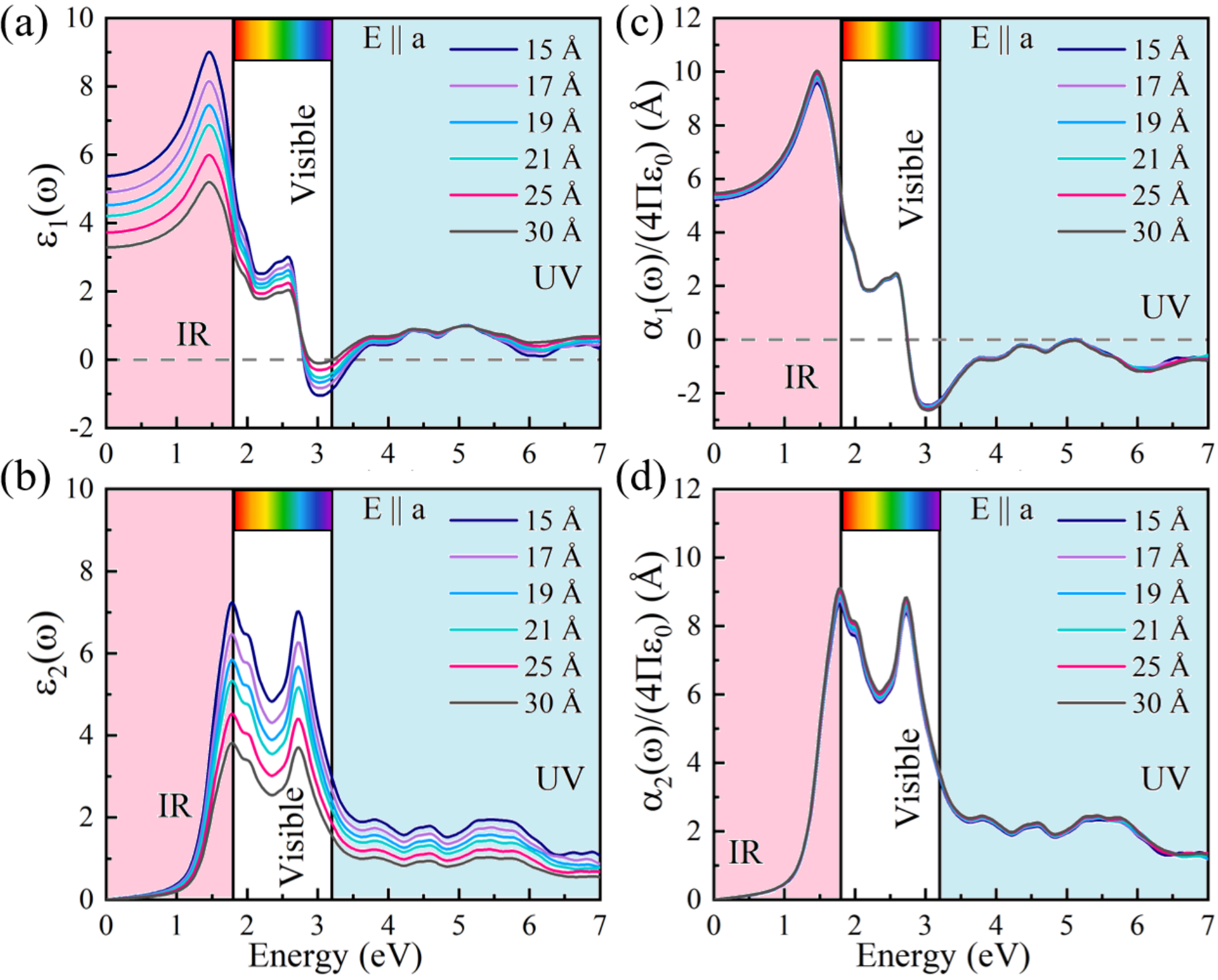}\caption{(a,b) Calculated real and imaginary parts of dielectric constant ($\epsilon_{1}(\omega)$, $\epsilon_{2}(\omega)$), and (c,d) electronic polarizability per unit area ($\alpha$$_{1}(\omega)$, $\alpha$$_{2}(\omega)$) with different vacuum levels for the unit cell of the pristine~\emph{p}-PdTe$_{2}$ monolayer at the GGA-PBE level of theory for $E$$\parallel$$a$. The calculations also include SOC.}
\label{fig:pris_optical} 
\end{figure}

\begin{figure}[ht]
\includegraphics[width=1\linewidth]{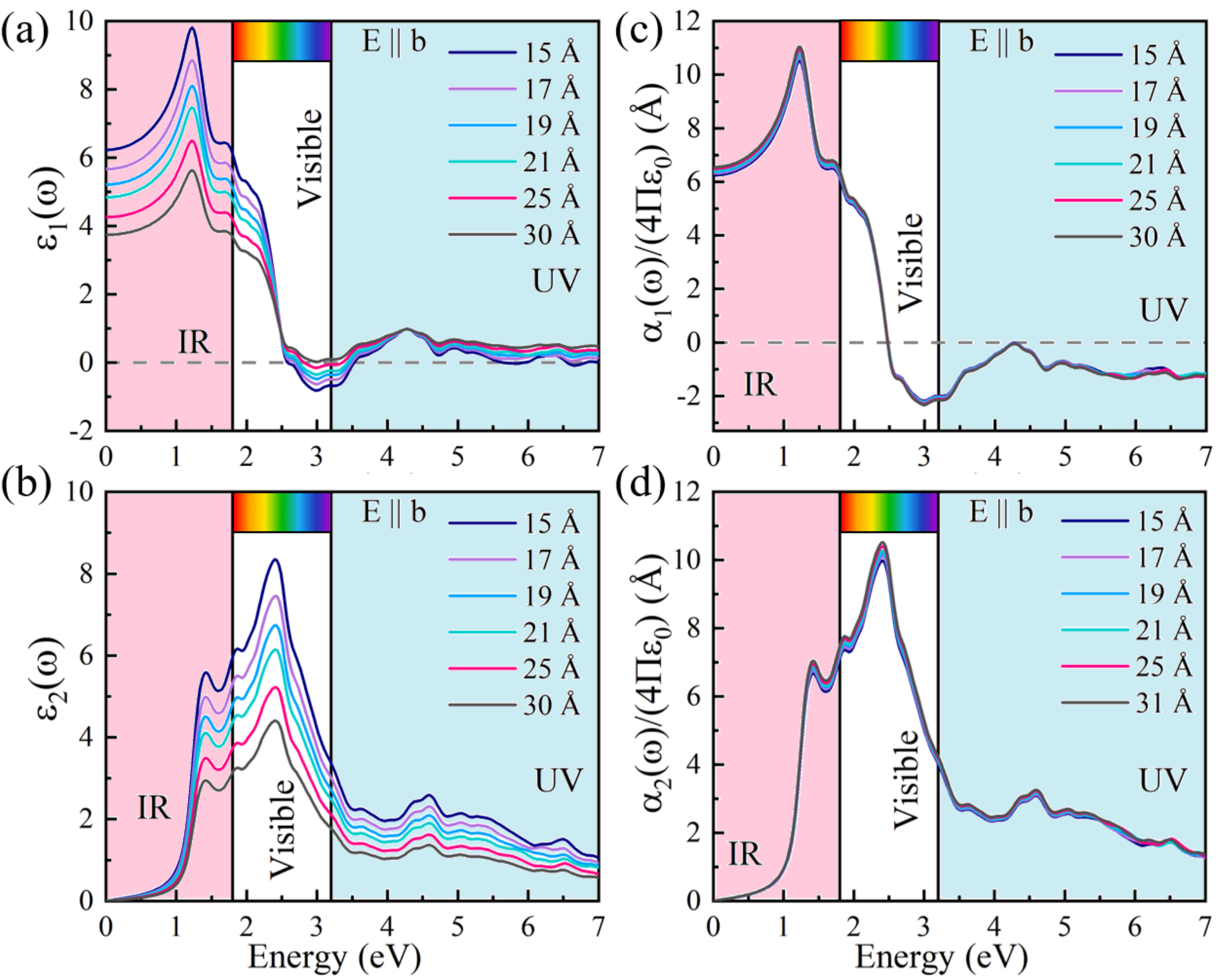}\caption{(a,b) Calculated real and imaginary parts of dielectric constant ($\epsilon_{1}(\omega)$, $\epsilon_{2}(\omega)$), and (c,d) electronic polarizability per unit area ($\alpha$$_{1}(\omega)$, $\alpha$$_{2}(\omega)$) with different vacuum levels for the unit cell of the pristine~\emph{p}-PdTe$_{2}$ monolayer at the GGA-PBE level of theory for $E\parallel b$. The calculations also include SOC.}
\label{fig:pris_optical2} 
\end{figure}

Moreover, information about the inter-band transition, i.e., from the valence band to the conduction band, can be obtained from the imaginary part of the dielectric constant.
However, it is found that both $\epsilon_{1}(\omega)$ and $\epsilon_{2}(\omega)$ are highly sensitive to the choice of the size of the vacuum in 2D materials \cite{YADAV20232711}. Both $\epsilon_{1}(\omega)$ and $\epsilon_{2}(\omega)$ vary significantly with the vacuum size, as shown in Figs. \ref{fig:pris_optical}(a,b) and \ref{fig:pris_optical2}(a,b) for the pristine~\emph{p}-PdTe$_{2}$ monolayer. Consequently, other quantities that depend on the dielectric constant, such as the refractive index,
electron energy loss spectrum, and optical absorption, will also exhibit
dependence on the vacuum size, leading to inaccurate results. To correct
for this dependence, Tian \emph{et al.} \cite{tian2019electronic} developed a formalism
in which the electronic polarizability $\alpha(\omega)=\alpha_{1}(\omega)+i\alpha_{2}(\omega)$
($\alpha_{1}(\omega)/\alpha_{2}(\omega)$ denote real/imaginary parts)
is treated as a more fundamental physical quantity as compared to the dielectric constant. Their work demonstrates that, unlike the dielectric constant, electronic polarizability remains independent of the varying vacuum lengths in 2D materials (Fig.~\ref{fig:pris_optical}(c,d) and Fig.~\ref{fig:pris_optical2}(c,d)), thus eliminating the length effect. Therefore, our discussion of the optical properties of the considered systems is based on electronic polarizability.

The calculations of electronic polarizability are performed using the Kohn-Sham orbitals, which do not take into account the electron correlation, as well as electron-hole interaction, effects. In order to obtain quantitatively accurate results, both of these effects must be taken into account using methods such as the GW and the Bethe–Salpeter equation, which will correct the energy gap as well as include excitonic effects in the absorption spectra. Furthermore, we have not taken into account electron-phonon coupling, which may also make small contributions to the absorption spectra of pristine and defective \emph{p}-PdTe$_{2}$ monolayers. However, we believe that our DFT-level results will provide a qualitative
understanding of the influence of vacancies on the optical absorption spectra of various defective configurations.

\begin{table*}[ht]
\caption{Calculated static electronic polarizability ($\alpha_{1}(0)$)
per unit area, and the locations of the most intense (MI) peaks (eV)
in $\alpha_{1}(\omega)$ (real part) and $\alpha_{2}(\omega)$ (imaginary
part) of the pristine \emph{p}-PdTe$_{2}$
monolayer, and those of various defective configurations, both for
$E\parallel a$ and $E\parallel b$.}

\begin{ruledtabular}
\begin{tabular}{c|c|c|c|c|c|c|c|c|c|c|c|ccc}
 & \multicolumn{1}{c}{} & \multicolumn{5}{c|}{$E\parallel a$ } & \multicolumn{1}{c}{} & \multicolumn{5}{c}{$E\parallel b$ } &  & \tabularnewline
\hline 
Vacancy types$\rightarrow$  & Pristine  & V$_{Pd}$  & V$_{Te}$  & V$_{Pd+Te}$  & V$_{Te+Te}$  & V$_{Pd+4Te}$  & Pristine  & V$_{Pd}$  & V$_{Te}$  & V$_{Pd+Te}$  & V$_{Te+Te}$  & V$_{Pd+4Te}$  &  & \tabularnewline
\hline 
$\alpha_{1}(0)$  & 5.34  & 5.88  & 5.96  & 6.56  & 7.60  & 8.62  & 6.40  & 7.10  & 6.96  & 7.38  & 9.49  & 10.24  &  & \tabularnewline
$\alpha_{1}(\omega)$ (MI, eV)  & 1.46  & 1.42  & 1.42  & 1.39  & 1.45  & 1.37  & 1.23  & 1.22  & 1.18  & 1.22  & 1.22  & 1.23  &  & \tabularnewline
$\alpha_{2}(\omega)$ (MI, eV)  & 1.77  & 1.80  & 1.78  & 1.77  & 1.77  & 1.75  & 2.41  & 2.38  & 2.35  & 2.35  & 2.35  & 2.33  &  & \tabularnewline
\end{tabular}\label{tab:def} 
\end{ruledtabular}

\end{table*}

\subsubsection{Dielectric Response of the Pristine Monolayer}
The static value of the electronic polarizability
($\alpha_{1}(0)$) of the pristine~\emph{p}-PdTe$_{2}$ monolayer corresponding to the directions of the external electric field $E\parallel a$ ($E\parallel b$) is found to be 5.34 (6.40),
indicating larger polarizability for the $E\parallel b$ case. Furthermore, as is evident from Figs. \ref{fig:pris_optical} and \ref{fig:pris_optical2}, both the components of the dynamic polarizability, $\alpha_{1}(\omega)$
and $\alpha_{2}(\omega)$, exhibit different responses for the two polarization directions of the incident photons. This direction-dependent response of the pristine~\emph{p}-PdTe$_{2}$ monolayer to the external fields demonstrates the inherently anisotropic nature of its dielectric properties. This clearly is a consequence of the structural anisotropy
of the \emph{p}-PdTe$_{2}$ monolayer, as discussed earlier. As far as $\alpha_{1}(\omega)$ is concerned, we note that beyond the photon energies of 2.74 eV (2.47 eV) for $E\parallel a$ ($E\parallel b$), it becomes negative, indicating the metallic behavior of its charge carriers. Moreover, to understand the absorption spectrum, we consider $\alpha_{2}(\omega)$, the peaks of which (see Table \ref{tab:def}) for the two considered photon polarizations at 1.77 eV ($E\parallel a$), and 1.42 eV ($E\parallel b$) are above the calculated indirect band gap of the material (1.24 eV), indicating the lack of optical absorption near the gap region. Additionally, in $\alpha_{2}(\omega)$, another peak for $E\parallel a$ is noticed at 2.72 eV (see Fig. \ref{fig:pris_optical}(d)), and for the $E\parallel b$ case, two more peaks at 1.86 and 2.41 eV are seen (see Fig.~\ref{fig:pris_optical2}(d)).
\begin{figure}[ht]
\includegraphics[width=1\linewidth]{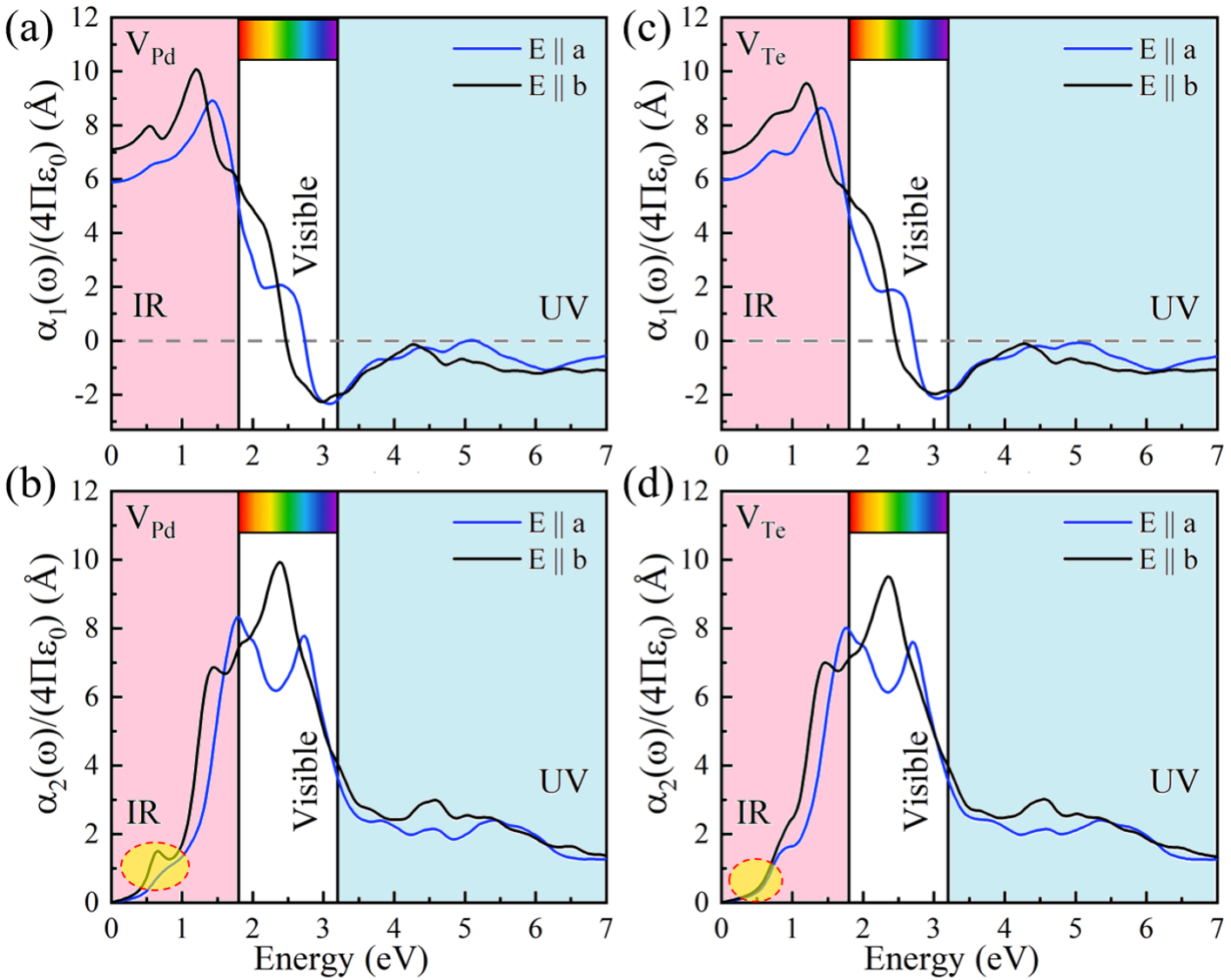}\caption{The (a,c) real ($\alpha$$_{1}(\omega)$) and (b,d) imaginary ($\alpha$$_{2}(\omega)$) parts of the electronic polarizability per unit area corresponding to V$_{Pd}$ and V$_{Te}$, respectively. The SOC is included in all the calculations.}
\label{fig:def_optical} 
\end{figure}

From Figs. \ref{fig:pris_optical}(d) and \ref{fig:pris_optical2}(d), it is also evident that the pristine monolayer has a strong optical response both in the infrared (IR) and visible regions of the spectrum. In particular, the presence of several strong peaks in $\alpha_{2}(\omega)$ both for the $E\parallel a$ and $E\parallel b$ components in the IR and visible regions suggests that the \emph{p}-PdTe2 monolayer has potential applications as an IR/visible absorber.

\subsubsection{Dielectric Response of the Defective Monolayers}

Now, we discuss the optical response of \emph{p}-PdTe$_{2}$ monolayers
with vacancies. The real and imaginary parts of the electronic polarizability
corresponding to V$_{Pd}$ and V$_{Te}$ are shown in Fig.~\ref{fig:def_optical}(a-d).
For the other vacancies, i.e., V$_{Pd+Te}$, V$_{Te+Te}$, and V$_{Pd+4Te}$,
we present the corresponding plots in Fig. S7~\cite{SI}. For each considered vacancy, finite peaks are noticed below the band edge in the imaginary part of the electronic polarization, and they are found to be absent in the pristine~\emph{p}-PdTe$_{2}$ monolayer. Moreover, the static value of electronic polarizability, as well as
the locations of the most intense (MI) peaks in $\alpha_{1}(\omega)$
and $\alpha_{2}(\omega)$ for the defective monolayers, are presented in Table~\ref{tab:def}. From the table, it is clear that the values of the static electronic polarizability are somewhat larger for mono-vacancies but significantly larger for complex vacancies when compared to the pristine ones. However, no significant changes in the locations of the MI peaks for the defective configurations are observed as compared to the pristine ones.

\begin{figure}[ht]
\includegraphics[width=1\linewidth]{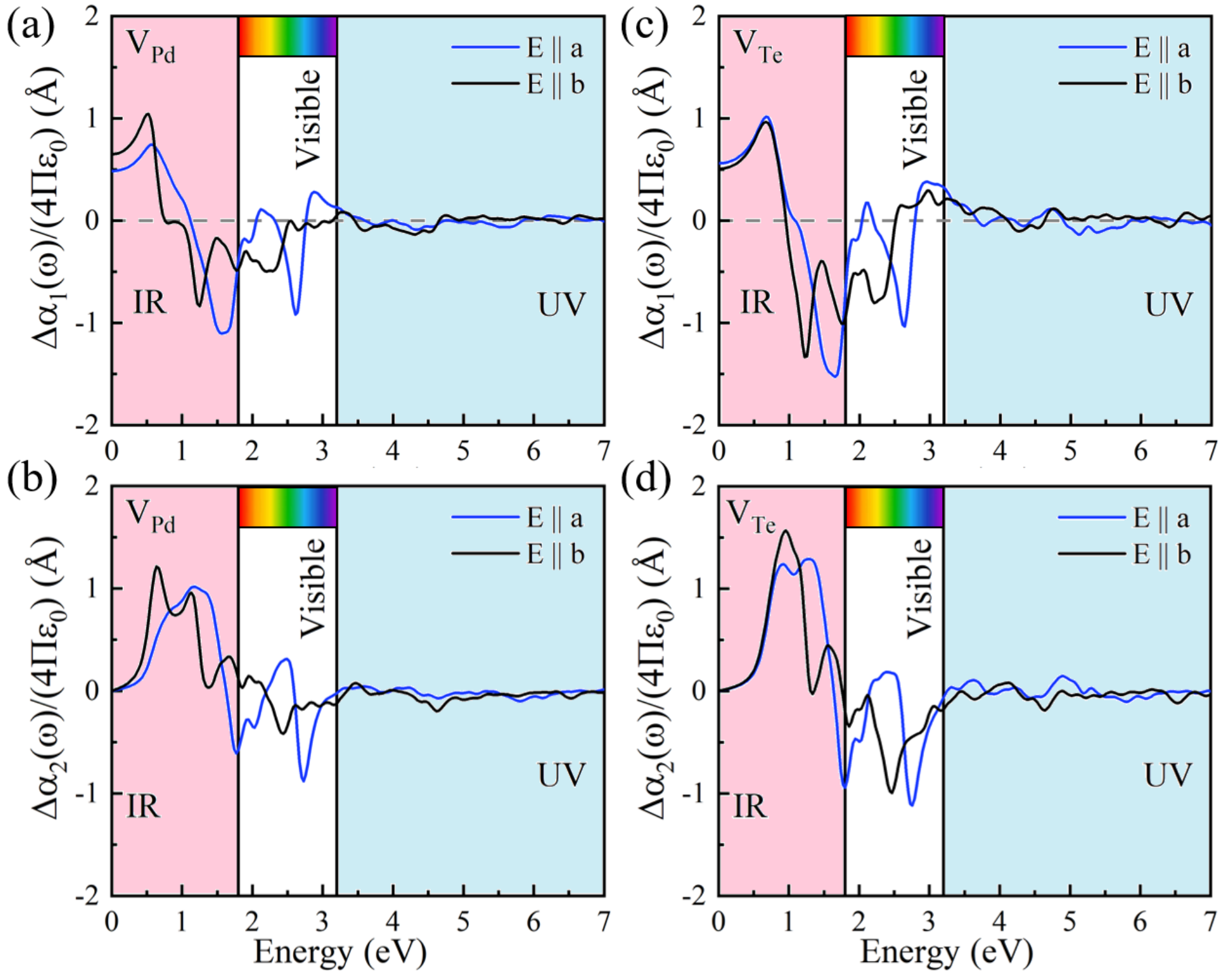}\caption{(a,c) calculated $\Delta$$\alpha$$_{1}(\omega)$, and (b,d) $\Delta$$\alpha$$_{2}(\omega)$ corresponding to V$_{Pd}$ and V$_{Te}$, respectively. The SOC is included in all the calculations.}
\label{fig:exc_def_optical} 
\end{figure}

Furthermore, we computed the real and imaginary parts
of the excess electronic polarizability per unit area ($\Delta\alpha_{1}(\omega)$
and $\Delta\alpha_{2}(\omega)$) for all the considered vacancy configurations,
which gives us a visual description of the influence of the defect
on the dielectric response. In Fig. \ref{fig:exc_def_optical}, we
have plotted $\Delta\alpha_{1}(\omega)$ and $\Delta\alpha_{2}(\omega)$
for V$_{Pd}$ and V$_{Te}$, while some selected data from the corresponding
figures is presented in Table \ref{tab:exc_def_table}. For the other
vacancies, i.e., V$_{Pd+Te}$, V$_{Te+Te}$, and V$_{Pd+4Te}$, we
present the corresponding information in Fig. S8 \cite{SI} and Table
S2~\cite{SI}. We note the following trends for all the vacancy configurations
considered in this paper: (a) in the IR and visible regions, $\Delta\alpha_{1}(\omega)$
and $\Delta\alpha_{2}(\omega)$ exhibit maximum variation in the values, while (b) in the UV region, the values hardly exhibit any variation and are close to zero. This implies that the defects considered in this paper change the dielectric response of the pristine monolayer the most in the IR-Visible region of the spectrum, also obvious from the locations of MI peaks in $\Delta\alpha_{1}(\omega)$ and $\Delta\alpha_{2}(\omega)$ as shown in Table \ref{tab:exc_def_table}. The fact that the location of these MI peaks is different for different defects suggests the interesting possibility of defect engineering to tune the optical response of the system in the desired region of the spectrum.

\begin{table}[ht]
\caption{Calculated excess static electronic polarizability ($\Delta\alpha_{1}(0)$), as well as the location of the most intense (MI) peaks (eV) of $\Delta\alpha_{1}(\omega)$ and $\Delta\alpha_{2}(\omega)$ for V$_{Pd}$ and V$_{Te}$.}

\begin{ruledtabular}
\begin{tabular}{c|c|c|c|cc}
 & \multicolumn{2}{c|}{$E\parallel a$ } & \multicolumn{2}{c}{$E\parallel b$ } & \tabularnewline
\hline 
Vacancy types  & V$_{Pd}$  & V$_{Te}$  & V$_{Pd}$  & V$_{Te}$  & \tabularnewline
\hline 
$\Delta\alpha_{1}$(0)  & 0.48  & 0.56  & 0.64  & 0.51  & \tabularnewline
$\Delta\alpha_{1}(\omega)$ (MI, eV)  & 0.58  & 0.66  & 0.52  & 0.96  & \tabularnewline
$\Delta\alpha_{2}(\omega)$ (MI, eV)  & 1.19  & 1.27  & 0.64  & 0.95  & \tabularnewline
\end{tabular}\label{tab:exc_def_table} 
\end{ruledtabular}

\end{table}

\section{CONCLUSIONS}

\label{sec:Conclusion}
Using the first-principles DFT, in this paper, we studied the structural, electronic, magnetic, and optical characteristics of the \emph{p}-PdTe$_{2}$ monolayer in the pristine form and in the presence of various simple and complex vacancies. From the formation energy calculations, it is found that V$_{Pd}$ and V$_{Te}$ are the most stable defects in the \emph{p}-PdTe$_{2}$ monolayer in Te-rich and Pd-rich conditions, respectively. In all considered defect configurations, the system is found to maintain its non-magnetic behavior as in the pristine \emph{p}-PdTe$_{2}$ monolayer except for V$_{Pd+4Te}$, where a magnetic moment of 1.87 $\mu_{B}$ is obtained. Further, in the charged defect configurations, negatively charged states are most likely to form as compared to positively charged and neutral ones close to the CBM. Bias-dependent STM images are also simulated, corresponding to all considered defective configurations, which can be verified in future experiments. Our calculations predict that the creation of V$_{Pd}$ and V$_{Te}$ requires electron beam energies of 73 and 111 keV, respectively, which are similar in magnitude to graphene, hexagonal boron nitride, and other 2D materials. From the CI-NEB calculations, we find that the diffusion energy barrier for V$_{Te}$ is larger in the in-plane direction as compared to the out-of-plane direction. However, both barriers can be overcome using an STM tip to achieve controlled migration of V$_{Te}$. As far as optical properties are considered, defects are found to activate new transitions in the monolayer as compared to its pristine structure. In all defective configurations, we notice finite peaks below the band edge, which are absent in the pristine \emph{p}-PdTe$_{2}$  monolayer. Furthermore, the pristine \emph{p}-PdTe$_{2}$ monolayer is found to be a good IR to visible absorber. Moreover, the examination of excess dynamic polarizabilities of the defective configurations suggests the possibility of defect engineering with the aim of tuning the optical response of the monolayer.

\section{Acknowledgement}
We thank Brahmananda Chakraborty at Bhabha Atomic Research Center (BARC), India, for fruitful discussions and help with computations. P.S. acknowledges the University Grants Commission (UGC), India, for providing the Ph.D. fellowship (Reference No. 1330) through the joint Council of Scientific and Industrial Research-University Grants Commission National Eligibility Test (CSIR-UGC NET), June 2018. Our calculations were performed using the computational facilities of I.I.T. Bombay and BARC.

\bibliographystyle{apsrev4-2}

\bibliography{pdte2}

\begin{thebibliography}{69}%
\makeatletter
\providecommand \@ifxundefined [1]{%
 \@ifx{#1\undefined}
}%
\providecommand \@ifnum [1]{%
 \ifnum #1\expandafter \@firstoftwo
 \else \expandafter \@secondoftwo
 \fi
}%
\providecommand \@ifx [1]{%
 \ifx #1\expandafter \@firstoftwo
 \else \expandafter \@secondoftwo
 \fi
}%
\providecommand \natexlab [1]{#1}%
\providecommand \enquote  [1]{``#1''}%
\providecommand \bibnamefont  [1]{#1}%
\providecommand \bibfnamefont [1]{#1}%
\providecommand \citenamefont [1]{#1}%
\providecommand \href@noop [0]{\@secondoftwo}%
\providecommand \href [0]{\begingroup \@sanitize@url \@href}%
\providecommand \@href[1]{\@@startlink{#1}\@@href}%
\providecommand \@@href[1]{\endgroup#1\@@endlink}%
\providecommand \@sanitize@url [0]{\catcode `\\12\catcode `\$12\catcode
  `\&12\catcode `\#12\catcode `\^12\catcode `\_12\catcode `\%12\relax}%
\providecommand \@@startlink[1]{}%
\providecommand \@@endlink[0]{}%
\providecommand \url  [0]{\begingroup\@sanitize@url \@url }%
\providecommand \@url [1]{\endgroup\@href {#1}{\urlprefix }}%
\providecommand \urlprefix  [0]{URL }%
\providecommand \Eprint [0]{\href }%
\providecommand \doibase [0]{https://doi.org/}%
\providecommand \selectlanguage [0]{\@gobble}%
\providecommand \bibinfo  [0]{\@secondoftwo}%
\providecommand \bibfield  [0]{\@secondoftwo}%
\providecommand \translation [1]{[#1]}%
\providecommand \BibitemOpen [0]{}%
\providecommand \bibitemStop [0]{}%
\providecommand \bibitemNoStop [0]{.\EOS\space}%
\providecommand \EOS [0]{\spacefactor3000\relax}%
\providecommand \BibitemShut  [1]{\csname bibitem#1\endcsname}%
\let\auto@bib@innerbib\@empty
%</preamble>
\bibitem [{\citenamefont {Henriques}\ \emph {et~al.}(2021)\citenamefont
  {Henriques}, \citenamefont {Kamban}, \citenamefont {Pedersen},\ and\
  \citenamefont {Peres}}]{henriques2021calculation}%
  \BibitemOpen
  \bibfield  {author} {\bibinfo {author} {\bibfnamefont {J.~C.~G.}\
  \bibnamefont {Henriques}}, \bibinfo {author} {\bibfnamefont {H.~C.}\
  \bibnamefont {Kamban}}, \bibinfo {author} {\bibfnamefont {T.~G.}\
  \bibnamefont {Pedersen}},\ and\ \bibinfo {author} {\bibfnamefont {N.~M.~R.}\
  \bibnamefont {Peres}},\ }\href@noop {} {\bibfield  {journal} {\bibinfo
  {journal} {Physical Review B}\ }\textbf {\bibinfo {volume} {\textbf{103}}},\
  \bibinfo {pages} {235412} (\bibinfo {year} {2021})}\BibitemShut {NoStop}%
\bibitem [{\citenamefont {Zibouche}\ \emph {et~al.}(2014)\citenamefont
  {Zibouche}, \citenamefont {Philipsen}, \citenamefont {Kuc},\ and\
  \citenamefont {Heine}}]{PhysRevB.90.125440}%
  \BibitemOpen
  \bibfield  {author} {\bibinfo {author} {\bibfnamefont {N.}~\bibnamefont
  {Zibouche}}, \bibinfo {author} {\bibfnamefont {P.}~\bibnamefont {Philipsen}},
  \bibinfo {author} {\bibfnamefont {A.}~\bibnamefont {Kuc}},\ and\ \bibinfo
  {author} {\bibfnamefont {T.}~\bibnamefont {Heine}},\ }\href@noop {}
  {\bibfield  {journal} {\bibinfo  {journal} {Phys. Rev. B}\ }\textbf {\bibinfo
  {volume} {\textbf{90}}},\ \bibinfo {pages} {125440} (\bibinfo {year}
  {2014})}\BibitemShut {NoStop}%
\bibitem [{\citenamefont {Wang}\ \emph {et~al.}(2024)\citenamefont {Wang},
  \citenamefont {Chen}, \citenamefont {Cheng}, \citenamefont {Su},
  \citenamefont {Liu}, \citenamefont {Jia}, \citenamefont {Li}, \citenamefont
  {Li},\ and\ \citenamefont {Xu}}]{PhysRevB.110.045425}%
  \BibitemOpen
  \bibfield  {author} {\bibinfo {author} {\bibfnamefont {Y.}~\bibnamefont
  {Wang}}, \bibinfo {author} {\bibfnamefont {Z.}~\bibnamefont {Chen}}, \bibinfo
  {author} {\bibfnamefont {X.}~\bibnamefont {Cheng}}, \bibinfo {author}
  {\bibfnamefont {X.}~\bibnamefont {Su}}, \bibinfo {author} {\bibfnamefont
  {P.}~\bibnamefont {Liu}}, \bibinfo {author} {\bibfnamefont {F.}~\bibnamefont
  {Jia}}, \bibinfo {author} {\bibfnamefont {R.}~\bibnamefont {Li}}, \bibinfo
  {author} {\bibfnamefont {Y.}~\bibnamefont {Li}},\ and\ \bibinfo {author}
  {\bibfnamefont {G.}~\bibnamefont {Xu}},\ }\href@noop {} {\bibfield  {journal}
  {\bibinfo  {journal} {Phys. Rev. B}\ }\textbf {\bibinfo {volume}
  {\textbf{110}}},\ \bibinfo {pages} {045425} (\bibinfo {year}
  {2024})}\BibitemShut {NoStop}%
\bibitem [{\citenamefont {He}\ \emph {et~al.}(2021)\citenamefont {He},
  \citenamefont {Peng}, \citenamefont {Feng}, \citenamefont {Xu}, \citenamefont
  {Dai}, \citenamefont {Huang},\ and\ \citenamefont {Ma}}]{he2021two}%
  \BibitemOpen
  \bibfield  {author} {\bibinfo {author} {\bibfnamefont {Z.}~\bibnamefont
  {He}}, \bibinfo {author} {\bibfnamefont {R.}~\bibnamefont {Peng}}, \bibinfo
  {author} {\bibfnamefont {X.}~\bibnamefont {Feng}}, \bibinfo {author}
  {\bibfnamefont {X.}~\bibnamefont {Xu}}, \bibinfo {author} {\bibfnamefont
  {Y.}~\bibnamefont {Dai}}, \bibinfo {author} {\bibfnamefont {B.}~\bibnamefont
  {Huang}},\ and\ \bibinfo {author} {\bibfnamefont {Y.}~\bibnamefont {Ma}},\
  }\href@noop {} {\bibfield  {journal} {\bibinfo  {journal} {Physical Review
  B}\ }\textbf {\bibinfo {volume} {\textbf{104}}},\ \bibinfo {pages} {075105}
  (\bibinfo {year} {2021})}\BibitemShut {NoStop}%
\bibitem [{\citenamefont {Liu}\ \emph {et~al.}(2018)\citenamefont {Liu},
  \citenamefont {Lian}, \citenamefont {Liao}, \citenamefont {Wang},
  \citenamefont {Zhong}, \citenamefont {Ding}, \citenamefont {Li},
  \citenamefont {Song}, \citenamefont {He}, \citenamefont {Ma}, \citenamefont
  {Duan}, \citenamefont {Zhang}, \citenamefont {Xu}, \citenamefont {Wang},\
  and\ \citenamefont {Xue}}]{PhysRevMaterials.2.094001}%
  \BibitemOpen
  \bibfield  {author} {\bibinfo {author} {\bibfnamefont {C.}~\bibnamefont
  {Liu}}, \bibinfo {author} {\bibfnamefont {C.-S.}\ \bibnamefont {Lian}},
  \bibinfo {author} {\bibfnamefont {M.-H.}\ \bibnamefont {Liao}}, \bibinfo
  {author} {\bibfnamefont {Y.}~\bibnamefont {Wang}}, \bibinfo {author}
  {\bibfnamefont {Y.}~\bibnamefont {Zhong}}, \bibinfo {author} {\bibfnamefont
  {C.}~\bibnamefont {Ding}}, \bibinfo {author} {\bibfnamefont {W.}~\bibnamefont
  {Li}}, \bibinfo {author} {\bibfnamefont {C.-L.}\ \bibnamefont {Song}},
  \bibinfo {author} {\bibfnamefont {K.}~\bibnamefont {He}}, \bibinfo {author}
  {\bibfnamefont {X.-C.}\ \bibnamefont {Ma}}, \bibinfo {author} {\bibfnamefont
  {W.}~\bibnamefont {Duan}}, \bibinfo {author} {\bibfnamefont {D.}~\bibnamefont
  {Zhang}}, \bibinfo {author} {\bibfnamefont {Y.}~\bibnamefont {Xu}}, \bibinfo
  {author} {\bibfnamefont {L.}~\bibnamefont {Wang}},\ and\ \bibinfo {author}
  {\bibfnamefont {Q.-K.}\ \bibnamefont {Xue}},\ }\href@noop {} {\bibfield
  {journal} {\bibinfo  {journal} {Phys. Rev. Mater.}\ }\textbf {\bibinfo
  {volume} {\textbf{2}}},\ \bibinfo {pages} {094001} (\bibinfo {year}
  {2018})}\BibitemShut {NoStop}%
\bibitem [{\citenamefont {Rossnagel}(2011)}]{rossnagel2011origin}%
  \BibitemOpen
  \bibfield  {author} {\bibinfo {author} {\bibfnamefont {K.}~\bibnamefont
  {Rossnagel}},\ }\href@noop {} {\bibfield  {journal} {\bibinfo  {journal}
  {Journal of Physics: Condensed Matter}\ }\textbf {\bibinfo {volume}
  {\textbf{23}}},\ \bibinfo {pages} {213001} (\bibinfo {year}
  {2011})}\BibitemShut {NoStop}%
\bibitem [{\citenamefont {Rahimi}\ \emph {et~al.}(2017)\citenamefont {Rahimi},
  \citenamefont {Moghaddam}, \citenamefont {Dykstra}, \citenamefont
  {Governale},\ and\ \citenamefont {Z\"{u}licke}}]{rahimi2017unconventional}%
  \BibitemOpen
  \bibfield  {author} {\bibinfo {author} {\bibfnamefont {M.}~\bibnamefont
  {Rahimi}}, \bibinfo {author} {\bibfnamefont {A.}~\bibnamefont {Moghaddam}},
  \bibinfo {author} {\bibfnamefont {C.}~\bibnamefont {Dykstra}}, \bibinfo
  {author} {\bibfnamefont {M.}~\bibnamefont {Governale}},\ and\ \bibinfo
  {author} {\bibfnamefont {U.}~\bibnamefont {Z\"{u}licke}},\ }\href@noop {}
  {\bibfield  {journal} {\bibinfo  {journal} {Physical Review B}\ }\textbf
  {\bibinfo {volume} {\textbf{95}}},\ \bibinfo {pages} {104515} (\bibinfo
  {year} {2017})}\BibitemShut {NoStop}%
\bibitem [{\citenamefont {Lin}\ \emph {et~al.}(2022)\citenamefont {Lin},
  \citenamefont {Liu}, \citenamefont {Wang}, \citenamefont {Xu}, \citenamefont
  {Chen}, \citenamefont {Duan},\ and\ \citenamefont
  {Monserrat}}]{lin2022phonon}%
  \BibitemOpen
  \bibfield  {author} {\bibinfo {author} {\bibfnamefont {Z.}~\bibnamefont
  {Lin}}, \bibinfo {author} {\bibfnamefont {Y.}~\bibnamefont {Liu}}, \bibinfo
  {author} {\bibfnamefont {Z.}~\bibnamefont {Wang}}, \bibinfo {author}
  {\bibfnamefont {S.}~\bibnamefont {Xu}}, \bibinfo {author} {\bibfnamefont
  {S.}~\bibnamefont {Chen}}, \bibinfo {author} {\bibfnamefont {W.}~\bibnamefont
  {Duan}},\ and\ \bibinfo {author} {\bibfnamefont {B.}~\bibnamefont
  {Monserrat}},\ }\href@noop {} {\bibfield  {journal} {\bibinfo  {journal}
  {Physical review letters}\ }\textbf {\bibinfo {volume} {\textbf{129}}},\
  \bibinfo {pages} {027401} (\bibinfo {year} {2022})}\BibitemShut {NoStop}%
\bibitem [{\citenamefont {Maguire}\ \emph {et~al.}(2018)\citenamefont
  {Maguire}, \citenamefont {Fox}, \citenamefont {Zhou}, \citenamefont {Wang},
  \citenamefont {O'Brien}, \citenamefont {Jadwiszczak}, \citenamefont {Cullen},
  \citenamefont {McManus}, \citenamefont {Bateman}, \citenamefont {McEvoy}
  \emph {et~al.}}]{maguire2018defect}%
  \BibitemOpen
  \bibfield  {author} {\bibinfo {author} {\bibfnamefont {P.}~\bibnamefont
  {Maguire}}, \bibinfo {author} {\bibfnamefont {D.~S.}\ \bibnamefont {Fox}},
  \bibinfo {author} {\bibfnamefont {Y.}~\bibnamefont {Zhou}}, \bibinfo {author}
  {\bibfnamefont {Q.}~\bibnamefont {Wang}}, \bibinfo {author} {\bibfnamefont
  {M.}~\bibnamefont {O'Brien}}, \bibinfo {author} {\bibfnamefont
  {J.}~\bibnamefont {Jadwiszczak}}, \bibinfo {author} {\bibfnamefont {C.~P.}\
  \bibnamefont {Cullen}}, \bibinfo {author} {\bibfnamefont {J.}~\bibnamefont
  {McManus}}, \bibinfo {author} {\bibfnamefont {S.}~\bibnamefont {Bateman}},
  \bibinfo {author} {\bibfnamefont {N.}~\bibnamefont {McEvoy}}, \emph
  {et~al.},\ }\href@noop {} {\bibfield  {journal} {\bibinfo  {journal}
  {Physical Review B}\ }\textbf {\bibinfo {volume} {\textbf{98}}},\ \bibinfo
  {pages} {134109} (\bibinfo {year} {2018})}\BibitemShut {NoStop}%
\bibitem [{\citenamefont {Wang}\ \emph {et~al.}(2019)\citenamefont {Wang},
  \citenamefont {March}, \citenamefont {Ponce},\ and\ \citenamefont
  {Rez}}]{PhysRevB.99.115312}%
  \BibitemOpen
  \bibfield  {author} {\bibinfo {author} {\bibfnamefont {S.}~\bibnamefont
  {Wang}}, \bibinfo {author} {\bibfnamefont {K.}~\bibnamefont {March}},
  \bibinfo {author} {\bibfnamefont {F.~A.}\ \bibnamefont {Ponce}},\ and\
  \bibinfo {author} {\bibfnamefont {P.}~\bibnamefont {Rez}},\ }\href@noop {}
  {\bibfield  {journal} {\bibinfo  {journal} {Phys. Rev. B}\ }\textbf {\bibinfo
  {volume} {\textbf{99}}},\ \bibinfo {pages} {115312} (\bibinfo {year}
  {2019})}\BibitemShut {NoStop}%
\bibitem [{\citenamefont {Shin}\ \emph {et~al.}(2021)\citenamefont {Shin},
  \citenamefont {Wang}, \citenamefont {Han}, \citenamefont {Lin}, \citenamefont
  {\'{O}Hara}, \citenamefont {Chen}, \citenamefont {Lin},\ and\ \citenamefont
  {Pantelides}}]{PhysRevMaterials.5.044002}%
  \BibitemOpen
  \bibfield  {author} {\bibinfo {author} {\bibfnamefont {D.}~\bibnamefont
  {Shin}}, \bibinfo {author} {\bibfnamefont {G.}~\bibnamefont {Wang}}, \bibinfo
  {author} {\bibfnamefont {M.}~\bibnamefont {Han}}, \bibinfo {author}
  {\bibfnamefont {Z.}~\bibnamefont {Lin}}, \bibinfo {author} {\bibfnamefont
  {A.}~\bibnamefont {\'{O}Hara}}, \bibinfo {author} {\bibfnamefont
  {F.}~\bibnamefont {Chen}}, \bibinfo {author} {\bibfnamefont {J.}~\bibnamefont
  {Lin}},\ and\ \bibinfo {author} {\bibfnamefont {S.~T.}\ \bibnamefont
  {Pantelides}},\ }\href@noop {} {\bibfield  {journal} {\bibinfo  {journal}
  {Phys. Rev. Mater.}\ }\textbf {\bibinfo {volume} {\textbf{5}}},\ \bibinfo
  {pages} {044002} (\bibinfo {year} {2021})}\BibitemShut {NoStop}%
\bibitem [{\citenamefont {Chen}\ \emph {et~al.}(2022)\citenamefont {Chen},
  \citenamefont {Kawakami}, \citenamefont {Lin}, \citenamefont {Huang},
  \citenamefont {Arafune}, \citenamefont {Takagi},\ and\ \citenamefont
  {Lin}}]{PhysRevB.106.075428}%
  \BibitemOpen
  \bibfield  {author} {\bibinfo {author} {\bibfnamefont {W.-H.}\ \bibnamefont
  {Chen}}, \bibinfo {author} {\bibfnamefont {N.}~\bibnamefont {Kawakami}},
  \bibinfo {author} {\bibfnamefont {J.-J.}\ \bibnamefont {Lin}}, \bibinfo
  {author} {\bibfnamefont {H.-I.}\ \bibnamefont {Huang}}, \bibinfo {author}
  {\bibfnamefont {R.}~\bibnamefont {Arafune}}, \bibinfo {author} {\bibfnamefont
  {N.}~\bibnamefont {Takagi}},\ and\ \bibinfo {author} {\bibfnamefont {C.-L.}\
  \bibnamefont {Lin}},\ }\href@noop {} {\bibfield  {journal} {\bibinfo
  {journal} {Phys. Rev. B}\ }\textbf {\bibinfo {volume} {\textbf{106}}},\
  \bibinfo {pages} {075428} (\bibinfo {year} {2022})}\BibitemShut {NoStop}%
\bibitem [{\citenamefont {Komsa}\ \emph
  {et~al.}(2012{\natexlab{a}})\citenamefont {Komsa}, \citenamefont {Kotakoski},
  \citenamefont {Kurasch}, \citenamefont {Lehtinen}, \citenamefont {Kaiser},\
  and\ \citenamefont {Krasheninnikov}}]{PhysRevLett.109.035503}%
  \BibitemOpen
  \bibfield  {author} {\bibinfo {author} {\bibfnamefont {H.-P.}\ \bibnamefont
  {Komsa}}, \bibinfo {author} {\bibfnamefont {J.}~\bibnamefont {Kotakoski}},
  \bibinfo {author} {\bibfnamefont {S.}~\bibnamefont {Kurasch}}, \bibinfo
  {author} {\bibfnamefont {O.}~\bibnamefont {Lehtinen}}, \bibinfo {author}
  {\bibfnamefont {U.}~\bibnamefont {Kaiser}},\ and\ \bibinfo {author}
  {\bibfnamefont {A.~V.}\ \bibnamefont {Krasheninnikov}},\ }\href@noop {}
  {\bibfield  {journal} {\bibinfo  {journal} {Phys. Rev. Lett.}\ }\textbf
  {\bibinfo {volume} {\textbf{109}}},\ \bibinfo {pages} {035503} (\bibinfo
  {year} {2012}{\natexlab{a}})}\BibitemShut {NoStop}%
\bibitem [{\citenamefont {Krasheninnikov}\ \emph {et~al.}(2002)\citenamefont
  {Krasheninnikov}, \citenamefont {Nordlund},\ and\ \citenamefont
  {Keinonen}}]{krasheninnikov2002production}%
  \BibitemOpen
  \bibfield  {author} {\bibinfo {author} {\bibfnamefont {A.}~\bibnamefont
  {Krasheninnikov}}, \bibinfo {author} {\bibfnamefont {K.}~\bibnamefont
  {Nordlund}},\ and\ \bibinfo {author} {\bibfnamefont {J.}~\bibnamefont
  {Keinonen}},\ }\href@noop {} {\bibfield  {journal} {\bibinfo  {journal}
  {Physical Review B}\ }\textbf {\bibinfo {volume} {\textbf{65}}},\ \bibinfo
  {pages} {165423} (\bibinfo {year} {2002})}\BibitemShut {NoStop}%
\bibitem [{\citenamefont {Lehtinen}\ \emph {et~al.}(2010)\citenamefont
  {Lehtinen}, \citenamefont {Kotakoski}, \citenamefont {Krasheninnikov},
  \citenamefont {Tolvanen}, \citenamefont {Nordlund},\ and\ \citenamefont
  {Keinonen}}]{PhysRevB.81.153401}%
  \BibitemOpen
  \bibfield  {author} {\bibinfo {author} {\bibfnamefont {O.}~\bibnamefont
  {Lehtinen}}, \bibinfo {author} {\bibfnamefont {J.}~\bibnamefont {Kotakoski}},
  \bibinfo {author} {\bibfnamefont {A.~V.}\ \bibnamefont {Krasheninnikov}},
  \bibinfo {author} {\bibfnamefont {A.}~\bibnamefont {Tolvanen}}, \bibinfo
  {author} {\bibfnamefont {K.}~\bibnamefont {Nordlund}},\ and\ \bibinfo
  {author} {\bibfnamefont {J.}~\bibnamefont {Keinonen}},\ }\href@noop {}
  {\bibfield  {journal} {\bibinfo  {journal} {Phys. Rev. B}\ }\textbf {\bibinfo
  {volume} {\textbf{81}}},\ \bibinfo {pages} {153401} (\bibinfo {year}
  {2010})}\BibitemShut {NoStop}%
\bibitem [{\citenamefont {Zheng}\ \emph {et~al.}(2013)\citenamefont {Zheng},
  \citenamefont {Weismann},\ and\ \citenamefont
  {Berndt}}]{zheng2013manipulation}%
  \BibitemOpen
  \bibfield  {author} {\bibinfo {author} {\bibfnamefont {H.}~\bibnamefont
  {Zheng}}, \bibinfo {author} {\bibfnamefont {A.}~\bibnamefont {Weismann}},\
  and\ \bibinfo {author} {\bibfnamefont {R.}~\bibnamefont {Berndt}},\
  }\href@noop {} {\bibfield  {journal} {\bibinfo  {journal} {Physical Review
  letters}\ }\textbf {\bibinfo {volume} {\textbf{110}}},\ \bibinfo {pages}
  {226101} (\bibinfo {year} {2013})}\BibitemShut {NoStop}%
\bibitem [{\citenamefont {Dagdeviren}\ \emph {et~al.}(2016)\citenamefont
  {Dagdeviren}, \citenamefont {Simon}, \citenamefont {Zou}, \citenamefont
  {Walker}, \citenamefont {Ahn}, \citenamefont {Altman},\ and\ \citenamefont
  {Schwarz}}]{dagdeviren2016surface}%
  \BibitemOpen
  \bibfield  {author} {\bibinfo {author} {\bibfnamefont {O.~E.}\ \bibnamefont
  {Dagdeviren}}, \bibinfo {author} {\bibfnamefont {G.~H.}\ \bibnamefont
  {Simon}}, \bibinfo {author} {\bibfnamefont {K.}~\bibnamefont {Zou}}, \bibinfo
  {author} {\bibfnamefont {F.~J.}\ \bibnamefont {Walker}}, \bibinfo {author}
  {\bibfnamefont {C.}~\bibnamefont {Ahn}}, \bibinfo {author} {\bibfnamefont
  {E.~I.}\ \bibnamefont {Altman}},\ and\ \bibinfo {author} {\bibfnamefont
  {U.~D.}\ \bibnamefont {Schwarz}},\ }\href@noop {} {\bibfield  {journal}
  {\bibinfo  {journal} {Physical Review B}\ }\textbf {\bibinfo {volume}
  {\textbf{93}}},\ \bibinfo {pages} {195303} (\bibinfo {year}
  {2016})}\BibitemShut {NoStop}%
\bibitem [{\citenamefont {Shen}\ and\ \citenamefont
  {Wang}(2022)}]{shen2022pentagon}%
  \BibitemOpen
  \bibfield  {author} {\bibinfo {author} {\bibfnamefont {Y.}~\bibnamefont
  {Shen}}\ and\ \bibinfo {author} {\bibfnamefont {Q.}~\bibnamefont {Wang}},\
  }\href@noop {} {\bibfield  {journal} {\bibinfo  {journal} {Physics Reports}\
  }\textbf {\bibinfo {volume} {\textbf{964}}},\ \bibinfo {pages} {1} (\bibinfo
  {year} {2022})}\BibitemShut {NoStop}%
\bibitem [{\citenamefont {Marfoua}\ and\ \citenamefont
  {Hong}(2019)}]{marfoua2019high}%
  \BibitemOpen
  \bibfield  {author} {\bibinfo {author} {\bibfnamefont {B.}~\bibnamefont
  {Marfoua}}\ and\ \bibinfo {author} {\bibfnamefont {J.}~\bibnamefont {Hong}},\
  }\href@noop {} {\bibfield  {journal} {\bibinfo  {journal} {ACS applied
  materials and interfaces}\ }\textbf {\bibinfo {volume} {\textbf{11}}},\
  \bibinfo {pages} {38819} (\bibinfo {year} {2019})}\BibitemShut {NoStop}%
\bibitem [{\citenamefont {Wang}\ \emph
  {et~al.}(2015{\natexlab{a}})\citenamefont {Wang}, \citenamefont {Li},\ and\
  \citenamefont {Chen}}]{wang2015not}%
  \BibitemOpen
  \bibfield  {author} {\bibinfo {author} {\bibfnamefont {Y.}~\bibnamefont
  {Wang}}, \bibinfo {author} {\bibfnamefont {Y.}~\bibnamefont {Li}},\ and\
  \bibinfo {author} {\bibfnamefont {Z.}~\bibnamefont {Chen}},\ }\href@noop {}
  {\bibfield  {journal} {\bibinfo  {journal} {Journal of Materials Chemistry
  C}\ }\textbf {\bibinfo {volume} {\textbf{3}}},\ \bibinfo {pages} {9603}
  (\bibinfo {year} {2015}{\natexlab{a}})}\BibitemShut {NoStop}%
\bibitem [{\citenamefont {Figueiredo}\ and\ \citenamefont
  {Seixas}(2022)}]{figueiredo2022hydrogen}%
  \BibitemOpen
  \bibfield  {author} {\bibinfo {author} {\bibfnamefont {R.~O.}\ \bibnamefont
  {Figueiredo}}\ and\ \bibinfo {author} {\bibfnamefont {L.}~\bibnamefont
  {Seixas}},\ }\href@noop {} {\bibfield  {journal} {\bibinfo  {journal}
  {Physical Review Applied}\ }\textbf {\bibinfo {volume} {\textbf{17}}},\
  \bibinfo {pages} {034035} (\bibinfo {year} {2022})}\BibitemShut {NoStop}%
\bibitem [{\citenamefont {Gonzalez}\ and\ \citenamefont
  {Oleynik}(2016)}]{gonzalez2016layer}%
  \BibitemOpen
  \bibfield  {author} {\bibinfo {author} {\bibfnamefont {J.~M.}\ \bibnamefont
  {Gonzalez}}\ and\ \bibinfo {author} {\bibfnamefont {I.~I.}\ \bibnamefont
  {Oleynik}},\ }\href@noop {} {\bibfield  {journal} {\bibinfo  {journal}
  {Physical Review B}\ }\textbf {\bibinfo {volume} {94}},\ \bibinfo {pages}
  {125443} (\bibinfo {year} {2016})}\BibitemShut {NoStop}%
\bibitem [{\citenamefont {Leng}\ \emph {et~al.}(2019)\citenamefont {Leng},
  \citenamefont {Orain}, \citenamefont {Amato}, \citenamefont {Huang},\ and\
  \citenamefont {de~Visser}}]{pdte2-superconductivity-PhysRevB.100.224501}%
  \BibitemOpen
  \bibfield  {author} {\bibinfo {author} {\bibfnamefont {H.}~\bibnamefont
  {Leng}}, \bibinfo {author} {\bibfnamefont {J.-C.}\ \bibnamefont {Orain}},
  \bibinfo {author} {\bibfnamefont {A.}~\bibnamefont {Amato}}, \bibinfo
  {author} {\bibfnamefont {Y.~K.}\ \bibnamefont {Huang}},\ and\ \bibinfo
  {author} {\bibfnamefont {A.}~\bibnamefont {de~Visser}},\ }\href@noop {}
  {\bibfield  {journal} {\bibinfo  {journal} {Phys. Rev. B}\ }\textbf {\bibinfo
  {volume} {\textbf{100}}},\ \bibinfo {pages} {224501} (\bibinfo {year}
  {2019})}\BibitemShut {NoStop}%
\bibitem [{\citenamefont {Lan}\ \emph {et~al.}(2019)\citenamefont {Lan},
  \citenamefont {Chen}, \citenamefont {Hu}, \citenamefont {Cheng},\ and\
  \citenamefont {Chen}}]{p-pdx2-thermoelectric-theory-2019}%
  \BibitemOpen
  \bibfield  {author} {\bibinfo {author} {\bibfnamefont {Y.-S.}\ \bibnamefont
  {Lan}}, \bibinfo {author} {\bibfnamefont {X.-R.}\ \bibnamefont {Chen}},
  \bibinfo {author} {\bibfnamefont {C.-E.}\ \bibnamefont {Hu}}, \bibinfo
  {author} {\bibfnamefont {Y.}~\bibnamefont {Cheng}},\ and\ \bibinfo {author}
  {\bibfnamefont {Q.-F.}\ \bibnamefont {Chen}},\ }\href@noop {} {\bibfield
  {journal} {\bibinfo  {journal} {J. Mater. Chem. A}\ }\textbf {\bibinfo
  {volume} {\textbf{7}}},\ \bibinfo {pages} {11134} (\bibinfo {year}
  {2019})}\BibitemShut {NoStop}%
\bibitem [{\citenamefont {Zuo}\ \emph {et~al.}(2023)\citenamefont {Zuo},
  \citenamefont {Antonatos}, \citenamefont {D\v{e}kanovsk\'{y}}, \citenamefont
  {Luxa}, \citenamefont {Elliott}, \citenamefont {Gianolio}, \citenamefont
  {\v{S}turala}, \citenamefont {Guzzetta}, \citenamefont {Mourdikoudis},
  \citenamefont {Regner} \emph {et~al.}}]{zuo2023defect}%
  \BibitemOpen
  \bibfield  {author} {\bibinfo {author} {\bibfnamefont {Y.}~\bibnamefont
  {Zuo}}, \bibinfo {author} {\bibfnamefont {N.}~\bibnamefont {Antonatos}},
  \bibinfo {author} {\bibfnamefont {L.}~\bibnamefont {D\v{e}kanovsk\'{y}}},
  \bibinfo {author} {\bibfnamefont {J.}~\bibnamefont {Luxa}}, \bibinfo {author}
  {\bibfnamefont {J.~D.}\ \bibnamefont {Elliott}}, \bibinfo {author}
  {\bibfnamefont {D.}~\bibnamefont {Gianolio}}, \bibinfo {author}
  {\bibfnamefont {J.}~\bibnamefont {\v{S}turala}}, \bibinfo {author}
  {\bibfnamefont {F.}~\bibnamefont {Guzzetta}}, \bibinfo {author}
  {\bibfnamefont {S.}~\bibnamefont {Mourdikoudis}}, \bibinfo {author}
  {\bibfnamefont {J.}~\bibnamefont {Regner}}, \emph {et~al.},\ }\href@noop {}
  {\bibfield  {journal} {\bibinfo  {journal} {ACS Catalysis}\ }\textbf
  {\bibinfo {volume} {\textbf{13}}},\ \bibinfo {pages} {2601} (\bibinfo {year}
  {2023})}\BibitemShut {NoStop}%
\bibitem [{\citenamefont {Liu}\ \emph {et~al.}(2024)\citenamefont {Liu},
  \citenamefont {Ji}, \citenamefont {Bianchi}, \citenamefont {Hus},
  \citenamefont {Li}, \citenamefont {Balog}, \citenamefont {Miwa},
  \citenamefont {Hofmann}, \citenamefont {Li}, \citenamefont {Zemlyanov},
  \citenamefont {Li},\ and\ \citenamefont {Chen}}]{liu2024metastable}%
  \BibitemOpen
  \bibfield  {author} {\bibinfo {author} {\bibfnamefont {L.}~\bibnamefont
  {Liu}}, \bibinfo {author} {\bibfnamefont {Y.}~\bibnamefont {Ji}}, \bibinfo
  {author} {\bibfnamefont {M.}~\bibnamefont {Bianchi}}, \bibinfo {author}
  {\bibfnamefont {S.~M.}\ \bibnamefont {Hus}}, \bibinfo {author} {\bibfnamefont
  {Z.}~\bibnamefont {Li}}, \bibinfo {author} {\bibfnamefont {R.}~\bibnamefont
  {Balog}}, \bibinfo {author} {\bibfnamefont {J.~A.}\ \bibnamefont {Miwa}},
  \bibinfo {author} {\bibfnamefont {P.}~\bibnamefont {Hofmann}}, \bibinfo
  {author} {\bibfnamefont {A.-P.}\ \bibnamefont {Li}}, \bibinfo {author}
  {\bibfnamefont {D.~Y.}\ \bibnamefont {Zemlyanov}}, \bibinfo {author}
  {\bibfnamefont {Y.}~\bibnamefont {Li}},\ and\ \bibinfo {author}
  {\bibfnamefont {Y.~P.}\ \bibnamefont {Chen}},\ }\href@noop {} {\bibfield
  {journal} {\bibinfo  {journal} {Nature Materials}\ }\textbf {\bibinfo
  {volume} {\textbf{23}}},\ \bibinfo {pages} {1339} (\bibinfo {year}
  {2024})}\BibitemShut {NoStop}%
\bibitem [{\citenamefont {Manchanda}\ \emph {et~al.}(2021)\citenamefont
  {Manchanda}, \citenamefont {Kumar},\ and\ \citenamefont
  {Dev}}]{manchanda2021defect}%
  \BibitemOpen
  \bibfield  {author} {\bibinfo {author} {\bibfnamefont {P.}~\bibnamefont
  {Manchanda}}, \bibinfo {author} {\bibfnamefont {P.}~\bibnamefont {Kumar}},\
  and\ \bibinfo {author} {\bibfnamefont {P.}~\bibnamefont {Dev}},\ }\href@noop
  {} {\bibfield  {journal} {\bibinfo  {journal} {Physical Review B}\ }\textbf
  {\bibinfo {volume} {\textbf{103}}},\ \bibinfo {pages} {144403} (\bibinfo
  {year} {2021})}\BibitemShut {NoStop}%
\bibitem [{\citenamefont {Manchanda}\ \emph {et~al.}(2016)\citenamefont
  {Manchanda}, \citenamefont {Enders}, \citenamefont {Sellmyer},\ and\
  \citenamefont {Skomski}}]{manchanda2016hydrogen}%
  \BibitemOpen
  \bibfield  {author} {\bibinfo {author} {\bibfnamefont {P.}~\bibnamefont
  {Manchanda}}, \bibinfo {author} {\bibfnamefont {A.}~\bibnamefont {Enders}},
  \bibinfo {author} {\bibfnamefont {D.~J.}\ \bibnamefont {Sellmyer}},\ and\
  \bibinfo {author} {\bibfnamefont {R.}~\bibnamefont {Skomski}},\ }\href@noop
  {} {\bibfield  {journal} {\bibinfo  {journal} {Physical Review B}\ }\textbf
  {\bibinfo {volume} {\textbf{94}}},\ \bibinfo {pages} {104426} (\bibinfo
  {year} {2016})}\BibitemShut {NoStop}%
\bibitem [{\citenamefont {Hohenberg}\ and\ \citenamefont
  {Kohn}(1964)}]{hohenberg1964inhomogeneous}%
  \BibitemOpen
  \bibfield  {author} {\bibinfo {author} {\bibfnamefont {P.}~\bibnamefont
  {Hohenberg}}\ and\ \bibinfo {author} {\bibfnamefont {W.}~\bibnamefont
  {Kohn}},\ }\href@noop {} {\bibfield  {journal} {\bibinfo  {journal} {Physical
  review}\ }\textbf {\bibinfo {volume} {\textbf{136}}},\ \bibinfo {pages}
  {B864} (\bibinfo {year} {1964})}\BibitemShut {NoStop}%
\bibitem [{\citenamefont {Kohn}\ and\ \citenamefont
  {Sham}(1965)}]{kohn1965self}%
  \BibitemOpen
  \bibfield  {author} {\bibinfo {author} {\bibfnamefont {W.}~\bibnamefont
  {Kohn}}\ and\ \bibinfo {author} {\bibfnamefont {L.~J.}\ \bibnamefont
  {Sham}},\ }\href@noop {} {\bibfield  {journal} {\bibinfo  {journal} {Physical
  review}\ }\textbf {\bibinfo {volume} {\textbf{140}}},\ \bibinfo {pages}
  {A1133} (\bibinfo {year} {1965})}\BibitemShut {NoStop}%
\bibitem [{\citenamefont {Kresse}\ and\ \citenamefont
  {Furthm{\"u}ller}(1996)}]{kresse1996efficient}%
  \BibitemOpen
  \bibfield  {author} {\bibinfo {author} {\bibfnamefont {G.}~\bibnamefont
  {Kresse}}\ and\ \bibinfo {author} {\bibfnamefont {J.}~\bibnamefont
  {Furthm{\"u}ller}},\ }\href@noop {} {\bibfield  {journal} {\bibinfo
  {journal} {Physical review B}\ }\textbf {\bibinfo {volume} {\textbf{54}}},\
  \bibinfo {pages} {11169} (\bibinfo {year} {1996})}\BibitemShut {NoStop}%
\bibitem [{\citenamefont {Perdew}\ \emph {et~al.}(1996)\citenamefont {Perdew},
  \citenamefont {Burke},\ and\ \citenamefont
  {Ernzerhof}}]{perdew1996generalized}%
  \BibitemOpen
  \bibfield  {author} {\bibinfo {author} {\bibfnamefont {J.~P.}\ \bibnamefont
  {Perdew}}, \bibinfo {author} {\bibfnamefont {K.}~\bibnamefont {Burke}},\ and\
  \bibinfo {author} {\bibfnamefont {M.}~\bibnamefont {Ernzerhof}},\ }\href@noop
  {} {\bibfield  {journal} {\bibinfo  {journal} {Physical review letters}\
  }\textbf {\bibinfo {volume} {\textbf{77}}},\ \bibinfo {pages} {3865}
  (\bibinfo {year} {1996})}\BibitemShut {NoStop}%
\bibitem [{\citenamefont {Perdew}\ \emph {et~al.}(1997)\citenamefont {Perdew},
  \citenamefont {Burke},\ and\ \citenamefont
  {Ernzerhof}}]{perdew1997generalized}%
  \BibitemOpen
  \bibfield  {author} {\bibinfo {author} {\bibfnamefont {J.~P.}\ \bibnamefont
  {Perdew}}, \bibinfo {author} {\bibfnamefont {K.}~\bibnamefont {Burke}},\ and\
  \bibinfo {author} {\bibfnamefont {M.}~\bibnamefont {Ernzerhof}},\ }\href@noop
  {} {\bibfield  {journal} {\bibinfo  {journal} {Physical review letters}\
  }\textbf {\bibinfo {volume} {\textbf{78}}},\ \bibinfo {pages} {1396}
  (\bibinfo {year} {1997})}\BibitemShut {NoStop}%
\bibitem [{\citenamefont {Monkhorst}\ and\ \citenamefont
  {Pack}(1976)}]{monkhorst1976special}%
  \BibitemOpen
  \bibfield  {author} {\bibinfo {author} {\bibfnamefont {H.~J.}\ \bibnamefont
  {Monkhorst}}\ and\ \bibinfo {author} {\bibfnamefont {J.~D.}\ \bibnamefont
  {Pack}},\ }\href@noop {} {\bibfield  {journal} {\bibinfo  {journal} {Physical
  review B}\ }\textbf {\bibinfo {volume} {\textbf{13}}},\ \bibinfo {pages}
  {5188} (\bibinfo {year} {1976})}\BibitemShut {NoStop}%
\bibitem [{\citenamefont {Tian}\ \emph {et~al.}(2019)\citenamefont {Tian},
  \citenamefont {Scullion}, \citenamefont {Hughes}, \citenamefont {Li},
  \citenamefont {Shih}, \citenamefont {Coleman}, \citenamefont {Chhowalla},\
  and\ \citenamefont {Santos}}]{tian2019electronic}%
  \BibitemOpen
  \bibfield  {author} {\bibinfo {author} {\bibfnamefont {T.}~\bibnamefont
  {Tian}}, \bibinfo {author} {\bibfnamefont {D.}~\bibnamefont {Scullion}},
  \bibinfo {author} {\bibfnamefont {D.}~\bibnamefont {Hughes}}, \bibinfo
  {author} {\bibfnamefont {L.~H.}\ \bibnamefont {Li}}, \bibinfo {author}
  {\bibfnamefont {C.-J.}\ \bibnamefont {Shih}}, \bibinfo {author}
  {\bibfnamefont {J.}~\bibnamefont {Coleman}}, \bibinfo {author} {\bibfnamefont
  {M.}~\bibnamefont {Chhowalla}},\ and\ \bibinfo {author} {\bibfnamefont
  {E.~J.}\ \bibnamefont {Santos}},\ }\href@noop {} {\bibfield  {journal}
  {\bibinfo  {journal} {Nano letters}\ }\textbf {\bibinfo {volume}
  {\textbf{20}}},\ \bibinfo {pages} {841} (\bibinfo {year} {2019})}\BibitemShut
  {NoStop}%
\bibitem [{\citenamefont {Tersoff}\ and\ \citenamefont
  {Hamann}(1985)}]{tersoff1985theory}%
  \BibitemOpen
  \bibfield  {author} {\bibinfo {author} {\bibfnamefont {J.}~\bibnamefont
  {Tersoff}}\ and\ \bibinfo {author} {\bibfnamefont {D.~R.}\ \bibnamefont
  {Hamann}},\ }\href@noop {} {\bibfield  {journal} {\bibinfo  {journal}
  {Physical Review B}\ }\textbf {\bibinfo {volume} {\textbf{31}}},\ \bibinfo
  {pages} {805} (\bibinfo {year} {1985})}\BibitemShut {NoStop}%
\bibitem [{\citenamefont {Vanpoucke}\ and\ \citenamefont
  {Brocks}(2008)}]{vanpoucke2008formation}%
  \BibitemOpen
  \bibfield  {author} {\bibinfo {author} {\bibfnamefont {D.~E.}\ \bibnamefont
  {Vanpoucke}}\ and\ \bibinfo {author} {\bibfnamefont {G.}~\bibnamefont
  {Brocks}},\ }\href@noop {} {\bibfield  {journal} {\bibinfo  {journal}
  {Physical Review B}\ }\textbf {\bibinfo {volume} {\textbf{77}}},\ \bibinfo
  {pages} {241308} (\bibinfo {year} {2008})}\BibitemShut {NoStop}%
\bibitem [{\citenamefont {Kirklin}\ \emph {et~al.}(2015)\citenamefont
  {Kirklin}, \citenamefont {Saal}, \citenamefont {Meredig}, \citenamefont
  {Thompson}, \citenamefont {Doak}, \citenamefont {Aykol}, \citenamefont
  {R{\"u}hl},\ and\ \citenamefont {Wolverton}}]{kirklin2015open}%
  \BibitemOpen
  \bibfield  {author} {\bibinfo {author} {\bibfnamefont {S.}~\bibnamefont
  {Kirklin}}, \bibinfo {author} {\bibfnamefont {J.~E.}\ \bibnamefont {Saal}},
  \bibinfo {author} {\bibfnamefont {B.}~\bibnamefont {Meredig}}, \bibinfo
  {author} {\bibfnamefont {A.}~\bibnamefont {Thompson}}, \bibinfo {author}
  {\bibfnamefont {J.~W.}\ \bibnamefont {Doak}}, \bibinfo {author}
  {\bibfnamefont {M.}~\bibnamefont {Aykol}}, \bibinfo {author} {\bibfnamefont
  {S.}~\bibnamefont {R{\"u}hl}},\ and\ \bibinfo {author} {\bibfnamefont
  {C.}~\bibnamefont {Wolverton}},\ }\href@noop {} {\bibfield  {journal}
  {\bibinfo  {journal} {npj Computational Materials}\ }\textbf {\bibinfo
  {volume} {\textbf{1}}},\ \bibinfo {pages} {1} (\bibinfo {year}
  {2015})}\BibitemShut {NoStop}%
\bibitem [{SI()}]{SI}%
  \BibitemOpen
  \href@noop {} {\bibinfo  {journal} {See Supplemental Material for optimized
  structure, partial and total density of states, GGA+SOC and HSE+SOC band
  gaps, electronic polarizability, charged defect formation ener gies, and
  thermal stability}\ }\BibitemShut {NoStop}%
\bibitem [{\citenamefont {Komsa}\ and\ \citenamefont
  {Krasheninnikov}(2015)}]{komsa2015native}%
  \BibitemOpen
\bibfield  {journal} {  }\bibfield  {author} {\bibinfo {author} {\bibfnamefont
  {H.-P.}\ \bibnamefont {Komsa}}\ and\ \bibinfo {author} {\bibfnamefont
  {A.~V.}\ \bibnamefont {Krasheninnikov}},\ }\href@noop {} {\bibfield
  {journal} {\bibinfo  {journal} {Physical Review B}\ }\textbf {\bibinfo
  {volume} {\textbf{91}}},\ \bibinfo {pages} {125304} (\bibinfo {year}
  {2015})}\BibitemShut {NoStop}%
\bibitem [{\citenamefont {Nguyen}\ \emph {et~al.}(2018)\citenamefont {Nguyen},
  \citenamefont {Liang}, \citenamefont {Zou}, \citenamefont {Fu}, \citenamefont
  {Oyedele}, \citenamefont {Sumpter}, \citenamefont {Liu}, \citenamefont {Gai},
  \citenamefont {Xiao},\ and\ \citenamefont {Li}}]{nguyen20183d}%
  \BibitemOpen
  \bibfield  {author} {\bibinfo {author} {\bibfnamefont {G.~D.}\ \bibnamefont
  {Nguyen}}, \bibinfo {author} {\bibfnamefont {L.}~\bibnamefont {Liang}},
  \bibinfo {author} {\bibfnamefont {Q.}~\bibnamefont {Zou}}, \bibinfo {author}
  {\bibfnamefont {M.}~\bibnamefont {Fu}}, \bibinfo {author} {\bibfnamefont
  {A.~D.}\ \bibnamefont {Oyedele}}, \bibinfo {author} {\bibfnamefont {B.~G.}\
  \bibnamefont {Sumpter}}, \bibinfo {author} {\bibfnamefont {Z.}~\bibnamefont
  {Liu}}, \bibinfo {author} {\bibfnamefont {Z.}~\bibnamefont {Gai}}, \bibinfo
  {author} {\bibfnamefont {K.}~\bibnamefont {Xiao}},\ and\ \bibinfo {author}
  {\bibfnamefont {A.-P.}\ \bibnamefont {Li}},\ }\href@noop {} {\bibfield
  {journal} {\bibinfo  {journal} {Physical Review Letters}\ }\textbf {\bibinfo
  {volume} {\textbf{121}}},\ \bibinfo {pages} {086101} (\bibinfo {year}
  {2018})}\BibitemShut {NoStop}%
\bibitem [{\citenamefont {Lin}\ \emph {et~al.}(2017)\citenamefont {Lin},
  \citenamefont {Zuluaga}, \citenamefont {Yu}, \citenamefont {Liu},
  \citenamefont {Pantelides},\ and\ \citenamefont {Suenaga}}]{lin2017novel}%
  \BibitemOpen
  \bibfield  {author} {\bibinfo {author} {\bibfnamefont {J.}~\bibnamefont
  {Lin}}, \bibinfo {author} {\bibfnamefont {S.}~\bibnamefont {Zuluaga}},
  \bibinfo {author} {\bibfnamefont {P.}~\bibnamefont {Yu}}, \bibinfo {author}
  {\bibfnamefont {Z.}~\bibnamefont {Liu}}, \bibinfo {author} {\bibfnamefont
  {S.~T.}\ \bibnamefont {Pantelides}},\ and\ \bibinfo {author} {\bibfnamefont
  {K.}~\bibnamefont {Suenaga}},\ }\href@noop {} {\bibfield  {journal} {\bibinfo
   {journal} {Physical review letters}\ }\textbf {\bibinfo {volume}
  {\textbf{119}}},\ \bibinfo {pages} {016101} (\bibinfo {year}
  {2017})}\BibitemShut {NoStop}%
\bibitem [{\citenamefont {Martin}(2004)}]{martin_2004}%
  \BibitemOpen
  \bibfield  {author} {\bibinfo {author} {\bibfnamefont {R.~M.}\ \bibnamefont
  {Martin}},\ }\href {https://doi.org/10.1017/CBO9780511805769} {}\ (\bibinfo
  {publisher} {Cambridge University Press},\ \bibinfo {year}
  {2004})\BibitemShut {NoStop}%
\bibitem [{\citenamefont {Zheng}\ \emph {et~al.}(2006)\citenamefont {Zheng},
  \citenamefont {Ceder}, \citenamefont {Maxisch}, \citenamefont {Chim},\ and\
  \citenamefont {Choi}}]{zheng2006native}%
  \BibitemOpen
  \bibfield  {author} {\bibinfo {author} {\bibfnamefont {J.}~\bibnamefont
  {Zheng}}, \bibinfo {author} {\bibfnamefont {G.}~\bibnamefont {Ceder}},
  \bibinfo {author} {\bibfnamefont {T.}~\bibnamefont {Maxisch}}, \bibinfo
  {author} {\bibfnamefont {W.~K.}\ \bibnamefont {Chim}},\ and\ \bibinfo
  {author} {\bibfnamefont {W.~K.}\ \bibnamefont {Choi}},\ }\href@noop {}
  {\bibfield  {journal} {\bibinfo  {journal} {Physical Review B}\ }\textbf
  {\bibinfo {volume} {\textbf{73}}},\ \bibinfo {pages} {104101} (\bibinfo
  {year} {2006})}\BibitemShut {NoStop}%
\bibitem [{\citenamefont {Kohan}\ \emph {et~al.}(2000)\citenamefont {Kohan},
  \citenamefont {Ceder}, \citenamefont {Morgan},\ and\ \citenamefont {Van~de
  Walle}}]{kohan2000first}%
  \BibitemOpen
  \bibfield  {author} {\bibinfo {author} {\bibfnamefont {A.}~\bibnamefont
  {Kohan}}, \bibinfo {author} {\bibfnamefont {G.}~\bibnamefont {Ceder}},
  \bibinfo {author} {\bibfnamefont {D.}~\bibnamefont {Morgan}},\ and\ \bibinfo
  {author} {\bibfnamefont {C.~G.}\ \bibnamefont {Van~de Walle}},\ }\href@noop
  {} {\bibfield  {journal} {\bibinfo  {journal} {Physical Review B}\ }\textbf
  {\bibinfo {volume} {\textbf{61}}},\ \bibinfo {pages} {15019} (\bibinfo {year}
  {2000})}\BibitemShut {NoStop}%
\bibitem [{\citenamefont {Zhang}\ \emph {et~al.}(2001)\citenamefont {Zhang},
  \citenamefont {Wei},\ and\ \citenamefont {Zunger}}]{zhang2001intrinsic}%
  \BibitemOpen
  \bibfield  {author} {\bibinfo {author} {\bibfnamefont {S.}~\bibnamefont
  {Zhang}}, \bibinfo {author} {\bibfnamefont {S.-H.}\ \bibnamefont {Wei}},\
  and\ \bibinfo {author} {\bibfnamefont {A.}~\bibnamefont {Zunger}},\
  }\href@noop {} {\bibfield  {journal} {\bibinfo  {journal} {Physical Review
  B}\ }\textbf {\bibinfo {volume} {\textbf{63}}},\ \bibinfo {pages} {075205}
  (\bibinfo {year} {2001})}\BibitemShut {NoStop}%
\bibitem [{\citenamefont {Noh}\ \emph {et~al.}(2014)\citenamefont {Noh},
  \citenamefont {Kim},\ and\ \citenamefont {Kim}}]{noh2014stability}%
  \BibitemOpen
  \bibfield  {author} {\bibinfo {author} {\bibfnamefont {J.-Y.}\ \bibnamefont
  {Noh}}, \bibinfo {author} {\bibfnamefont {H.}~\bibnamefont {Kim}},\ and\
  \bibinfo {author} {\bibfnamefont {Y.-S.}\ \bibnamefont {Kim}},\ }\href@noop
  {} {\bibfield  {journal} {\bibinfo  {journal} {Physical Review B}\ }\textbf
  {\bibinfo {volume} {\textbf{89}}},\ \bibinfo {pages} {205417} (\bibinfo
  {year} {2014})}\BibitemShut {NoStop}%
\bibitem [{\citenamefont {Sharma}\ \emph {et~al.}(2023)\citenamefont {Sharma},
  \citenamefont {Mishra},\ and\ \citenamefont {Shukla}}]{sharma2023influence}%
  \BibitemOpen
  \bibfield  {author} {\bibinfo {author} {\bibfnamefont {P.}~\bibnamefont
  {Sharma}}, \bibinfo {author} {\bibfnamefont {V.}~\bibnamefont {Mishra}},\
  and\ \bibinfo {author} {\bibfnamefont {A.}~\bibnamefont {Shukla}},\
  }\href@noop {} {\bibfield  {journal} {\bibinfo  {journal} {Journal of
  Physics: Condensed Matter}\ }\textbf {\bibinfo {volume} {\textbf{35}}},\
  \bibinfo {pages} {345501} (\bibinfo {year} {2023})}\BibitemShut {NoStop}%
\bibitem [{\citenamefont {Kuklin}\ \emph {et~al.}(2021)\citenamefont {Kuklin},
  \citenamefont {Begunovich}, \citenamefont {Gao}, \citenamefont {Zhang},\ and\
  \citenamefont {\AA{}gren}}]{kuklin2021point}%
  \BibitemOpen
  \bibfield  {author} {\bibinfo {author} {\bibfnamefont {A.~V.}\ \bibnamefont
  {Kuklin}}, \bibinfo {author} {\bibfnamefont {L.~V.}\ \bibnamefont
  {Begunovich}}, \bibinfo {author} {\bibfnamefont {L.}~\bibnamefont {Gao}},
  \bibinfo {author} {\bibfnamefont {H.}~\bibnamefont {Zhang}},\ and\ \bibinfo
  {author} {\bibfnamefont {H.}~\bibnamefont {\AA{}gren}},\ }\href@noop {}
  {\bibfield  {journal} {\bibinfo  {journal} {Phys. Rev. B}\ }\textbf {\bibinfo
  {volume} {\textbf{104}}},\ \bibinfo {pages} {134109} (\bibinfo {year}
  {2021})}\BibitemShut {NoStop}%
\bibitem [{\citenamefont {Horzum}\ \emph {et~al.}(2014)\citenamefont {Horzum},
  \citenamefont {\ifmmode \mbox{\c{C}}\else \c{C}\fi{}ak\ifmmode \imath
  \else~\i \fi{}r}, \citenamefont {Suh}, \citenamefont {Tongay}, \citenamefont
  {Huang}, \citenamefont {Ho}, \citenamefont {Wu}, \citenamefont {Sahin},\ and\
  \citenamefont {Peeters}}]{horzum2014formation}%
  \BibitemOpen
  \bibfield  {author} {\bibinfo {author} {\bibfnamefont {S.}~\bibnamefont
  {Horzum}}, \bibinfo {author} {\bibfnamefont {D.}~\bibnamefont {\ifmmode
  \mbox{\c{C}}\else \c{C}\fi{}ak\ifmmode \imath \else~\i \fi{}r}}, \bibinfo
  {author} {\bibfnamefont {J.}~\bibnamefont {Suh}}, \bibinfo {author}
  {\bibfnamefont {S.}~\bibnamefont {Tongay}}, \bibinfo {author} {\bibfnamefont
  {Y.-S.}\ \bibnamefont {Huang}}, \bibinfo {author} {\bibfnamefont {C.-H.}\
  \bibnamefont {Ho}}, \bibinfo {author} {\bibfnamefont {J.}~\bibnamefont {Wu}},
  \bibinfo {author} {\bibfnamefont {H.}~\bibnamefont {Sahin}},\ and\ \bibinfo
  {author} {\bibfnamefont {F.~M.}\ \bibnamefont {Peeters}},\ }\href@noop {}
  {\bibfield  {journal} {\bibinfo  {journal} {Phys. Rev. B}\ }\textbf {\bibinfo
  {volume} {\textbf{89}}},\ \bibinfo {pages} {155433} (\bibinfo {year}
  {2014})}\BibitemShut {NoStop}%
\bibitem [{\citenamefont {B{\"o}rner}\ \emph {et~al.}(2016)\citenamefont
  {B{\"o}rner}, \citenamefont {Kaiser},\ and\ \citenamefont
  {Lehtinen}}]{borner2016evidence}%
  \BibitemOpen
  \bibfield  {author} {\bibinfo {author} {\bibfnamefont {P.}~\bibnamefont
  {B{\"o}rner}}, \bibinfo {author} {\bibfnamefont {U.}~\bibnamefont {Kaiser}},\
  and\ \bibinfo {author} {\bibfnamefont {O.}~\bibnamefont {Lehtinen}},\
  }\href@noop {} {\bibfield  {journal} {\bibinfo  {journal} {Physical Review
  B}\ }\textbf {\bibinfo {volume} {\textbf{93}}},\ \bibinfo {pages} {134104}
  (\bibinfo {year} {2016})}\BibitemShut {NoStop}%
\bibitem [{\citenamefont {Ugeda}\ \emph {et~al.}(2010)\citenamefont {Ugeda},
  \citenamefont {Brihuega}, \citenamefont {Guinea},\ and\ \citenamefont
  {G\'{o}mez-Rodr\'{\i}guez}}]{ugeda2010missing}%
  \BibitemOpen
  \bibfield  {author} {\bibinfo {author} {\bibfnamefont {M.~M.}\ \bibnamefont
  {Ugeda}}, \bibinfo {author} {\bibfnamefont {I.}~\bibnamefont {Brihuega}},
  \bibinfo {author} {\bibfnamefont {F.}~\bibnamefont {Guinea}},\ and\ \bibinfo
  {author} {\bibfnamefont {J.~M.}\ \bibnamefont {G\'{o}mez-Rodr\'{\i}guez}},\
  }\href@noop {} {\bibfield  {journal} {\bibinfo  {journal} {Physical Review
  Letters}\ }\textbf {\bibinfo {volume} {\textbf{104}}},\ \bibinfo {pages}
  {096804} (\bibinfo {year} {2010})}\BibitemShut {NoStop}%
\bibitem [{\citenamefont {Freysoldt}\ \emph {et~al.}(2014)\citenamefont
  {Freysoldt}, \citenamefont {Grabowski}, \citenamefont {Hickel}, \citenamefont
  {Neugebauer}, \citenamefont {Kresse}, \citenamefont {Janotti},\ and\
  \citenamefont {Van~de Walle}}]{RevModPhys.86.253}%
  \BibitemOpen
  \bibfield  {author} {\bibinfo {author} {\bibfnamefont {C.}~\bibnamefont
  {Freysoldt}}, \bibinfo {author} {\bibfnamefont {B.}~\bibnamefont
  {Grabowski}}, \bibinfo {author} {\bibfnamefont {T.}~\bibnamefont {Hickel}},
  \bibinfo {author} {\bibfnamefont {J.}~\bibnamefont {Neugebauer}}, \bibinfo
  {author} {\bibfnamefont {G.}~\bibnamefont {Kresse}}, \bibinfo {author}
  {\bibfnamefont {A.}~\bibnamefont {Janotti}},\ and\ \bibinfo {author}
  {\bibfnamefont {C.~G.}\ \bibnamefont {Van~de Walle}},\ }\href@noop {}
  {\bibfield  {journal} {\bibinfo  {journal} {Rev. Mod. Phys.}\ }\textbf
  {\bibinfo {volume} {\textbf{86}}},\ \bibinfo {pages} {253} (\bibinfo {year}
  {2014})}\BibitemShut {NoStop}%
\bibitem [{\citenamefont {Tan}\ \emph {et~al.}(2020)\citenamefont {Tan},
  \citenamefont {Freysoldt},\ and\ \citenamefont {Hennig}}]{tan2020stability}%
  \BibitemOpen
  \bibfield  {author} {\bibinfo {author} {\bibfnamefont {A.~M.~Z.}\
  \bibnamefont {Tan}}, \bibinfo {author} {\bibfnamefont {C.}~\bibnamefont
  {Freysoldt}},\ and\ \bibinfo {author} {\bibfnamefont {R.~G.}\ \bibnamefont
  {Hennig}},\ }\href@noop {} {\bibfield  {journal} {\bibinfo  {journal}
  {Physical Review Materials}\ }\textbf {\bibinfo {volume} {\textbf{4}}},\
  \bibinfo {pages} {064004} (\bibinfo {year} {2020})}\BibitemShut {NoStop}%
\bibitem [{\citenamefont {Komsa}\ \emph
  {et~al.}(2012{\natexlab{b}})\citenamefont {Komsa}, \citenamefont {Rantala},\
  and\ \citenamefont {Pasquarello}}]{komsa2012finite}%
  \BibitemOpen
  \bibfield  {author} {\bibinfo {author} {\bibfnamefont {H.-P.}\ \bibnamefont
  {Komsa}}, \bibinfo {author} {\bibfnamefont {T.~T.}\ \bibnamefont {Rantala}},\
  and\ \bibinfo {author} {\bibfnamefont {A.}~\bibnamefont {Pasquarello}},\
  }\href@noop {} {\bibfield  {journal} {\bibinfo  {journal} {Physical Review
  B}\ }\textbf {\bibinfo {volume} {\textbf{86}}},\ \bibinfo {pages} {045112}
  (\bibinfo {year} {2012}{\natexlab{b}})}\BibitemShut {NoStop}%
\bibitem [{\citenamefont {Fu}\ \emph {et~al.}(2019)\citenamefont {Fu},
  \citenamefont {Liang}, \citenamefont {Zou}, \citenamefont {Nguyen},
  \citenamefont {Xiao}, \citenamefont {Li}, \citenamefont {Kang}, \citenamefont
  {Wu},\ and\ \citenamefont {Gai}}]{fu2019defects}%
  \BibitemOpen
  \bibfield  {author} {\bibinfo {author} {\bibfnamefont {M.}~\bibnamefont
  {Fu}}, \bibinfo {author} {\bibfnamefont {L.}~\bibnamefont {Liang}}, \bibinfo
  {author} {\bibfnamefont {Q.}~\bibnamefont {Zou}}, \bibinfo {author}
  {\bibfnamefont {G.~D.}\ \bibnamefont {Nguyen}}, \bibinfo {author}
  {\bibfnamefont {K.}~\bibnamefont {Xiao}}, \bibinfo {author} {\bibfnamefont
  {A.-P.}\ \bibnamefont {Li}}, \bibinfo {author} {\bibfnamefont
  {J.}~\bibnamefont {Kang}}, \bibinfo {author} {\bibfnamefont {Z.}~\bibnamefont
  {Wu}},\ and\ \bibinfo {author} {\bibfnamefont {Z.}~\bibnamefont {Gai}},\
  }\href@noop {} {\bibfield  {journal} {\bibinfo  {journal} {The Journal of
  Physical Chemistry Letters}\ }\textbf {\bibinfo {volume} {\textbf{11}}},\
  \bibinfo {pages} {740} (\bibinfo {year} {2019})}\BibitemShut {NoStop}%
\bibitem [{\citenamefont {Hla}\ \emph {et~al.}(2000)\citenamefont {Hla},
  \citenamefont {Bartels}, \citenamefont {Meyer},\ and\ \citenamefont
  {Rieder}}]{hla2000inducing}%
  \BibitemOpen
  \bibfield  {author} {\bibinfo {author} {\bibfnamefont {S.-W.}\ \bibnamefont
  {Hla}}, \bibinfo {author} {\bibfnamefont {L.}~\bibnamefont {Bartels}},
  \bibinfo {author} {\bibfnamefont {G.}~\bibnamefont {Meyer}},\ and\ \bibinfo
  {author} {\bibfnamefont {K.-H.}\ \bibnamefont {Rieder}},\ }\href@noop {}
  {\bibfield  {journal} {\bibinfo  {journal} {Physical review letters}\
  }\textbf {\bibinfo {volume} {\textbf{85}}},\ \bibinfo {pages} {2777}
  (\bibinfo {year} {2000})}\BibitemShut {NoStop}%
\bibitem [{\citenamefont {Komsa}\ and\ \citenamefont
  {Pasquarello}(2013)}]{komsa2013finite}%
  \BibitemOpen
  \bibfield  {author} {\bibinfo {author} {\bibfnamefont {H.-P.}\ \bibnamefont
  {Komsa}}\ and\ \bibinfo {author} {\bibfnamefont {A.}~\bibnamefont
  {Pasquarello}},\ }\href@noop {} {\bibfield  {journal} {\bibinfo  {journal}
  {Physical review letters}\ }\textbf {\bibinfo {volume} {\textbf{110}}},\
  \bibinfo {pages} {095505} (\bibinfo {year} {2013})}\BibitemShut {NoStop}%
\bibitem [{\citenamefont {Wang}\ \emph
  {et~al.}(2015{\natexlab{b}})\citenamefont {Wang}, \citenamefont {Han},
  \citenamefont {Li}, \citenamefont {Xie}, \citenamefont {Chen}, \citenamefont
  {Tian}, \citenamefont {West}, \citenamefont {Sun},\ and\ \citenamefont
  {Zhang}}]{wang2015determination}%
  \BibitemOpen
  \bibfield  {author} {\bibinfo {author} {\bibfnamefont {D.}~\bibnamefont
  {Wang}}, \bibinfo {author} {\bibfnamefont {D.}~\bibnamefont {Han}}, \bibinfo
  {author} {\bibfnamefont {X.-B.}\ \bibnamefont {Li}}, \bibinfo {author}
  {\bibfnamefont {S.-Y.}\ \bibnamefont {Xie}}, \bibinfo {author} {\bibfnamefont
  {N.-K.}\ \bibnamefont {Chen}}, \bibinfo {author} {\bibfnamefont {W.~Q.}\
  \bibnamefont {Tian}}, \bibinfo {author} {\bibfnamefont {D.}~\bibnamefont
  {West}}, \bibinfo {author} {\bibfnamefont {H.-B.}\ \bibnamefont {Sun}},\ and\
  \bibinfo {author} {\bibfnamefont {S.}~\bibnamefont {Zhang}},\ }\href@noop {}
  {\bibfield  {journal} {\bibinfo  {journal} {Physical review letters}\
  }\textbf {\bibinfo {volume} {\textbf{114}}},\ \bibinfo {pages} {196801}
  (\bibinfo {year} {2015}{\natexlab{b}})}\BibitemShut {NoStop}%
\bibitem [{\citenamefont {Komsa}\ \emph {et~al.}(2018)\citenamefont {Komsa},
  \citenamefont {Berseneva}, \citenamefont {Krasheninnikov},\ and\
  \citenamefont {Nieminen}}]{komsa2018erratum}%
  \BibitemOpen
  \bibfield  {author} {\bibinfo {author} {\bibfnamefont {H.-P.}\ \bibnamefont
  {Komsa}}, \bibinfo {author} {\bibfnamefont {N.}~\bibnamefont {Berseneva}},
  \bibinfo {author} {\bibfnamefont {A.~V.}\ \bibnamefont {Krasheninnikov}},\
  and\ \bibinfo {author} {\bibfnamefont {R.~M.}\ \bibnamefont {Nieminen}},\
  }\href@noop {} {\bibfield  {journal} {\bibinfo  {journal} {Physical Review
  X}\ }\textbf {\bibinfo {volume} {\textbf{8}}},\ \bibinfo {pages} {039902}
  (\bibinfo {year} {2018})}\BibitemShut {NoStop}%
\bibitem [{\citenamefont {Henkelman}\ \emph {et~al.}(2000)\citenamefont
  {Henkelman}, \citenamefont {Uberuaga},\ and\ \citenamefont
  {J{\'o}nsson}}]{henkelman2000climbing}%
  \BibitemOpen
  \bibfield  {author} {\bibinfo {author} {\bibfnamefont {G.}~\bibnamefont
  {Henkelman}}, \bibinfo {author} {\bibfnamefont {B.~P.}\ \bibnamefont
  {Uberuaga}},\ and\ \bibinfo {author} {\bibfnamefont {H.}~\bibnamefont
  {J{\'o}nsson}},\ }\href@noop {} {\bibfield  {journal} {\bibinfo  {journal}
  {The Journal of chemical physics}\ }\textbf {\bibinfo {volume}
  {\textbf{113}}},\ \bibinfo {pages} {9901} (\bibinfo {year}
  {2000})}\BibitemShut {NoStop}%
\bibitem [{\citenamefont {Komsa}\ \emph {et~al.}(2013)\citenamefont {Komsa},
  \citenamefont {Kurasch}, \citenamefont {Lehtinen}, \citenamefont {Kaiser},\
  and\ \citenamefont {Krasheninnikov}}]{komsa2013point}%
  \BibitemOpen
  \bibfield  {author} {\bibinfo {author} {\bibfnamefont {H.-P.}\ \bibnamefont
  {Komsa}}, \bibinfo {author} {\bibfnamefont {S.}~\bibnamefont {Kurasch}},
  \bibinfo {author} {\bibfnamefont {O.}~\bibnamefont {Lehtinen}}, \bibinfo
  {author} {\bibfnamefont {U.}~\bibnamefont {Kaiser}},\ and\ \bibinfo {author}
  {\bibfnamefont {A.~V.}\ \bibnamefont {Krasheninnikov}},\ }\href@noop {}
  {\bibfield  {journal} {\bibinfo  {journal} {Physical Review B}\ }\textbf
  {\bibinfo {volume} {\textbf{88}}},\ \bibinfo {pages} {035301} (\bibinfo
  {year} {2013})}\BibitemShut {NoStop}%
\bibitem [{\citenamefont {Wang}\ and\ \citenamefont
  {Seebauer}(2001)}]{WANG2001111}%
  \BibitemOpen
  \bibfield  {author} {\bibinfo {author} {\bibfnamefont {Z.}~\bibnamefont
  {Wang}}\ and\ \bibinfo {author} {\bibfnamefont {E.}~\bibnamefont
  {Seebauer}},\ }\href@noop {} {\bibfield  {journal} {\bibinfo  {journal}
  {Applied Surface Science}\ }\textbf {\bibinfo {volume} {\textbf{181}}},\
  \bibinfo {pages} {111} (\bibinfo {year} {2001})}\BibitemShut {NoStop}%
\bibitem [{\citenamefont {Krasheninnikov}\ and\ \citenamefont
  {Nordlund}(2010)}]{krasheninnikov2010ion}%
  \BibitemOpen
  \bibfield  {author} {\bibinfo {author} {\bibfnamefont {A.}~\bibnamefont
  {Krasheninnikov}}\ and\ \bibinfo {author} {\bibfnamefont {K.}~\bibnamefont
  {Nordlund}},\ }\href@noop {} {\bibfield  {journal} {\bibinfo  {journal}
  {Journal of applied physics}\ }\textbf {\bibinfo {volume} {\textbf{107}}},\
  \bibinfo {pages} {3} (\bibinfo {year} {2010})}\BibitemShut {NoStop}%
\bibitem [{\citenamefont {Kotakoski}\ \emph {et~al.}(2010)\citenamefont
  {Kotakoski}, \citenamefont {Jin}, \citenamefont {Lehtinen}, \citenamefont
  {Suenaga},\ and\ \citenamefont {Krasheninnikov}}]{kotakoski2010electron}%
  \BibitemOpen
  \bibfield  {author} {\bibinfo {author} {\bibfnamefont {J.}~\bibnamefont
  {Kotakoski}}, \bibinfo {author} {\bibfnamefont {C.~H.}\ \bibnamefont {Jin}},
  \bibinfo {author} {\bibfnamefont {O.}~\bibnamefont {Lehtinen}}, \bibinfo
  {author} {\bibfnamefont {K.}~\bibnamefont {Suenaga}},\ and\ \bibinfo {author}
  {\bibfnamefont {A.~V.}\ \bibnamefont {Krasheninnikov}},\ }\href@noop {}
  {\bibfield  {journal} {\bibinfo  {journal} {Physical Review B}\ }\textbf
  {\bibinfo {volume} {\textbf{82}}},\ \bibinfo {pages} {113404} (\bibinfo
  {year} {2010})}\BibitemShut {NoStop}%
\bibitem [{\citenamefont {Zobelli}\ \emph {et~al.}(2007)\citenamefont
  {Zobelli}, \citenamefont {Gloter}, \citenamefont {Ewels}, \citenamefont
  {Seifert},\ and\ \citenamefont {Colliex}}]{zobelli2007electron}%
  \BibitemOpen
  \bibfield  {author} {\bibinfo {author} {\bibfnamefont {A.}~\bibnamefont
  {Zobelli}}, \bibinfo {author} {\bibfnamefont {A.}~\bibnamefont {Gloter}},
  \bibinfo {author} {\bibfnamefont {C.}~\bibnamefont {Ewels}}, \bibinfo
  {author} {\bibfnamefont {G.}~\bibnamefont {Seifert}},\ and\ \bibinfo {author}
  {\bibfnamefont {C.}~\bibnamefont {Colliex}},\ }\href@noop {} {\bibfield
  {journal} {\bibinfo  {journal} {Physical Review B}\ }\textbf {\bibinfo
  {volume} {\textbf{75}}},\ \bibinfo {pages} {245402} (\bibinfo {year}
  {2007})}\BibitemShut {NoStop}%
\bibitem [{\citenamefont {Garcia}\ \emph {et~al.}(2014)\citenamefont {Garcia},
  \citenamefont {Raya}, \citenamefont {Mariscal}, \citenamefont {Esparza},
  \citenamefont {Herrera}, \citenamefont {Molina}, \citenamefont {Scavello},
  \citenamefont {Galindo}, \citenamefont {Jose-Yacaman},\ and\ \citenamefont
  {Ponce}}]{garcia2014analysis}%
  \BibitemOpen
  \bibfield  {author} {\bibinfo {author} {\bibfnamefont {A.}~\bibnamefont
  {Garcia}}, \bibinfo {author} {\bibfnamefont {A.~M.}\ \bibnamefont {Raya}},
  \bibinfo {author} {\bibfnamefont {M.~M.}\ \bibnamefont {Mariscal}}, \bibinfo
  {author} {\bibfnamefont {R.}~\bibnamefont {Esparza}}, \bibinfo {author}
  {\bibfnamefont {M.}~\bibnamefont {Herrera}}, \bibinfo {author} {\bibfnamefont
  {S.~I.}\ \bibnamefont {Molina}}, \bibinfo {author} {\bibfnamefont
  {G.}~\bibnamefont {Scavello}}, \bibinfo {author} {\bibfnamefont {P.~L.}\
  \bibnamefont {Galindo}}, \bibinfo {author} {\bibfnamefont {M.}~\bibnamefont
  {Jose-Yacaman}},\ and\ \bibinfo {author} {\bibfnamefont {A.}~\bibnamefont
  {Ponce}},\ }\href@noop {} {\bibfield  {journal} {\bibinfo  {journal}
  {Ultramicroscopy}\ }\textbf {\bibinfo {volume} {\textbf{146}}},\ \bibinfo
  {pages} {33} (\bibinfo {year} {2014})}\BibitemShut {NoStop}%
\bibitem [{\citenamefont {Kretschmer}\ \emph {et~al.}(2020)\citenamefont
  {Kretschmer}, \citenamefont {Lehnert}, \citenamefont {Kaiser},\ and\
  \citenamefont {Krasheninnikov}}]{kretschmer2020formation}%
  \BibitemOpen
  \bibfield  {author} {\bibinfo {author} {\bibfnamefont {S.}~\bibnamefont
  {Kretschmer}}, \bibinfo {author} {\bibfnamefont {T.}~\bibnamefont {Lehnert}},
  \bibinfo {author} {\bibfnamefont {U.}~\bibnamefont {Kaiser}},\ and\ \bibinfo
  {author} {\bibfnamefont {A.~V.}\ \bibnamefont {Krasheninnikov}},\ }\href@noop
  {} {\bibfield  {journal} {\bibinfo  {journal} {Nano letters}\ }\textbf
  {\bibinfo {volume} {\textbf{20}}},\ \bibinfo {pages} {2865} (\bibinfo {year}
  {2020})}\BibitemShut {NoStop}%
\bibitem [{\citenamefont {Yadav}\ \emph {et~al.}(2023)\citenamefont {Yadav},
  \citenamefont {Acosta}, \citenamefont {Dalpian},\ and\ \citenamefont
  {Malyi}}]{YADAV20232711}%
  \BibitemOpen
  \bibfield  {author} {\bibinfo {author} {\bibfnamefont {A.}~\bibnamefont
  {Yadav}}, \bibinfo {author} {\bibfnamefont {C.~M.}\ \bibnamefont {Acosta}},
  \bibinfo {author} {\bibfnamefont {G.~M.}\ \bibnamefont {Dalpian}},\ and\
  \bibinfo {author} {\bibfnamefont {O.~I.}\ \bibnamefont {Malyi}},\ }\href@noop
  {} {\bibfield  {journal} {\bibinfo  {journal} {Matter}\ }\textbf {\bibinfo
  {volume} {\textbf{6}}},\ \bibinfo {pages} {2711} (\bibinfo {year}
  {2023})}\BibitemShut {NoStop}%
\end{thebibliography}%


\begin{thebibliography}{81}%
\makeatletter
\providecommand \@ifxundefined [1]{%
 \@ifx{#1\undefined}
}%
\providecommand \@ifnum [1]{%
 \ifnum #1\expandafter \@firstoftwo
 \else \expandafter \@secondoftwo
 \fi
}%
\providecommand \@ifx [1]{%
 \ifx #1\expandafter \@firstoftwo
 \else \expandafter \@secondoftwo
 \fi
}%
\providecommand \natexlab [1]{#1}%
\providecommand \enquote  [1]{``#1''}%
\providecommand \bibnamefont  [1]{#1}%
\providecommand \bibfnamefont [1]{#1}%
\providecommand \citenamefont [1]{#1}%
\providecommand \href@noop [0]{\@secondoftwo}%
\providecommand \href [0]{\begingroup \@sanitize@url \@href}%
\providecommand \@href[1]{\@@startlink{#1}\@@href}%
\providecommand \@@href[1]{\endgroup#1\@@endlink}%
\providecommand \@sanitize@url [0]{\catcode `\\12\catcode `\$12\catcode
  `\&12\catcode `\#12\catcode `\^12\catcode `\_12\catcode `\%12\relax}%
\providecommand \@@startlink[1]{}%
\providecommand \@@endlink[0]{}%
\providecommand \url  [0]{\begingroup\@sanitize@url \@url }%
\providecommand \@url [1]{\endgroup\@href {#1}{\urlprefix }}%
\providecommand \urlprefix  [0]{URL }%
\providecommand \Eprint [0]{\href }%
\providecommand \doibase [0]{https://doi.org/}%
\providecommand \selectlanguage [0]{\@gobble}%
\providecommand \bibinfo  [0]{\@secondoftwo}%
\providecommand \bibfield  [0]{\@secondoftwo}%
\providecommand \translation [1]{[#1]}%
\providecommand \BibitemOpen [0]{}%
\providecommand \bibitemStop [0]{}%
\providecommand \bibitemNoStop [0]{.\EOS\space}%
\providecommand \EOS [0]{\spacefactor3000\relax}%
\providecommand \BibitemShut  [1]{\csname bibitem#1\endcsname}%
\let\auto@bib@innerbib\@empty
%</preamble>
\bibitem [{\citenamefont {Manzeli}\ \emph {et~al.}(2017)\citenamefont
  {Manzeli}, \citenamefont {Ovchinnikov}, \citenamefont {Pasquier},
  \citenamefont {Yazyev},\ and\ \citenamefont {Kis}}]{manzeli20172d}%
  \BibitemOpen
  \bibfield  {author} {\bibinfo {author} {\bibfnamefont {S.}~\bibnamefont
  {Manzeli}}, \bibinfo {author} {\bibfnamefont {D.}~\bibnamefont
  {Ovchinnikov}}, \bibinfo {author} {\bibfnamefont {D.}~\bibnamefont
  {Pasquier}}, \bibinfo {author} {\bibfnamefont {O.~V.}\ \bibnamefont
  {Yazyev}},\ and\ \bibinfo {author} {\bibfnamefont {A.}~\bibnamefont {Kis}},\
  }\href@noop {} {\bibfield  {journal} {\bibinfo  {journal} {Nature Reviews
  Materials}\ }\textbf {\bibinfo {volume} {2}},\ \bibinfo {pages} {1} (\bibinfo
  {year} {2017})}\BibitemShut {NoStop}%
\bibitem [{\citenamefont {Mak}\ and\ \citenamefont
  {Shan}(2016)}]{mak2016photonics}%
  \BibitemOpen
  \bibfield  {author} {\bibinfo {author} {\bibfnamefont {K.~F.}\ \bibnamefont
  {Mak}}\ and\ \bibinfo {author} {\bibfnamefont {J.}~\bibnamefont {Shan}},\
  }\href@noop {} {\bibfield  {journal} {\bibinfo  {journal} {Nature Photonics}\
  }\textbf {\bibinfo {volume} {10}},\ \bibinfo {pages} {216} (\bibinfo {year}
  {2016})}\BibitemShut {NoStop}%
\bibitem [{\citenamefont {Wang}\ \emph {et~al.}(2012)\citenamefont {Wang},
  \citenamefont {Kalantar-Zadeh}, \citenamefont {Kis}, \citenamefont
  {Coleman},\ and\ \citenamefont {Strano}}]{wang2012electronics}%
  \BibitemOpen
  \bibfield  {author} {\bibinfo {author} {\bibfnamefont {Q.~H.}\ \bibnamefont
  {Wang}}, \bibinfo {author} {\bibfnamefont {K.}~\bibnamefont
  {Kalantar-Zadeh}}, \bibinfo {author} {\bibfnamefont {A.}~\bibnamefont {Kis}},
  \bibinfo {author} {\bibfnamefont {J.~N.}\ \bibnamefont {Coleman}},\ and\
  \bibinfo {author} {\bibfnamefont {M.~S.}\ \bibnamefont {Strano}},\
  }\href@noop {} {\bibfield  {journal} {\bibinfo  {journal} {Nature
  nanotechnology}\ }\textbf {\bibinfo {volume} {7}},\ \bibinfo {pages} {699}
  (\bibinfo {year} {2012})}\BibitemShut {NoStop}%
\bibitem [{\citenamefont {Radisavljevic}\ \emph {et~al.}(2011)\citenamefont
  {Radisavljevic}, \citenamefont {Radenovic}, \citenamefont {Brivio},
  \citenamefont {Giacometti},\ and\ \citenamefont
  {Kis}}]{radisavljevic2011single}%
  \BibitemOpen
  \bibfield  {author} {\bibinfo {author} {\bibfnamefont {B.}~\bibnamefont
  {Radisavljevic}}, \bibinfo {author} {\bibfnamefont {A.}~\bibnamefont
  {Radenovic}}, \bibinfo {author} {\bibfnamefont {J.}~\bibnamefont {Brivio}},
  \bibinfo {author} {\bibfnamefont {V.}~\bibnamefont {Giacometti}},\ and\
  \bibinfo {author} {\bibfnamefont {A.}~\bibnamefont {Kis}},\ }\href@noop {}
  {\bibfield  {journal} {\bibinfo  {journal} {Nature nanotechnology}\ }\textbf
  {\bibinfo {volume} {6}},\ \bibinfo {pages} {147} (\bibinfo {year}
  {2011})}\BibitemShut {NoStop}%
\bibitem [{\citenamefont {Chatterjee}\ \emph {et~al.}(2015)\citenamefont
  {Chatterjee}, \citenamefont {Zhao}, \citenamefont {Iavarone}, \citenamefont
  {Di~Capua}, \citenamefont {Castellan}, \citenamefont {Karapetrov},
  \citenamefont {Malliakas}, \citenamefont {Kanatzidis}, \citenamefont {Claus},
  \citenamefont {Ruff} \emph {et~al.}}]{chatterjee2015emergence}%
  \BibitemOpen
  \bibfield  {author} {\bibinfo {author} {\bibfnamefont {U.}~\bibnamefont
  {Chatterjee}}, \bibinfo {author} {\bibfnamefont {J.}~\bibnamefont {Zhao}},
  \bibinfo {author} {\bibfnamefont {M.}~\bibnamefont {Iavarone}}, \bibinfo
  {author} {\bibfnamefont {R.}~\bibnamefont {Di~Capua}}, \bibinfo {author}
  {\bibfnamefont {J.}~\bibnamefont {Castellan}}, \bibinfo {author}
  {\bibfnamefont {G.}~\bibnamefont {Karapetrov}}, \bibinfo {author}
  {\bibfnamefont {C.}~\bibnamefont {Malliakas}}, \bibinfo {author}
  {\bibfnamefont {M.~G.}\ \bibnamefont {Kanatzidis}}, \bibinfo {author}
  {\bibfnamefont {H.}~\bibnamefont {Claus}}, \bibinfo {author} {\bibfnamefont
  {J.}~\bibnamefont {Ruff}}, \emph {et~al.},\ }\href@noop {} {\bibfield
  {journal} {\bibinfo  {journal} {Nature communications}\ }\textbf {\bibinfo
  {volume} {6}},\ \bibinfo {pages} {6313} (\bibinfo {year} {2015})}\BibitemShut
  {NoStop}%
\bibitem [{\citenamefont {Liu}\ \emph {et~al.}(2018)\citenamefont {Liu},
  \citenamefont {Lian}, \citenamefont {Liao}, \citenamefont {Wang},
  \citenamefont {Zhong}, \citenamefont {Ding}, \citenamefont {Li},
  \citenamefont {Song}, \citenamefont {He}, \citenamefont {Ma} \emph
  {et~al.}}]{liu2018two}%
  \BibitemOpen
  \bibfield  {author} {\bibinfo {author} {\bibfnamefont {C.}~\bibnamefont
  {Liu}}, \bibinfo {author} {\bibfnamefont {C.-S.}\ \bibnamefont {Lian}},
  \bibinfo {author} {\bibfnamefont {M.-H.}\ \bibnamefont {Liao}}, \bibinfo
  {author} {\bibfnamefont {Y.}~\bibnamefont {Wang}}, \bibinfo {author}
  {\bibfnamefont {Y.}~\bibnamefont {Zhong}}, \bibinfo {author} {\bibfnamefont
  {C.}~\bibnamefont {Ding}}, \bibinfo {author} {\bibfnamefont {W.}~\bibnamefont
  {Li}}, \bibinfo {author} {\bibfnamefont {C.-L.}\ \bibnamefont {Song}},
  \bibinfo {author} {\bibfnamefont {K.}~\bibnamefont {He}}, \bibinfo {author}
  {\bibfnamefont {X.-C.}\ \bibnamefont {Ma}}, \emph {et~al.},\ }\href@noop {}
  {\bibfield  {journal} {\bibinfo  {journal} {Physical Review Materials}\
  }\textbf {\bibinfo {volume} {2}},\ \bibinfo {pages} {094001} (\bibinfo {year}
  {2018})}\BibitemShut {NoStop}%
\bibitem [{\citenamefont {Rossnagel}(2011)}]{rossnagel2011origin}%
  \BibitemOpen
  \bibfield  {author} {\bibinfo {author} {\bibfnamefont {K.}~\bibnamefont
  {Rossnagel}},\ }\href@noop {} {\bibfield  {journal} {\bibinfo  {journal}
  {Journal of Physics: Condensed Matter}\ }\textbf {\bibinfo {volume} {23}},\
  \bibinfo {pages} {213001} (\bibinfo {year} {2011})}\BibitemShut {NoStop}%
\bibitem [{\citenamefont {Rahimi}\ \emph {et~al.}(2017)\citenamefont {Rahimi},
  \citenamefont {Moghaddam}, \citenamefont {Dykstra}, \citenamefont
  {Governale},\ and\ \citenamefont {Z\"{u}licke}}]{rahimi2017unconventional}%
  \BibitemOpen
  \bibfield  {author} {\bibinfo {author} {\bibfnamefont {M.}~\bibnamefont
  {Rahimi}}, \bibinfo {author} {\bibfnamefont {A.}~\bibnamefont {Moghaddam}},
  \bibinfo {author} {\bibfnamefont {C.}~\bibnamefont {Dykstra}}, \bibinfo
  {author} {\bibfnamefont {M.}~\bibnamefont {Governale}},\ and\ \bibinfo
  {author} {\bibfnamefont {U.}~\bibnamefont {Z\"{u}licke}},\ }\href@noop {}
  {\bibfield  {journal} {\bibinfo  {journal} {Physical Review B}\ }\textbf
  {\bibinfo {volume} {95}},\ \bibinfo {pages} {104515} (\bibinfo {year}
  {2017})}\BibitemShut {NoStop}%
\bibitem [{\citenamefont {Lin}\ \emph {et~al.}(2022)\citenamefont {Lin},
  \citenamefont {Liu}, \citenamefont {Wang}, \citenamefont {Xu}, \citenamefont
  {Chen}, \citenamefont {Duan},\ and\ \citenamefont
  {Monserrat}}]{lin2022phonon}%
  \BibitemOpen
  \bibfield  {author} {\bibinfo {author} {\bibfnamefont {Z.}~\bibnamefont
  {Lin}}, \bibinfo {author} {\bibfnamefont {Y.}~\bibnamefont {Liu}}, \bibinfo
  {author} {\bibfnamefont {Z.}~\bibnamefont {Wang}}, \bibinfo {author}
  {\bibfnamefont {S.}~\bibnamefont {Xu}}, \bibinfo {author} {\bibfnamefont
  {S.}~\bibnamefont {Chen}}, \bibinfo {author} {\bibfnamefont {W.}~\bibnamefont
  {Duan}},\ and\ \bibinfo {author} {\bibfnamefont {B.}~\bibnamefont
  {Monserrat}},\ }\href@noop {} {\bibfield  {journal} {\bibinfo  {journal}
  {Physical review letters}\ }\textbf {\bibinfo {volume} {129}},\ \bibinfo
  {pages} {027401} (\bibinfo {year} {2022})}\BibitemShut {NoStop}%
\bibitem [{\citenamefont {Maguire}\ \emph {et~al.}(2018)\citenamefont
  {Maguire}, \citenamefont {Fox}, \citenamefont {Zhou}, \citenamefont {Wang},
  \citenamefont {O'Brien}, \citenamefont {Jadwiszczak}, \citenamefont {Cullen},
  \citenamefont {McManus}, \citenamefont {Bateman}, \citenamefont {McEvoy}
  \emph {et~al.}}]{maguire2018defect}%
  \BibitemOpen
  \bibfield  {author} {\bibinfo {author} {\bibfnamefont {P.}~\bibnamefont
  {Maguire}}, \bibinfo {author} {\bibfnamefont {D.~S.}\ \bibnamefont {Fox}},
  \bibinfo {author} {\bibfnamefont {Y.}~\bibnamefont {Zhou}}, \bibinfo {author}
  {\bibfnamefont {Q.}~\bibnamefont {Wang}}, \bibinfo {author} {\bibfnamefont
  {M.}~\bibnamefont {O'Brien}}, \bibinfo {author} {\bibfnamefont
  {J.}~\bibnamefont {Jadwiszczak}}, \bibinfo {author} {\bibfnamefont {C.~P.}\
  \bibnamefont {Cullen}}, \bibinfo {author} {\bibfnamefont {J.}~\bibnamefont
  {McManus}}, \bibinfo {author} {\bibfnamefont {S.}~\bibnamefont {Bateman}},
  \bibinfo {author} {\bibfnamefont {N.}~\bibnamefont {McEvoy}}, \emph
  {et~al.},\ }\href@noop {} {\bibfield  {journal} {\bibinfo  {journal}
  {Physical Review B}\ }\textbf {\bibinfo {volume} {98}},\ \bibinfo {pages}
  {134109} (\bibinfo {year} {2018})}\BibitemShut {NoStop}%
\bibitem [{\citenamefont {Wang}\ \emph {et~al.}(2019)\citenamefont {Wang},
  \citenamefont {March}, \citenamefont {Ponce},\ and\ \citenamefont
  {Rez}}]{PhysRevB.99.115312}%
  \BibitemOpen
  \bibfield  {author} {\bibinfo {author} {\bibfnamefont {S.}~\bibnamefont
  {Wang}}, \bibinfo {author} {\bibfnamefont {K.}~\bibnamefont {March}},
  \bibinfo {author} {\bibfnamefont {F.~A.}\ \bibnamefont {Ponce}},\ and\
  \bibinfo {author} {\bibfnamefont {P.}~\bibnamefont {Rez}},\ }\href
  {https://doi.org/10.1103/PhysRevB.99.115312} {\bibfield  {journal} {\bibinfo
  {journal} {Phys. Rev. B}\ }\textbf {\bibinfo {volume} {99}},\ \bibinfo
  {pages} {115312} (\bibinfo {year} {2019})}\BibitemShut {NoStop}%
\bibitem [{\citenamefont {Shin}\ \emph {et~al.}(2021)\citenamefont {Shin},
  \citenamefont {Wang}, \citenamefont {Han}, \citenamefont {Lin}, \citenamefont
  {\'{O}Hara}, \citenamefont {Chen}, \citenamefont {Lin},\ and\ \citenamefont
  {Pantelides}}]{PhysRevMaterials.5.044002}%
  \BibitemOpen
  \bibfield  {author} {\bibinfo {author} {\bibfnamefont {D.}~\bibnamefont
  {Shin}}, \bibinfo {author} {\bibfnamefont {G.}~\bibnamefont {Wang}}, \bibinfo
  {author} {\bibfnamefont {M.}~\bibnamefont {Han}}, \bibinfo {author}
  {\bibfnamefont {Z.}~\bibnamefont {Lin}}, \bibinfo {author} {\bibfnamefont
  {A.}~\bibnamefont {\'{O}Hara}}, \bibinfo {author} {\bibfnamefont
  {F.}~\bibnamefont {Chen}}, \bibinfo {author} {\bibfnamefont {J.}~\bibnamefont
  {Lin}},\ and\ \bibinfo {author} {\bibfnamefont {S.~T.}\ \bibnamefont
  {Pantelides}},\ }\href {https://doi.org/10.1103/PhysRevMaterials.5.044002}
  {\bibfield  {journal} {\bibinfo  {journal} {Phys. Rev. Mater.}\ }\textbf
  {\bibinfo {volume} {5}},\ \bibinfo {pages} {044002} (\bibinfo {year}
  {2021})}\BibitemShut {NoStop}%
\bibitem [{\citenamefont {Chen}\ \emph {et~al.}(2022)\citenamefont {Chen},
  \citenamefont {Kawakami}, \citenamefont {Lin}, \citenamefont {Huang},
  \citenamefont {Arafune}, \citenamefont {Takagi},\ and\ \citenamefont
  {Lin}}]{PhysRevB.106.075428}%
  \BibitemOpen
  \bibfield  {author} {\bibinfo {author} {\bibfnamefont {W.-H.}\ \bibnamefont
  {Chen}}, \bibinfo {author} {\bibfnamefont {N.}~\bibnamefont {Kawakami}},
  \bibinfo {author} {\bibfnamefont {J.-J.}\ \bibnamefont {Lin}}, \bibinfo
  {author} {\bibfnamefont {H.-I.}\ \bibnamefont {Huang}}, \bibinfo {author}
  {\bibfnamefont {R.}~\bibnamefont {Arafune}}, \bibinfo {author} {\bibfnamefont
  {N.}~\bibnamefont {Takagi}},\ and\ \bibinfo {author} {\bibfnamefont {C.-L.}\
  \bibnamefont {Lin}},\ }\href {https://doi.org/10.1103/PhysRevB.106.075428}
  {\bibfield  {journal} {\bibinfo  {journal} {Phys. Rev. B}\ }\textbf {\bibinfo
  {volume} {106}},\ \bibinfo {pages} {075428} (\bibinfo {year}
  {2022})}\BibitemShut {NoStop}%
\bibitem [{\citenamefont {Komsa}\ \emph
  {et~al.}(2012{\natexlab{a}})\citenamefont {Komsa}, \citenamefont {Kotakoski},
  \citenamefont {Kurasch}, \citenamefont {Lehtinen}, \citenamefont {Kaiser},\
  and\ \citenamefont {Krasheninnikov}}]{PhysRevLett.109.035503}%
  \BibitemOpen
  \bibfield  {author} {\bibinfo {author} {\bibfnamefont {H.-P.}\ \bibnamefont
  {Komsa}}, \bibinfo {author} {\bibfnamefont {J.}~\bibnamefont {Kotakoski}},
  \bibinfo {author} {\bibfnamefont {S.}~\bibnamefont {Kurasch}}, \bibinfo
  {author} {\bibfnamefont {O.}~\bibnamefont {Lehtinen}}, \bibinfo {author}
  {\bibfnamefont {U.}~\bibnamefont {Kaiser}},\ and\ \bibinfo {author}
  {\bibfnamefont {A.~V.}\ \bibnamefont {Krasheninnikov}},\ }\href
  {https://doi.org/10.1103/PhysRevLett.109.035503} {\bibfield  {journal}
  {\bibinfo  {journal} {Phys. Rev. Lett.}\ }\textbf {\bibinfo {volume} {109}},\
  \bibinfo {pages} {035503} (\bibinfo {year} {2012}{\natexlab{a}})}\BibitemShut
  {NoStop}%
\bibitem [{\citenamefont {Krasheninnikov}\ \emph {et~al.}(2002)\citenamefont
  {Krasheninnikov}, \citenamefont {Nordlund},\ and\ \citenamefont
  {Keinonen}}]{krasheninnikov2002production}%
  \BibitemOpen
  \bibfield  {author} {\bibinfo {author} {\bibfnamefont {A.}~\bibnamefont
  {Krasheninnikov}}, \bibinfo {author} {\bibfnamefont {K.}~\bibnamefont
  {Nordlund}},\ and\ \bibinfo {author} {\bibfnamefont {J.}~\bibnamefont
  {Keinonen}},\ }\href@noop {} {\bibfield  {journal} {\bibinfo  {journal}
  {Physical Review B}\ }\textbf {\bibinfo {volume} {65}},\ \bibinfo {pages}
  {165423} (\bibinfo {year} {2002})}\BibitemShut {NoStop}%
\bibitem [{\citenamefont {Lehtinen}\ \emph {et~al.}(2010)\citenamefont
  {Lehtinen}, \citenamefont {Kotakoski}, \citenamefont {Krasheninnikov},
  \citenamefont {Tolvanen}, \citenamefont {Nordlund},\ and\ \citenamefont
  {Keinonen}}]{PhysRevB.81.153401}%
  \BibitemOpen
  \bibfield  {author} {\bibinfo {author} {\bibfnamefont {O.}~\bibnamefont
  {Lehtinen}}, \bibinfo {author} {\bibfnamefont {J.}~\bibnamefont {Kotakoski}},
  \bibinfo {author} {\bibfnamefont {A.~V.}\ \bibnamefont {Krasheninnikov}},
  \bibinfo {author} {\bibfnamefont {A.}~\bibnamefont {Tolvanen}}, \bibinfo
  {author} {\bibfnamefont {K.}~\bibnamefont {Nordlund}},\ and\ \bibinfo
  {author} {\bibfnamefont {J.}~\bibnamefont {Keinonen}},\ }\href
  {https://doi.org/10.1103/PhysRevB.81.153401} {\bibfield  {journal} {\bibinfo
  {journal} {Phys. Rev. B}\ }\textbf {\bibinfo {volume} {81}},\ \bibinfo
  {pages} {153401} (\bibinfo {year} {2010})}\BibitemShut {NoStop}%
\bibitem [{\citenamefont {Komsa}\ \emph
  {et~al.}(2012{\natexlab{b}})\citenamefont {Komsa}, \citenamefont {Kotakoski},
  \citenamefont {Kurasch}, \citenamefont {Lehtinen}, \citenamefont {Kaiser},\
  and\ \citenamefont {Krasheninnikov}}]{komsa2012two}%
  \BibitemOpen
  \bibfield  {author} {\bibinfo {author} {\bibfnamefont {H.-P.}\ \bibnamefont
  {Komsa}}, \bibinfo {author} {\bibfnamefont {J.}~\bibnamefont {Kotakoski}},
  \bibinfo {author} {\bibfnamefont {S.}~\bibnamefont {Kurasch}}, \bibinfo
  {author} {\bibfnamefont {O.}~\bibnamefont {Lehtinen}}, \bibinfo {author}
  {\bibfnamefont {U.}~\bibnamefont {Kaiser}},\ and\ \bibinfo {author}
  {\bibfnamefont {A.~V.}\ \bibnamefont {Krasheninnikov}},\ }\href@noop {}
  {\bibfield  {journal} {\bibinfo  {journal} {Physical review letters}\
  }\textbf {\bibinfo {volume} {109}},\ \bibinfo {pages} {035503} (\bibinfo
  {year} {2012}{\natexlab{b}})}\BibitemShut {NoStop}%
\bibitem [{\citenamefont {Zheng}\ \emph {et~al.}(2013)\citenamefont {Zheng},
  \citenamefont {Weismann},\ and\ \citenamefont
  {Berndt}}]{zheng2013manipulation}%
  \BibitemOpen
  \bibfield  {author} {\bibinfo {author} {\bibfnamefont {H.}~\bibnamefont
  {Zheng}}, \bibinfo {author} {\bibfnamefont {A.}~\bibnamefont {Weismann}},\
  and\ \bibinfo {author} {\bibfnamefont {R.}~\bibnamefont {Berndt}},\
  }\href@noop {} {\bibfield  {journal} {\bibinfo  {journal} {Physical Review
  letters}\ }\textbf {\bibinfo {volume} {110}},\ \bibinfo {pages} {226101}
  (\bibinfo {year} {2013})}\BibitemShut {NoStop}%
\bibitem [{\citenamefont {Qiu}\ \emph {et~al.}(2013)\citenamefont {Qiu},
  \citenamefont {Xu}, \citenamefont {Wang}, \citenamefont {Ren}, \citenamefont
  {Nan}, \citenamefont {Ni}, \citenamefont {Chen}, \citenamefont {Yuan},
  \citenamefont {Miao}, \citenamefont {Song} \emph {et~al.}}]{qiu2013hopping}%
  \BibitemOpen
  \bibfield  {author} {\bibinfo {author} {\bibfnamefont {H.}~\bibnamefont
  {Qiu}}, \bibinfo {author} {\bibfnamefont {T.}~\bibnamefont {Xu}}, \bibinfo
  {author} {\bibfnamefont {Z.}~\bibnamefont {Wang}}, \bibinfo {author}
  {\bibfnamefont {W.}~\bibnamefont {Ren}}, \bibinfo {author} {\bibfnamefont
  {H.}~\bibnamefont {Nan}}, \bibinfo {author} {\bibfnamefont {Z.}~\bibnamefont
  {Ni}}, \bibinfo {author} {\bibfnamefont {Q.}~\bibnamefont {Chen}}, \bibinfo
  {author} {\bibfnamefont {S.}~\bibnamefont {Yuan}}, \bibinfo {author}
  {\bibfnamefont {F.}~\bibnamefont {Miao}}, \bibinfo {author} {\bibfnamefont
  {F.}~\bibnamefont {Song}}, \emph {et~al.},\ }\href@noop {} {\bibfield
  {journal} {\bibinfo  {journal} {Nature communications}\ }\textbf {\bibinfo
  {volume} {4}},\ \bibinfo {pages} {2642} (\bibinfo {year} {2013})}\BibitemShut
  {NoStop}%
\bibitem [{\citenamefont {Dagdeviren}\ \emph {et~al.}(2016)\citenamefont
  {Dagdeviren}, \citenamefont {Simon}, \citenamefont {Zou}, \citenamefont
  {Walker}, \citenamefont {Ahn}, \citenamefont {Altman},\ and\ \citenamefont
  {Schwarz}}]{dagdeviren2016surface}%
  \BibitemOpen
  \bibfield  {author} {\bibinfo {author} {\bibfnamefont {O.~E.}\ \bibnamefont
  {Dagdeviren}}, \bibinfo {author} {\bibfnamefont {G.~H.}\ \bibnamefont
  {Simon}}, \bibinfo {author} {\bibfnamefont {K.}~\bibnamefont {Zou}}, \bibinfo
  {author} {\bibfnamefont {F.~J.}\ \bibnamefont {Walker}}, \bibinfo {author}
  {\bibfnamefont {C.}~\bibnamefont {Ahn}}, \bibinfo {author} {\bibfnamefont
  {E.~I.}\ \bibnamefont {Altman}},\ and\ \bibinfo {author} {\bibfnamefont
  {U.~D.}\ \bibnamefont {Schwarz}},\ }\href@noop {} {\bibfield  {journal}
  {\bibinfo  {journal} {Physical Review B}\ }\textbf {\bibinfo {volume} {93}},\
  \bibinfo {pages} {195303} (\bibinfo {year} {2016})}\BibitemShut {NoStop}%
\bibitem [{\citenamefont {Shen}\ and\ \citenamefont
  {Wang}(2022)}]{shen2022pentagon}%
  \BibitemOpen
  \bibfield  {author} {\bibinfo {author} {\bibfnamefont {Y.}~\bibnamefont
  {Shen}}\ and\ \bibinfo {author} {\bibfnamefont {Q.}~\bibnamefont {Wang}},\
  }\href@noop {} {\bibfield  {journal} {\bibinfo  {journal} {Physics Reports}\
  }\textbf {\bibinfo {volume} {964}},\ \bibinfo {pages} {1} (\bibinfo {year}
  {2022})}\BibitemShut {NoStop}%
\bibitem [{\citenamefont {Marfoua}\ and\ \citenamefont
  {Hong}(2019)}]{marfoua2019high}%
  \BibitemOpen
  \bibfield  {author} {\bibinfo {author} {\bibfnamefont {B.}~\bibnamefont
  {Marfoua}}\ and\ \bibinfo {author} {\bibfnamefont {J.}~\bibnamefont {Hong}},\
  }\href@noop {} {\bibfield  {journal} {\bibinfo  {journal} {ACS applied
  materials and interfaces}\ }\textbf {\bibinfo {volume} {11}},\ \bibinfo
  {pages} {38819} (\bibinfo {year} {2019})}\BibitemShut {NoStop}%
\bibitem [{\citenamefont {Wang}\ \emph
  {et~al.}(2015{\natexlab{a}})\citenamefont {Wang}, \citenamefont {Li},\ and\
  \citenamefont {Chen}}]{wang2015not}%
  \BibitemOpen
  \bibfield  {author} {\bibinfo {author} {\bibfnamefont {Y.}~\bibnamefont
  {Wang}}, \bibinfo {author} {\bibfnamefont {Y.}~\bibnamefont {Li}},\ and\
  \bibinfo {author} {\bibfnamefont {Z.}~\bibnamefont {Chen}},\ }\href@noop {}
  {\bibfield  {journal} {\bibinfo  {journal} {Journal of Materials Chemistry
  C}\ }\textbf {\bibinfo {volume} {3}},\ \bibinfo {pages} {9603} (\bibinfo
  {year} {2015}{\natexlab{a}})}\BibitemShut {NoStop}%
\bibitem [{\citenamefont {Zhussupbekov}\ \emph {et~al.}(2021)\citenamefont
  {Zhussupbekov}, \citenamefont {Ansari}, \citenamefont {McManus},
  \citenamefont {Zhussupbekova}, \citenamefont {Shvets}, \citenamefont
  {Duesberg}, \citenamefont {Hurley}, \citenamefont {Gity}, \citenamefont
  {Ó~Coileáin},\ and\ \citenamefont {McEvoy}}]{zhussupbekov2021imaging}%
  \BibitemOpen
  \bibfield  {author} {\bibinfo {author} {\bibfnamefont {K.}~\bibnamefont
  {Zhussupbekov}}, \bibinfo {author} {\bibfnamefont {L.}~\bibnamefont
  {Ansari}}, \bibinfo {author} {\bibfnamefont {J.~B.}\ \bibnamefont {McManus}},
  \bibinfo {author} {\bibfnamefont {A.}~\bibnamefont {Zhussupbekova}}, \bibinfo
  {author} {\bibfnamefont {I.~V.}\ \bibnamefont {Shvets}}, \bibinfo {author}
  {\bibfnamefont {G.~S.}\ \bibnamefont {Duesberg}}, \bibinfo {author}
  {\bibfnamefont {P.~K.}\ \bibnamefont {Hurley}}, \bibinfo {author}
  {\bibfnamefont {F.}~\bibnamefont {Gity}}, \bibinfo {author} {\bibfnamefont
  {C.}~\bibnamefont {Ó~Coileáin}},\ and\ \bibinfo {author} {\bibfnamefont
  {N.}~\bibnamefont {McEvoy}},\ }\href@noop {} {\bibfield  {journal} {\bibinfo
  {journal} {npj 2D Materials and Applications}\ }\textbf {\bibinfo {volume}
  {5}},\ \bibinfo {pages} {14} (\bibinfo {year} {2021})}\BibitemShut {NoStop}%
\bibitem [{\citenamefont {Ma}\ \emph {et~al.}(2016)\citenamefont {Ma},
  \citenamefont {Kou}, \citenamefont {Li}, \citenamefont {Dai},\ and\
  \citenamefont {Heine}}]{ma2016room}%
  \BibitemOpen
  \bibfield  {author} {\bibinfo {author} {\bibfnamefont {Y.}~\bibnamefont
  {Ma}}, \bibinfo {author} {\bibfnamefont {L.}~\bibnamefont {Kou}}, \bibinfo
  {author} {\bibfnamefont {X.}~\bibnamefont {Li}}, \bibinfo {author}
  {\bibfnamefont {Y.}~\bibnamefont {Dai}},\ and\ \bibinfo {author}
  {\bibfnamefont {T.}~\bibnamefont {Heine}},\ }\href@noop {} {\bibfield
  {journal} {\bibinfo  {journal} {NPG Asia Materials}\ }\textbf {\bibinfo
  {volume} {8}},\ \bibinfo {pages} {e264} (\bibinfo {year} {2016})}\BibitemShut
  {NoStop}%
\bibitem [{\citenamefont {Hoffman}\ \emph {et~al.}(2019)\citenamefont
  {Hoffman}, \citenamefont {Gu}, \citenamefont {Liang}, \citenamefont
  {Fowlkes}, \citenamefont {Xiao},\ and\ \citenamefont
  {Rack}}]{hoffman2019exploring}%
  \BibitemOpen
  \bibfield  {author} {\bibinfo {author} {\bibfnamefont {A.~N.}\ \bibnamefont
  {Hoffman}}, \bibinfo {author} {\bibfnamefont {Y.}~\bibnamefont {Gu}},
  \bibinfo {author} {\bibfnamefont {L.}~\bibnamefont {Liang}}, \bibinfo
  {author} {\bibfnamefont {J.~D.}\ \bibnamefont {Fowlkes}}, \bibinfo {author}
  {\bibfnamefont {K.}~\bibnamefont {Xiao}},\ and\ \bibinfo {author}
  {\bibfnamefont {P.~D.}\ \bibnamefont {Rack}},\ }\href@noop {} {\bibfield
  {journal} {\bibinfo  {journal} {npj 2D Materials and Applications}\ }\textbf
  {\bibinfo {volume} {3}},\ \bibinfo {pages} {50} (\bibinfo {year}
  {2019})}\BibitemShut {NoStop}%
\bibitem [{\citenamefont {Leng}\ \emph {et~al.}(2019)\citenamefont {Leng},
  \citenamefont {Orain}, \citenamefont {Amato}, \citenamefont {Huang},\ and\
  \citenamefont {de~Visser}}]{pdte2-superconductivity-PhysRevB.100.224501}%
  \BibitemOpen
  \bibfield  {author} {\bibinfo {author} {\bibfnamefont {H.}~\bibnamefont
  {Leng}}, \bibinfo {author} {\bibfnamefont {J.-C.}\ \bibnamefont {Orain}},
  \bibinfo {author} {\bibfnamefont {A.}~\bibnamefont {Amato}}, \bibinfo
  {author} {\bibfnamefont {Y.~K.}\ \bibnamefont {Huang}},\ and\ \bibinfo
  {author} {\bibfnamefont {A.}~\bibnamefont {de~Visser}},\ }\href
  {https://doi.org/10.1103/PhysRevB.100.224501} {\bibfield  {journal} {\bibinfo
   {journal} {Phys. Rev. B}\ }\textbf {\bibinfo {volume} {100}},\ \bibinfo
  {pages} {224501} (\bibinfo {year} {2019})}\BibitemShut {NoStop}%
\bibitem [{\citenamefont {Anemone}\ \emph {et~al.}(2021)\citenamefont
  {Anemone}, \citenamefont {Casado~Aguilar}, \citenamefont {Garnica},
  \citenamefont {Calleja}, \citenamefont {Al~Taleb}, \citenamefont {Kuo},
  \citenamefont {Lue}, \citenamefont {Politano}, \citenamefont {V{\'a}zquez~de
  Parga}, \citenamefont {Benedek}, \citenamefont {Far{\'i}as},\ and\
  \citenamefont {Miranda}}]{t-pdte2-superconductivity-Anemone2021}%
  \BibitemOpen
  \bibfield  {author} {\bibinfo {author} {\bibfnamefont {G.}~\bibnamefont
  {Anemone}}, \bibinfo {author} {\bibfnamefont {P.}~\bibnamefont
  {Casado~Aguilar}}, \bibinfo {author} {\bibfnamefont {M.}~\bibnamefont
  {Garnica}}, \bibinfo {author} {\bibfnamefont {F.}~\bibnamefont {Calleja}},
  \bibinfo {author} {\bibfnamefont {A.}~\bibnamefont {Al~Taleb}}, \bibinfo
  {author} {\bibfnamefont {C.-N.}\ \bibnamefont {Kuo}}, \bibinfo {author}
  {\bibfnamefont {C.~S.}\ \bibnamefont {Lue}}, \bibinfo {author} {\bibfnamefont
  {A.}~\bibnamefont {Politano}}, \bibinfo {author} {\bibfnamefont {A.~L.}\
  \bibnamefont {V{\'a}zquez~de Parga}}, \bibinfo {author} {\bibfnamefont
  {G.}~\bibnamefont {Benedek}}, \bibinfo {author} {\bibfnamefont
  {D.}~\bibnamefont {Far{\'i}as}},\ and\ \bibinfo {author} {\bibfnamefont
  {R.}~\bibnamefont {Miranda}},\ }\href
  {https://doi.org/10.1038/s41699-021-00204-5} {\bibfield  {journal} {\bibinfo
  {journal} {npj 2D Materials and Applications}\ }\textbf {\bibinfo {volume}
  {5}},\ \bibinfo {pages} {25} (\bibinfo {year} {2021})}\BibitemShut {NoStop}%
\bibitem [{\citenamefont {Lan}\ \emph {et~al.}(2019{\natexlab{a}})\citenamefont
  {Lan}, \citenamefont {Chen}, \citenamefont {Hu}, \citenamefont {Cheng},\ and\
  \citenamefont {Chen}}]{p-pdx2-thermoelectric-theory-2019}%
  \BibitemOpen
  \bibfield  {author} {\bibinfo {author} {\bibfnamefont {Y.-S.}\ \bibnamefont
  {Lan}}, \bibinfo {author} {\bibfnamefont {X.-R.}\ \bibnamefont {Chen}},
  \bibinfo {author} {\bibfnamefont {C.-E.}\ \bibnamefont {Hu}}, \bibinfo
  {author} {\bibfnamefont {Y.}~\bibnamefont {Cheng}},\ and\ \bibinfo {author}
  {\bibfnamefont {Q.-F.}\ \bibnamefont {Chen}},\ }\href
  {https://doi.org/10.1039/C9TA02138H} {\bibfield  {journal} {\bibinfo
  {journal} {J. Mater. Chem. A}\ }\textbf {\bibinfo {volume} {7}},\ \bibinfo
  {pages} {11134} (\bibinfo {year} {2019}{\natexlab{a}})}\BibitemShut {NoStop}%
\bibitem [{\citenamefont {Li}\ \emph {et~al.}(2022)\citenamefont {Li},
  \citenamefont {Huang}, \citenamefont {Xu},\ and\ \citenamefont
  {Huang}}]{p-pdte2-frontiers-in-2022}%
  \BibitemOpen
  \bibfield  {author} {\bibinfo {author} {\bibfnamefont {L.}~\bibnamefont
  {Li}}, \bibinfo {author} {\bibfnamefont {Z.}~\bibnamefont {Huang}}, \bibinfo
  {author} {\bibfnamefont {J.}~\bibnamefont {Xu}},\ and\ \bibinfo {author}
  {\bibfnamefont {H.}~\bibnamefont {Huang}},\ }\href
  {https://www.frontiersin.org/articles/10.3389/fchem.2022.1061703} {\bibfield
  {journal} {\bibinfo  {journal} {Frontiers in Chemistry}\ }\textbf {\bibinfo
  {volume} {10}} (\bibinfo {year} {2022})}\BibitemShut {NoStop}%
\bibitem [{\citenamefont {Zuo}\ \emph {et~al.}(2023)\citenamefont {Zuo},
  \citenamefont {Antonatos}, \citenamefont {D\v{e}kanovsk\'{y}}, \citenamefont
  {Luxa}, \citenamefont {Elliott}, \citenamefont {Gianolio}, \citenamefont
  {\v{S}turala}, \citenamefont {Guzzetta}, \citenamefont {Mourdikoudis},
  \citenamefont {Regner} \emph {et~al.}}]{zuo2023defect}%
  \BibitemOpen
  \bibfield  {author} {\bibinfo {author} {\bibfnamefont {Y.}~\bibnamefont
  {Zuo}}, \bibinfo {author} {\bibfnamefont {N.}~\bibnamefont {Antonatos}},
  \bibinfo {author} {\bibfnamefont {L.}~\bibnamefont {D\v{e}kanovsk\'{y}}},
  \bibinfo {author} {\bibfnamefont {J.}~\bibnamefont {Luxa}}, \bibinfo {author}
  {\bibfnamefont {J.~D.}\ \bibnamefont {Elliott}}, \bibinfo {author}
  {\bibfnamefont {D.}~\bibnamefont {Gianolio}}, \bibinfo {author}
  {\bibfnamefont {J.}~\bibnamefont {\v{S}turala}}, \bibinfo {author}
  {\bibfnamefont {F.}~\bibnamefont {Guzzetta}}, \bibinfo {author}
  {\bibfnamefont {S.}~\bibnamefont {Mourdikoudis}}, \bibinfo {author}
  {\bibfnamefont {J.}~\bibnamefont {Regner}}, \emph {et~al.},\ }\href@noop {}
  {\bibfield  {journal} {\bibinfo  {journal} {ACS Catalysis}\ }\textbf
  {\bibinfo {volume} {13}},\ \bibinfo {pages} {2601} (\bibinfo {year}
  {2023})}\BibitemShut {NoStop}%
\bibitem [{\citenamefont {Liu}\ \emph {et~al.}(2023)\citenamefont {Liu},
  \citenamefont {Ji}, \citenamefont {Bianchi}, \citenamefont {Hus},
  \citenamefont {Balog}, \citenamefont {Miwa}, \citenamefont {Hofmann},
  \citenamefont {ping Li}, \citenamefont {Zemlyanov}, \citenamefont {Li},\ and\
  \citenamefont {Chen}}]{pdte2-synthesis}%
  \BibitemOpen
  \bibfield  {author} {\bibinfo {author} {\bibfnamefont {L.}~\bibnamefont
  {Liu}}, \bibinfo {author} {\bibfnamefont {Y.}~\bibnamefont {Ji}}, \bibinfo
  {author} {\bibfnamefont {M.}~\bibnamefont {Bianchi}}, \bibinfo {author}
  {\bibfnamefont {S.~M.}\ \bibnamefont {Hus}}, \bibinfo {author} {\bibfnamefont
  {R.}~\bibnamefont {Balog}}, \bibinfo {author} {\bibfnamefont {J.~A.}\
  \bibnamefont {Miwa}}, \bibinfo {author} {\bibfnamefont {P.}~\bibnamefont
  {Hofmann}}, \bibinfo {author} {\bibfnamefont {A.}~\bibnamefont {ping Li}},
  \bibinfo {author} {\bibfnamefont {D.}~\bibnamefont {Zemlyanov}}, \bibinfo
  {author} {\bibfnamefont {Y.}~\bibnamefont {Li}},\ and\ \bibinfo {author}
  {\bibfnamefont {Y.~P.}\ \bibnamefont {Chen}},\ }in\ \href
  {https://phantomsfoundation.com/GRAPHENECONF/2023/Abstracts/Grapheneconf2023_Liu_Lina_242.pdf}
  {\emph {\bibinfo {booktitle} {Graphene2023}}}\ (\bibinfo {year}
  {2023})\BibitemShut {NoStop}%
\bibitem [{\citenamefont {Hong}\ \emph {et~al.}(2015)\citenamefont {Hong},
  \citenamefont {Hu}, \citenamefont {Probert}, \citenamefont {Li},
  \citenamefont {Lv}, \citenamefont {Yang}, \citenamefont {Gu}, \citenamefont
  {Mao}, \citenamefont {Feng}, \citenamefont {Xie} \emph
  {et~al.}}]{hong2015exploring}%
  \BibitemOpen
  \bibfield  {author} {\bibinfo {author} {\bibfnamefont {J.}~\bibnamefont
  {Hong}}, \bibinfo {author} {\bibfnamefont {Z.}~\bibnamefont {Hu}}, \bibinfo
  {author} {\bibfnamefont {M.}~\bibnamefont {Probert}}, \bibinfo {author}
  {\bibfnamefont {K.}~\bibnamefont {Li}}, \bibinfo {author} {\bibfnamefont
  {D.}~\bibnamefont {Lv}}, \bibinfo {author} {\bibfnamefont {X.}~\bibnamefont
  {Yang}}, \bibinfo {author} {\bibfnamefont {L.}~\bibnamefont {Gu}}, \bibinfo
  {author} {\bibfnamefont {N.}~\bibnamefont {Mao}}, \bibinfo {author}
  {\bibfnamefont {Q.}~\bibnamefont {Feng}}, \bibinfo {author} {\bibfnamefont
  {L.}~\bibnamefont {Xie}}, \emph {et~al.},\ }\href@noop {} {\bibfield
  {journal} {\bibinfo  {journal} {Nature communications}\ }\textbf {\bibinfo
  {volume} {6}},\ \bibinfo {pages} {6293} (\bibinfo {year} {2015})}\BibitemShut
  {NoStop}%
\bibitem [{\citenamefont {Bonilla}\ \emph {et~al.}(2018)\citenamefont
  {Bonilla}, \citenamefont {Kolekar}, \citenamefont {Ma}, \citenamefont {Diaz},
  \citenamefont {Kalappattil}, \citenamefont {Das}, \citenamefont {Eggers},
  \citenamefont {Gutierrez}, \citenamefont {Phan},\ and\ \citenamefont
  {Batzill}}]{bonilla2018strong}%
  \BibitemOpen
  \bibfield  {author} {\bibinfo {author} {\bibfnamefont {M.}~\bibnamefont
  {Bonilla}}, \bibinfo {author} {\bibfnamefont {S.}~\bibnamefont {Kolekar}},
  \bibinfo {author} {\bibfnamefont {Y.}~\bibnamefont {Ma}}, \bibinfo {author}
  {\bibfnamefont {H.~C.}\ \bibnamefont {Diaz}}, \bibinfo {author}
  {\bibfnamefont {V.}~\bibnamefont {Kalappattil}}, \bibinfo {author}
  {\bibfnamefont {R.}~\bibnamefont {Das}}, \bibinfo {author} {\bibfnamefont
  {T.}~\bibnamefont {Eggers}}, \bibinfo {author} {\bibfnamefont {H.~R.}\
  \bibnamefont {Gutierrez}}, \bibinfo {author} {\bibfnamefont {M.-H.}\
  \bibnamefont {Phan}},\ and\ \bibinfo {author} {\bibfnamefont
  {M.}~\bibnamefont {Batzill}},\ }\href@noop {} {\bibfield  {journal} {\bibinfo
   {journal} {Nature nanotechnology}\ }\textbf {\bibinfo {volume} {13}},\
  \bibinfo {pages} {289} (\bibinfo {year} {2018})}\BibitemShut {NoStop}%
\bibitem [{\citenamefont {Hohenberg}\ and\ \citenamefont
  {Kohn}(1964)}]{hohenberg1964inhomogeneous}%
  \BibitemOpen
  \bibfield  {author} {\bibinfo {author} {\bibfnamefont {P.}~\bibnamefont
  {Hohenberg}}\ and\ \bibinfo {author} {\bibfnamefont {W.}~\bibnamefont
  {Kohn}},\ }\href@noop {} {\bibfield  {journal} {\bibinfo  {journal} {Physical
  review}\ }\textbf {\bibinfo {volume} {136}},\ \bibinfo {pages} {B864}
  (\bibinfo {year} {1964})}\BibitemShut {NoStop}%
\bibitem [{\citenamefont {Kohn}\ and\ \citenamefont
  {Sham}(1965)}]{kohn1965self}%
  \BibitemOpen
  \bibfield  {author} {\bibinfo {author} {\bibfnamefont {W.}~\bibnamefont
  {Kohn}}\ and\ \bibinfo {author} {\bibfnamefont {L.~J.}\ \bibnamefont
  {Sham}},\ }\href@noop {} {\bibfield  {journal} {\bibinfo  {journal} {Physical
  review}\ }\textbf {\bibinfo {volume} {140}},\ \bibinfo {pages} {A1133}
  (\bibinfo {year} {1965})}\BibitemShut {NoStop}%
\bibitem [{\citenamefont {Kresse}\ and\ \citenamefont
  {Furthm{\"u}ller}(1996)}]{kresse1996efficient}%
  \BibitemOpen
  \bibfield  {author} {\bibinfo {author} {\bibfnamefont {G.}~\bibnamefont
  {Kresse}}\ and\ \bibinfo {author} {\bibfnamefont {J.}~\bibnamefont
  {Furthm{\"u}ller}},\ }\href@noop {} {\bibfield  {journal} {\bibinfo
  {journal} {Physical review B}\ }\textbf {\bibinfo {volume} {54}},\ \bibinfo
  {pages} {11169} (\bibinfo {year} {1996})}\BibitemShut {NoStop}%
\bibitem [{\citenamefont {Perdew}\ \emph {et~al.}(1996)\citenamefont {Perdew},
  \citenamefont {Burke},\ and\ \citenamefont
  {Ernzerhof}}]{perdew1996generalized}%
  \BibitemOpen
  \bibfield  {author} {\bibinfo {author} {\bibfnamefont {J.~P.}\ \bibnamefont
  {Perdew}}, \bibinfo {author} {\bibfnamefont {K.}~\bibnamefont {Burke}},\ and\
  \bibinfo {author} {\bibfnamefont {M.}~\bibnamefont {Ernzerhof}},\ }\href@noop
  {} {\bibfield  {journal} {\bibinfo  {journal} {Physical review letters}\
  }\textbf {\bibinfo {volume} {77}},\ \bibinfo {pages} {3865} (\bibinfo {year}
  {1996})}\BibitemShut {NoStop}%
\bibitem [{\citenamefont {Perdew}\ \emph {et~al.}(1997)\citenamefont {Perdew},
  \citenamefont {Burke},\ and\ \citenamefont
  {Ernzerhof}}]{perdew1997generalized}%
  \BibitemOpen
  \bibfield  {author} {\bibinfo {author} {\bibfnamefont {J.~P.}\ \bibnamefont
  {Perdew}}, \bibinfo {author} {\bibfnamefont {K.}~\bibnamefont {Burke}},\ and\
  \bibinfo {author} {\bibfnamefont {M.}~\bibnamefont {Ernzerhof}},\ }\href@noop
  {} {\bibfield  {journal} {\bibinfo  {journal} {Physical review letters}\
  }\textbf {\bibinfo {volume} {78}},\ \bibinfo {pages} {1396} (\bibinfo {year}
  {1997})}\BibitemShut {NoStop}%
\bibitem [{\citenamefont {Monkhorst}\ and\ \citenamefont
  {Pack}(1976)}]{monkhorst1976special}%
  \BibitemOpen
  \bibfield  {author} {\bibinfo {author} {\bibfnamefont {H.~J.}\ \bibnamefont
  {Monkhorst}}\ and\ \bibinfo {author} {\bibfnamefont {J.~D.}\ \bibnamefont
  {Pack}},\ }\href@noop {} {\bibfield  {journal} {\bibinfo  {journal} {Physical
  review B}\ }\textbf {\bibinfo {volume} {13}},\ \bibinfo {pages} {5188}
  (\bibinfo {year} {1976})}\BibitemShut {NoStop}%
\bibitem [{\citenamefont {Tian}\ \emph {et~al.}(2019)\citenamefont {Tian},
  \citenamefont {Scullion}, \citenamefont {Hughes}, \citenamefont {Li},
  \citenamefont {Shih}, \citenamefont {Coleman}, \citenamefont {Chhowalla},\
  and\ \citenamefont {Santos}}]{tian2019electronic}%
  \BibitemOpen
  \bibfield  {author} {\bibinfo {author} {\bibfnamefont {T.}~\bibnamefont
  {Tian}}, \bibinfo {author} {\bibfnamefont {D.}~\bibnamefont {Scullion}},
  \bibinfo {author} {\bibfnamefont {D.}~\bibnamefont {Hughes}}, \bibinfo
  {author} {\bibfnamefont {L.~H.}\ \bibnamefont {Li}}, \bibinfo {author}
  {\bibfnamefont {C.-J.}\ \bibnamefont {Shih}}, \bibinfo {author}
  {\bibfnamefont {J.}~\bibnamefont {Coleman}}, \bibinfo {author} {\bibfnamefont
  {M.}~\bibnamefont {Chhowalla}},\ and\ \bibinfo {author} {\bibfnamefont
  {E.~J.}\ \bibnamefont {Santos}},\ }\href@noop {} {\bibfield  {journal}
  {\bibinfo  {journal} {Nano letters}\ }\textbf {\bibinfo {volume} {20}},\
  \bibinfo {pages} {841} (\bibinfo {year} {2019})}\BibitemShut {NoStop}%
\bibitem [{\citenamefont {Tersoff}\ and\ \citenamefont
  {Hamann}(1985)}]{tersoff1985theory}%
  \BibitemOpen
  \bibfield  {author} {\bibinfo {author} {\bibfnamefont {J.}~\bibnamefont
  {Tersoff}}\ and\ \bibinfo {author} {\bibfnamefont {D.~R.}\ \bibnamefont
  {Hamann}},\ }\href@noop {} {\bibfield  {journal} {\bibinfo  {journal}
  {Physical Review B}\ }\textbf {\bibinfo {volume} {31}},\ \bibinfo {pages}
  {805} (\bibinfo {year} {1985})}\BibitemShut {NoStop}%
\bibitem [{\citenamefont {Vanpoucke}\ and\ \citenamefont
  {Brocks}(2008)}]{vanpoucke2008formation}%
  \BibitemOpen
  \bibfield  {author} {\bibinfo {author} {\bibfnamefont {D.~E.}\ \bibnamefont
  {Vanpoucke}}\ and\ \bibinfo {author} {\bibfnamefont {G.}~\bibnamefont
  {Brocks}},\ }\href@noop {} {\bibfield  {journal} {\bibinfo  {journal}
  {Physical Review B}\ }\textbf {\bibinfo {volume} {77}},\ \bibinfo {pages}
  {241308} (\bibinfo {year} {2008})}\BibitemShut {NoStop}%
\bibitem [{\citenamefont {Lan}\ \emph {et~al.}(2019{\natexlab{b}})\citenamefont
  {Lan}, \citenamefont {Chen}, \citenamefont {Hu}, \citenamefont {Cheng},\ and\
  \citenamefont {Chen}}]{lan2019penta}%
  \BibitemOpen
  \bibfield  {author} {\bibinfo {author} {\bibfnamefont {Y.-S.}\ \bibnamefont
  {Lan}}, \bibinfo {author} {\bibfnamefont {X.-R.}\ \bibnamefont {Chen}},
  \bibinfo {author} {\bibfnamefont {C.-E.}\ \bibnamefont {Hu}}, \bibinfo
  {author} {\bibfnamefont {Y.}~\bibnamefont {Cheng}},\ and\ \bibinfo {author}
  {\bibfnamefont {Q.-F.}\ \bibnamefont {Chen}},\ }\href@noop {} {\bibfield
  {journal} {\bibinfo  {journal} {Journal of Materials Chemistry A}\ }\textbf
  {\bibinfo {volume} {7}},\ \bibinfo {pages} {11134} (\bibinfo {year}
  {2019}{\natexlab{b}})}\BibitemShut {NoStop}%
\bibitem [{\citenamefont {Kirklin}\ \emph {et~al.}(2015)\citenamefont
  {Kirklin}, \citenamefont {Saal}, \citenamefont {Meredig}, \citenamefont
  {Thompson}, \citenamefont {Doak}, \citenamefont {Aykol}, \citenamefont
  {R{\"u}hl},\ and\ \citenamefont {Wolverton}}]{kirklin2015open}%
  \BibitemOpen
  \bibfield  {author} {\bibinfo {author} {\bibfnamefont {S.}~\bibnamefont
  {Kirklin}}, \bibinfo {author} {\bibfnamefont {J.~E.}\ \bibnamefont {Saal}},
  \bibinfo {author} {\bibfnamefont {B.}~\bibnamefont {Meredig}}, \bibinfo
  {author} {\bibfnamefont {A.}~\bibnamefont {Thompson}}, \bibinfo {author}
  {\bibfnamefont {J.~W.}\ \bibnamefont {Doak}}, \bibinfo {author}
  {\bibfnamefont {M.}~\bibnamefont {Aykol}}, \bibinfo {author} {\bibfnamefont
  {S.}~\bibnamefont {R{\"u}hl}},\ and\ \bibinfo {author} {\bibfnamefont
  {C.}~\bibnamefont {Wolverton}},\ }\href@noop {} {\bibfield  {journal}
  {\bibinfo  {journal} {npj Computational Materials}\ }\textbf {\bibinfo
  {volume} {1}},\ \bibinfo {pages} {1} (\bibinfo {year} {2015})}\BibitemShut
  {NoStop}%
\bibitem [{\citenamefont {Stevanovi{\'c}}\ \emph {et~al.}(2012)\citenamefont
  {Stevanovi{\'c}}, \citenamefont {Lany}, \citenamefont {Zhang},\ and\
  \citenamefont {Zunger}}]{stevanovic2012correcting}%
  \BibitemOpen
  \bibfield  {author} {\bibinfo {author} {\bibfnamefont {V.}~\bibnamefont
  {Stevanovi{\'c}}}, \bibinfo {author} {\bibfnamefont {S.}~\bibnamefont
  {Lany}}, \bibinfo {author} {\bibfnamefont {X.}~\bibnamefont {Zhang}},\ and\
  \bibinfo {author} {\bibfnamefont {A.}~\bibnamefont {Zunger}},\ }\href@noop {}
  {\bibfield  {journal} {\bibinfo  {journal} {Physical Review B}\ }\textbf
  {\bibinfo {volume} {85}},\ \bibinfo {pages} {115104} (\bibinfo {year}
  {2012})}\BibitemShut {NoStop}%
\bibitem [{SI()}]{SI}%
  \BibitemOpen
  \href@noop {} {\bibinfo  {journal} {See Supplemental Material at [URL to be
  inserted by publisher]}\ }\BibitemShut {NoStop}%
\bibitem [{\citenamefont {Barja}\ \emph {et~al.}(2019)\citenamefont {Barja},
  \citenamefont {Refaely-Abramson}, \citenamefont {Schuler}, \citenamefont
  {Qiu}, \citenamefont {Pulkin}, \citenamefont {Wickenburg}, \citenamefont
  {Ryu}, \citenamefont {Ugeda}, \citenamefont {Kastl}, \citenamefont {Chen}
  \emph {et~al.}}]{barja2019identifying}%
  \BibitemOpen
\bibfield  {journal} {  }\bibfield  {author} {\bibinfo {author} {\bibfnamefont
  {S.}~\bibnamefont {Barja}}, \bibinfo {author} {\bibfnamefont
  {S.}~\bibnamefont {Refaely-Abramson}}, \bibinfo {author} {\bibfnamefont
  {B.}~\bibnamefont {Schuler}}, \bibinfo {author} {\bibfnamefont {D.~Y.}\
  \bibnamefont {Qiu}}, \bibinfo {author} {\bibfnamefont {A.}~\bibnamefont
  {Pulkin}}, \bibinfo {author} {\bibfnamefont {S.}~\bibnamefont {Wickenburg}},
  \bibinfo {author} {\bibfnamefont {H.}~\bibnamefont {Ryu}}, \bibinfo {author}
  {\bibfnamefont {M.~M.}\ \bibnamefont {Ugeda}}, \bibinfo {author}
  {\bibfnamefont {C.}~\bibnamefont {Kastl}}, \bibinfo {author} {\bibfnamefont
  {C.}~\bibnamefont {Chen}}, \emph {et~al.},\ }\href@noop {} {\bibfield
  {journal} {\bibinfo  {journal} {Nature communications}\ }\textbf {\bibinfo
  {volume} {10}},\ \bibinfo {pages} {3382} (\bibinfo {year}
  {2019})}\BibitemShut {NoStop}%
\bibitem [{\citenamefont {Komsa}\ and\ \citenamefont
  {Krasheninnikov}(2015)}]{komsa2015native}%
  \BibitemOpen
  \bibfield  {author} {\bibinfo {author} {\bibfnamefont {H.-P.}\ \bibnamefont
  {Komsa}}\ and\ \bibinfo {author} {\bibfnamefont {A.~V.}\ \bibnamefont
  {Krasheninnikov}},\ }\href@noop {} {\bibfield  {journal} {\bibinfo  {journal}
  {Physical Review B}\ }\textbf {\bibinfo {volume} {91}},\ \bibinfo {pages}
  {125304} (\bibinfo {year} {2015})}\BibitemShut {NoStop}%
\bibitem [{\citenamefont {Nguyen}\ \emph {et~al.}(2018)\citenamefont {Nguyen},
  \citenamefont {Liang}, \citenamefont {Zou}, \citenamefont {Fu}, \citenamefont
  {Oyedele}, \citenamefont {Sumpter}, \citenamefont {Liu}, \citenamefont {Gai},
  \citenamefont {Xiao},\ and\ \citenamefont {Li}}]{nguyen20183d}%
  \BibitemOpen
  \bibfield  {author} {\bibinfo {author} {\bibfnamefont {G.~D.}\ \bibnamefont
  {Nguyen}}, \bibinfo {author} {\bibfnamefont {L.}~\bibnamefont {Liang}},
  \bibinfo {author} {\bibfnamefont {Q.}~\bibnamefont {Zou}}, \bibinfo {author}
  {\bibfnamefont {M.}~\bibnamefont {Fu}}, \bibinfo {author} {\bibfnamefont
  {A.~D.}\ \bibnamefont {Oyedele}}, \bibinfo {author} {\bibfnamefont {B.~G.}\
  \bibnamefont {Sumpter}}, \bibinfo {author} {\bibfnamefont {Z.}~\bibnamefont
  {Liu}}, \bibinfo {author} {\bibfnamefont {Z.}~\bibnamefont {Gai}}, \bibinfo
  {author} {\bibfnamefont {K.}~\bibnamefont {Xiao}},\ and\ \bibinfo {author}
  {\bibfnamefont {A.-P.}\ \bibnamefont {Li}},\ }\href@noop {} {\bibfield
  {journal} {\bibinfo  {journal} {Physical Review Letters}\ }\textbf {\bibinfo
  {volume} {121}},\ \bibinfo {pages} {086101} (\bibinfo {year}
  {2018})}\BibitemShut {NoStop}%
\bibitem [{\citenamefont {Lin}\ \emph {et~al.}(2017)\citenamefont {Lin},
  \citenamefont {Zuluaga}, \citenamefont {Yu}, \citenamefont {Liu},
  \citenamefont {Pantelides},\ and\ \citenamefont {Suenaga}}]{lin2017novel}%
  \BibitemOpen
  \bibfield  {author} {\bibinfo {author} {\bibfnamefont {J.}~\bibnamefont
  {Lin}}, \bibinfo {author} {\bibfnamefont {S.}~\bibnamefont {Zuluaga}},
  \bibinfo {author} {\bibfnamefont {P.}~\bibnamefont {Yu}}, \bibinfo {author}
  {\bibfnamefont {Z.}~\bibnamefont {Liu}}, \bibinfo {author} {\bibfnamefont
  {S.~T.}\ \bibnamefont {Pantelides}},\ and\ \bibinfo {author} {\bibfnamefont
  {K.}~\bibnamefont {Suenaga}},\ }\href@noop {} {\bibfield  {journal} {\bibinfo
   {journal} {Physical review letters}\ }\textbf {\bibinfo {volume} {119}},\
  \bibinfo {pages} {016101} (\bibinfo {year} {2017})}\BibitemShut {NoStop}%
\bibitem [{\citenamefont {Martin}(2004)}]{martin_2004}%
  \BibitemOpen
  \bibfield  {author} {\bibinfo {author} {\bibfnamefont {R.~M.}\ \bibnamefont
  {Martin}},\ }\href {https://doi.org/10.1017/CBO9780511805769} {\emph
  {\bibinfo {title} {Electronic Structure: Basic Theory and Practical
  Methods}}}\ (\bibinfo  {publisher} {Cambridge University Press},\ \bibinfo
  {year} {2004})\BibitemShut {NoStop}%
\bibitem [{\citenamefont {Zheng}\ \emph {et~al.}(2006)\citenamefont {Zheng},
  \citenamefont {Ceder}, \citenamefont {Maxisch}, \citenamefont {Chim},\ and\
  \citenamefont {Choi}}]{zheng2006native}%
  \BibitemOpen
  \bibfield  {author} {\bibinfo {author} {\bibfnamefont {J.}~\bibnamefont
  {Zheng}}, \bibinfo {author} {\bibfnamefont {G.}~\bibnamefont {Ceder}},
  \bibinfo {author} {\bibfnamefont {T.}~\bibnamefont {Maxisch}}, \bibinfo
  {author} {\bibfnamefont {W.~K.}\ \bibnamefont {Chim}},\ and\ \bibinfo
  {author} {\bibfnamefont {W.~K.}\ \bibnamefont {Choi}},\ }\href@noop {}
  {\bibfield  {journal} {\bibinfo  {journal} {Physical Review B}\ }\textbf
  {\bibinfo {volume} {73}},\ \bibinfo {pages} {104101} (\bibinfo {year}
  {2006})}\BibitemShut {NoStop}%
\bibitem [{\citenamefont {Kohan}\ \emph {et~al.}(2000)\citenamefont {Kohan},
  \citenamefont {Ceder}, \citenamefont {Morgan},\ and\ \citenamefont {Van~de
  Walle}}]{kohan2000first}%
  \BibitemOpen
  \bibfield  {author} {\bibinfo {author} {\bibfnamefont {A.}~\bibnamefont
  {Kohan}}, \bibinfo {author} {\bibfnamefont {G.}~\bibnamefont {Ceder}},
  \bibinfo {author} {\bibfnamefont {D.}~\bibnamefont {Morgan}},\ and\ \bibinfo
  {author} {\bibfnamefont {C.~G.}\ \bibnamefont {Van~de Walle}},\ }\href@noop
  {} {\bibfield  {journal} {\bibinfo  {journal} {Physical Review B}\ }\textbf
  {\bibinfo {volume} {61}},\ \bibinfo {pages} {15019} (\bibinfo {year}
  {2000})}\BibitemShut {NoStop}%
\bibitem [{\citenamefont {Zhang}\ \emph {et~al.}(2001)\citenamefont {Zhang},
  \citenamefont {Wei},\ and\ \citenamefont {Zunger}}]{zhang2001intrinsic}%
  \BibitemOpen
  \bibfield  {author} {\bibinfo {author} {\bibfnamefont {S.}~\bibnamefont
  {Zhang}}, \bibinfo {author} {\bibfnamefont {S.-H.}\ \bibnamefont {Wei}},\
  and\ \bibinfo {author} {\bibfnamefont {A.}~\bibnamefont {Zunger}},\
  }\href@noop {} {\bibfield  {journal} {\bibinfo  {journal} {Physical Review
  B}\ }\textbf {\bibinfo {volume} {63}},\ \bibinfo {pages} {075205} (\bibinfo
  {year} {2001})}\BibitemShut {NoStop}%
\bibitem [{\citenamefont {Noh}\ \emph {et~al.}(2014)\citenamefont {Noh},
  \citenamefont {Kim},\ and\ \citenamefont {Kim}}]{noh2014stability}%
  \BibitemOpen
  \bibfield  {author} {\bibinfo {author} {\bibfnamefont {J.-Y.}\ \bibnamefont
  {Noh}}, \bibinfo {author} {\bibfnamefont {H.}~\bibnamefont {Kim}},\ and\
  \bibinfo {author} {\bibfnamefont {Y.-S.}\ \bibnamefont {Kim}},\ }\href@noop
  {} {\bibfield  {journal} {\bibinfo  {journal} {Physical Review B}\ }\textbf
  {\bibinfo {volume} {89}},\ \bibinfo {pages} {205417} (\bibinfo {year}
  {2014})}\BibitemShut {NoStop}%
\bibitem [{\citenamefont {Sharma}\ \emph {et~al.}(2023)\citenamefont {Sharma},
  \citenamefont {Mishra},\ and\ \citenamefont {Shukla}}]{sharma2023influence}%
  \BibitemOpen
  \bibfield  {author} {\bibinfo {author} {\bibfnamefont {P.}~\bibnamefont
  {Sharma}}, \bibinfo {author} {\bibfnamefont {V.}~\bibnamefont {Mishra}},\
  and\ \bibinfo {author} {\bibfnamefont {A.}~\bibnamefont {Shukla}},\
  }\href@noop {} {\bibfield  {journal} {\bibinfo  {journal} {Journal of
  Physics: Condensed Matter}\ }\textbf {\bibinfo {volume} {35}},\ \bibinfo
  {pages} {345501} (\bibinfo {year} {2023})}\BibitemShut {NoStop}%
\bibitem [{\citenamefont {Zhao}\ \emph {et~al.}(2020)\citenamefont {Zhao},
  \citenamefont {Yang}, \citenamefont {Guo}, \citenamefont {Hu}, \citenamefont
  {Yue}, \citenamefont {Yuan},\ and\ \citenamefont {Ren}}]{zhao2020tuning}%
  \BibitemOpen
  \bibfield  {author} {\bibinfo {author} {\bibfnamefont {X.}~\bibnamefont
  {Zhao}}, \bibinfo {author} {\bibfnamefont {Z.}~\bibnamefont {Yang}}, \bibinfo
  {author} {\bibfnamefont {J.}~\bibnamefont {Guo}}, \bibinfo {author}
  {\bibfnamefont {G.}~\bibnamefont {Hu}}, \bibinfo {author} {\bibfnamefont
  {W.}~\bibnamefont {Yue}}, \bibinfo {author} {\bibfnamefont {X.}~\bibnamefont
  {Yuan}},\ and\ \bibinfo {author} {\bibfnamefont {J.}~\bibnamefont {Ren}},\
  }\href@noop {} {\bibfield  {journal} {\bibinfo  {journal} {Scientific
  Reports}\ }\textbf {\bibinfo {volume} {10}},\ \bibinfo {pages} {4028}
  (\bibinfo {year} {2020})}\BibitemShut {NoStop}%
\bibitem [{\citenamefont {Kuklin}\ \emph {et~al.}(2021)\citenamefont {Kuklin},
  \citenamefont {Begunovich}, \citenamefont {Gao}, \citenamefont {Zhang},\ and\
  \citenamefont {\AA{}gren}}]{kuklin2021point}%
  \BibitemOpen
  \bibfield  {author} {\bibinfo {author} {\bibfnamefont {A.~V.}\ \bibnamefont
  {Kuklin}}, \bibinfo {author} {\bibfnamefont {L.~V.}\ \bibnamefont
  {Begunovich}}, \bibinfo {author} {\bibfnamefont {L.}~\bibnamefont {Gao}},
  \bibinfo {author} {\bibfnamefont {H.}~\bibnamefont {Zhang}},\ and\ \bibinfo
  {author} {\bibfnamefont {H.}~\bibnamefont {\AA{}gren}},\ }\href
  {https://doi.org/10.1103/PhysRevB.104.134109} {\bibfield  {journal} {\bibinfo
   {journal} {Phys. Rev. B}\ }\textbf {\bibinfo {volume} {104}},\ \bibinfo
  {pages} {134109} (\bibinfo {year} {2021})}\BibitemShut {NoStop}%
\bibitem [{\citenamefont {Horzum}\ \emph {et~al.}(2014)\citenamefont {Horzum},
  \citenamefont {\ifmmode \mbox{\c{C}}\else \c{C}\fi{}ak\ifmmode \imath
  \else~\i \fi{}r}, \citenamefont {Suh}, \citenamefont {Tongay}, \citenamefont
  {Huang}, \citenamefont {Ho}, \citenamefont {Wu}, \citenamefont {Sahin},\ and\
  \citenamefont {Peeters}}]{horzum2014formation}%
  \BibitemOpen
  \bibfield  {author} {\bibinfo {author} {\bibfnamefont {S.}~\bibnamefont
  {Horzum}}, \bibinfo {author} {\bibfnamefont {D.}~\bibnamefont {\ifmmode
  \mbox{\c{C}}\else \c{C}\fi{}ak\ifmmode \imath \else~\i \fi{}r}}, \bibinfo
  {author} {\bibfnamefont {J.}~\bibnamefont {Suh}}, \bibinfo {author}
  {\bibfnamefont {S.}~\bibnamefont {Tongay}}, \bibinfo {author} {\bibfnamefont
  {Y.-S.}\ \bibnamefont {Huang}}, \bibinfo {author} {\bibfnamefont {C.-H.}\
  \bibnamefont {Ho}}, \bibinfo {author} {\bibfnamefont {J.}~\bibnamefont {Wu}},
  \bibinfo {author} {\bibfnamefont {H.}~\bibnamefont {Sahin}},\ and\ \bibinfo
  {author} {\bibfnamefont {F.~M.}\ \bibnamefont {Peeters}},\ }\href
  {https://doi.org/10.1103/PhysRevB.89.155433} {\bibfield  {journal} {\bibinfo
  {journal} {Phys. Rev. B}\ }\textbf {\bibinfo {volume} {89}},\ \bibinfo
  {pages} {155433} (\bibinfo {year} {2014})}\BibitemShut {NoStop}%
\bibitem [{\citenamefont {Hashimoto}\ \emph {et~al.}(2004)\citenamefont
  {Hashimoto}, \citenamefont {Suenaga}, \citenamefont {Gloter}, \citenamefont
  {Urita},\ and\ \citenamefont {Iijima}}]{hashimoto2004direct}%
  \BibitemOpen
  \bibfield  {author} {\bibinfo {author} {\bibfnamefont {A.}~\bibnamefont
  {Hashimoto}}, \bibinfo {author} {\bibfnamefont {K.}~\bibnamefont {Suenaga}},
  \bibinfo {author} {\bibfnamefont {A.}~\bibnamefont {Gloter}}, \bibinfo
  {author} {\bibfnamefont {K.}~\bibnamefont {Urita}},\ and\ \bibinfo {author}
  {\bibfnamefont {S.}~\bibnamefont {Iijima}},\ }\href@noop {} {\bibfield
  {journal} {\bibinfo  {journal} {nature}\ }\textbf {\bibinfo {volume} {430}},\
  \bibinfo {pages} {870} (\bibinfo {year} {2004})}\BibitemShut {NoStop}%
\bibitem [{\citenamefont {Ugeda}\ \emph {et~al.}(2010)\citenamefont {Ugeda},
  \citenamefont {Brihuega}, \citenamefont {Guinea},\ and\ \citenamefont
  {G\'{o}mez-Rodr\'{\i}guez}}]{ugeda2010missing}%
  \BibitemOpen
  \bibfield  {author} {\bibinfo {author} {\bibfnamefont {M.~M.}\ \bibnamefont
  {Ugeda}}, \bibinfo {author} {\bibfnamefont {I.}~\bibnamefont {Brihuega}},
  \bibinfo {author} {\bibfnamefont {F.}~\bibnamefont {Guinea}},\ and\ \bibinfo
  {author} {\bibfnamefont {J.~M.}\ \bibnamefont {G\'{o}mez-Rodr\'{\i}guez}},\
  }\href@noop {} {\bibfield  {journal} {\bibinfo  {journal} {Physical Review
  Letters}\ }\textbf {\bibinfo {volume} {104}},\ \bibinfo {pages} {096804}
  (\bibinfo {year} {2010})}\BibitemShut {NoStop}%
\bibitem [{\citenamefont {Freysoldt}\ \emph {et~al.}(2014)\citenamefont
  {Freysoldt}, \citenamefont {Grabowski}, \citenamefont {Hickel}, \citenamefont
  {Neugebauer}, \citenamefont {Kresse}, \citenamefont {Janotti},\ and\
  \citenamefont {Van~de Walle}}]{RevModPhys.86.253}%
  \BibitemOpen
  \bibfield  {author} {\bibinfo {author} {\bibfnamefont {C.}~\bibnamefont
  {Freysoldt}}, \bibinfo {author} {\bibfnamefont {B.}~\bibnamefont
  {Grabowski}}, \bibinfo {author} {\bibfnamefont {T.}~\bibnamefont {Hickel}},
  \bibinfo {author} {\bibfnamefont {J.}~\bibnamefont {Neugebauer}}, \bibinfo
  {author} {\bibfnamefont {G.}~\bibnamefont {Kresse}}, \bibinfo {author}
  {\bibfnamefont {A.}~\bibnamefont {Janotti}},\ and\ \bibinfo {author}
  {\bibfnamefont {C.~G.}\ \bibnamefont {Van~de Walle}},\ }\href
  {https://doi.org/10.1103/RevModPhys.86.253} {\bibfield  {journal} {\bibinfo
  {journal} {Rev. Mod. Phys.}\ }\textbf {\bibinfo {volume} {86}},\ \bibinfo
  {pages} {253} (\bibinfo {year} {2014})}\BibitemShut {NoStop}%
\bibitem [{\citenamefont {Ko\'{o}s}\ \emph {et~al.}(2019)\citenamefont
  {Ko\'{o}s}, \citenamefont {Vancs\'{o}}, \citenamefont {Szendr\~{o}},
  \citenamefont {Dobrik}, \citenamefont {Antognini~Silva}, \citenamefont
  {Popov}, \citenamefont {Sorokin}, \citenamefont {Henrard}, \citenamefont
  {Hwang}, \citenamefont {Bir\'{o}},\ and\ \citenamefont
  {Tapaszt\'{o}}}]{koos_influence_2019}%
  \BibitemOpen
  \bibfield  {author} {\bibinfo {author} {\bibfnamefont {A.~A.}\ \bibnamefont
  {Ko\'{o}s}}, \bibinfo {author} {\bibfnamefont {P.}~\bibnamefont
  {Vancs\'{o}}}, \bibinfo {author} {\bibfnamefont {M.}~\bibnamefont
  {Szendr\~{o}}}, \bibinfo {author} {\bibfnamefont {G.}~\bibnamefont {Dobrik}},
  \bibinfo {author} {\bibfnamefont {D.}~\bibnamefont {Antognini~Silva}},
  \bibinfo {author} {\bibfnamefont {Z.~I.}\ \bibnamefont {Popov}}, \bibinfo
  {author} {\bibfnamefont {P.~B.}\ \bibnamefont {Sorokin}}, \bibinfo {author}
  {\bibfnamefont {L.}~\bibnamefont {Henrard}}, \bibinfo {author} {\bibfnamefont
  {C.}~\bibnamefont {Hwang}}, \bibinfo {author} {\bibfnamefont {L.~P.}\
  \bibnamefont {Bir\'{o}}},\ and\ \bibinfo {author} {\bibfnamefont
  {L.}~\bibnamefont {Tapaszt\'{o}}},\ }\href@noop {} {\bibfield  {journal}
  {\bibinfo  {journal} {The Journal of Physical Chemistry C}\ }\textbf
  {\bibinfo {volume} {123}},\ \bibinfo {pages} {24855} (\bibinfo {year}
  {2019})},\ \bibinfo {note} {publisher: ACS Publications}\BibitemShut
  {NoStop}%
\bibitem [{\citenamefont {Lin}\ \emph {et~al.}(2016)\citenamefont {Lin},
  \citenamefont {Carvalho}, \citenamefont {Kahn}, \citenamefont {Lv},
  \citenamefont {Rao}, \citenamefont {Terrones}, \citenamefont {Pimenta},\ and\
  \citenamefont {Terrones}}]{Lin_2016-2d-defect-eng}%
  \BibitemOpen
  \bibfield  {author} {\bibinfo {author} {\bibfnamefont {Z.}~\bibnamefont
  {Lin}}, \bibinfo {author} {\bibfnamefont {B.~R.}\ \bibnamefont {Carvalho}},
  \bibinfo {author} {\bibfnamefont {E.}~\bibnamefont {Kahn}}, \bibinfo {author}
  {\bibfnamefont {R.}~\bibnamefont {Lv}}, \bibinfo {author} {\bibfnamefont
  {R.}~\bibnamefont {Rao}}, \bibinfo {author} {\bibfnamefont {H.}~\bibnamefont
  {Terrones}}, \bibinfo {author} {\bibfnamefont {M.~A.}\ \bibnamefont
  {Pimenta}},\ and\ \bibinfo {author} {\bibfnamefont {M.}~\bibnamefont
  {Terrones}},\ }\href {https://doi.org/10.1088/2053-1583/3/2/022002}
  {\bibfield  {journal} {\bibinfo  {journal} {2D Materials}\ }\textbf {\bibinfo
  {volume} {3}},\ \bibinfo {pages} {022002} (\bibinfo {year}
  {2016})}\BibitemShut {NoStop}%
\bibitem [{\citenamefont {Fu}\ \emph {et~al.}(2019)\citenamefont {Fu},
  \citenamefont {Liang}, \citenamefont {Zou}, \citenamefont {Nguyen},
  \citenamefont {Xiao}, \citenamefont {Li}, \citenamefont {Kang}, \citenamefont
  {Wu},\ and\ \citenamefont {Gai}}]{fu2019defects}%
  \BibitemOpen
  \bibfield  {author} {\bibinfo {author} {\bibfnamefont {M.}~\bibnamefont
  {Fu}}, \bibinfo {author} {\bibfnamefont {L.}~\bibnamefont {Liang}}, \bibinfo
  {author} {\bibfnamefont {Q.}~\bibnamefont {Zou}}, \bibinfo {author}
  {\bibfnamefont {G.~D.}\ \bibnamefont {Nguyen}}, \bibinfo {author}
  {\bibfnamefont {K.}~\bibnamefont {Xiao}}, \bibinfo {author} {\bibfnamefont
  {A.-P.}\ \bibnamefont {Li}}, \bibinfo {author} {\bibfnamefont
  {J.}~\bibnamefont {Kang}}, \bibinfo {author} {\bibfnamefont {Z.}~\bibnamefont
  {Wu}},\ and\ \bibinfo {author} {\bibfnamefont {Z.}~\bibnamefont {Gai}},\
  }\href@noop {} {\bibfield  {journal} {\bibinfo  {journal} {The Journal of
  Physical Chemistry Letters}\ }\textbf {\bibinfo {volume} {11}},\ \bibinfo
  {pages} {740} (\bibinfo {year} {2019})}\BibitemShut {NoStop}%
\bibitem [{\citenamefont {Hla}\ \emph {et~al.}(2000)\citenamefont {Hla},
  \citenamefont {Bartels}, \citenamefont {Meyer},\ and\ \citenamefont
  {Rieder}}]{hla2000inducing}%
  \BibitemOpen
  \bibfield  {author} {\bibinfo {author} {\bibfnamefont {S.-W.}\ \bibnamefont
  {Hla}}, \bibinfo {author} {\bibfnamefont {L.}~\bibnamefont {Bartels}},
  \bibinfo {author} {\bibfnamefont {G.}~\bibnamefont {Meyer}},\ and\ \bibinfo
  {author} {\bibfnamefont {K.-H.}\ \bibnamefont {Rieder}},\ }\href@noop {}
  {\bibfield  {journal} {\bibinfo  {journal} {Physical review letters}\
  }\textbf {\bibinfo {volume} {85}},\ \bibinfo {pages} {2777} (\bibinfo {year}
  {2000})}\BibitemShut {NoStop}%
\bibitem [{\citenamefont {Eigler}\ and\ \citenamefont
  {Schweizer}(1990)}]{eigler1990positioning}%
  \BibitemOpen
  \bibfield  {author} {\bibinfo {author} {\bibfnamefont {D.~M.}\ \bibnamefont
  {Eigler}}\ and\ \bibinfo {author} {\bibfnamefont {E.~K.}\ \bibnamefont
  {Schweizer}},\ }\href@noop {} {\bibfield  {journal} {\bibinfo  {journal}
  {Nature}\ }\textbf {\bibinfo {volume} {344}},\ \bibinfo {pages} {524}
  (\bibinfo {year} {1990})}\BibitemShut {NoStop}%
\bibitem [{\citenamefont {Komsa}\ and\ \citenamefont
  {Pasquarello}(2013)}]{komsa2013finite}%
  \BibitemOpen
  \bibfield  {author} {\bibinfo {author} {\bibfnamefont {H.-P.}\ \bibnamefont
  {Komsa}}\ and\ \bibinfo {author} {\bibfnamefont {A.}~\bibnamefont
  {Pasquarello}},\ }\href@noop {} {\bibfield  {journal} {\bibinfo  {journal}
  {Physical review letters}\ }\textbf {\bibinfo {volume} {110}},\ \bibinfo
  {pages} {095505} (\bibinfo {year} {2013})}\BibitemShut {NoStop}%
\bibitem [{\citenamefont {Wang}\ \emph
  {et~al.}(2015{\natexlab{b}})\citenamefont {Wang}, \citenamefont {Han},
  \citenamefont {Li}, \citenamefont {Xie}, \citenamefont {Chen}, \citenamefont
  {Tian}, \citenamefont {West}, \citenamefont {Sun},\ and\ \citenamefont
  {Zhang}}]{wang2015determination}%
  \BibitemOpen
  \bibfield  {author} {\bibinfo {author} {\bibfnamefont {D.}~\bibnamefont
  {Wang}}, \bibinfo {author} {\bibfnamefont {D.}~\bibnamefont {Han}}, \bibinfo
  {author} {\bibfnamefont {X.-B.}\ \bibnamefont {Li}}, \bibinfo {author}
  {\bibfnamefont {S.-Y.}\ \bibnamefont {Xie}}, \bibinfo {author} {\bibfnamefont
  {N.-K.}\ \bibnamefont {Chen}}, \bibinfo {author} {\bibfnamefont {W.~Q.}\
  \bibnamefont {Tian}}, \bibinfo {author} {\bibfnamefont {D.}~\bibnamefont
  {West}}, \bibinfo {author} {\bibfnamefont {H.-B.}\ \bibnamefont {Sun}},\ and\
  \bibinfo {author} {\bibfnamefont {S.}~\bibnamefont {Zhang}},\ }\href@noop {}
  {\bibfield  {journal} {\bibinfo  {journal} {Physical review letters}\
  }\textbf {\bibinfo {volume} {114}},\ \bibinfo {pages} {196801} (\bibinfo
  {year} {2015}{\natexlab{b}})}\BibitemShut {NoStop}%
\bibitem [{\citenamefont {Komsa}\ \emph {et~al.}(2018)\citenamefont {Komsa},
  \citenamefont {Berseneva}, \citenamefont {Krasheninnikov},\ and\
  \citenamefont {Nieminen}}]{komsa2018erratum}%
  \BibitemOpen
  \bibfield  {author} {\bibinfo {author} {\bibfnamefont {H.-P.}\ \bibnamefont
  {Komsa}}, \bibinfo {author} {\bibfnamefont {N.}~\bibnamefont {Berseneva}},
  \bibinfo {author} {\bibfnamefont {A.~V.}\ \bibnamefont {Krasheninnikov}},\
  and\ \bibinfo {author} {\bibfnamefont {R.~M.}\ \bibnamefont {Nieminen}},\
  }\href@noop {} {\bibfield  {journal} {\bibinfo  {journal} {Physical Review
  X}\ }\textbf {\bibinfo {volume} {8}},\ \bibinfo {pages} {039902} (\bibinfo
  {year} {2018})}\BibitemShut {NoStop}%
\bibitem [{\citenamefont {Komsa}\ \emph
  {et~al.}(2012{\natexlab{c}})\citenamefont {Komsa}, \citenamefont {Rantala},\
  and\ \citenamefont {Pasquarello}}]{komsa2012finite}%
  \BibitemOpen
  \bibfield  {author} {\bibinfo {author} {\bibfnamefont {H.-P.}\ \bibnamefont
  {Komsa}}, \bibinfo {author} {\bibfnamefont {T.~T.}\ \bibnamefont {Rantala}},\
  and\ \bibinfo {author} {\bibfnamefont {A.}~\bibnamefont {Pasquarello}},\
  }\href@noop {} {\bibfield  {journal} {\bibinfo  {journal} {Physical Review
  B}\ }\textbf {\bibinfo {volume} {86}},\ \bibinfo {pages} {045112} (\bibinfo
  {year} {2012}{\natexlab{c}})}\BibitemShut {NoStop}%
\bibitem [{\citenamefont {Henkelman}\ \emph {et~al.}(2000)\citenamefont
  {Henkelman}, \citenamefont {Uberuaga},\ and\ \citenamefont
  {J{\'o}nsson}}]{henkelman2000climbing}%
  \BibitemOpen
  \bibfield  {author} {\bibinfo {author} {\bibfnamefont {G.}~\bibnamefont
  {Henkelman}}, \bibinfo {author} {\bibfnamefont {B.~P.}\ \bibnamefont
  {Uberuaga}},\ and\ \bibinfo {author} {\bibfnamefont {H.}~\bibnamefont
  {J{\'o}nsson}},\ }\href@noop {} {\bibfield  {journal} {\bibinfo  {journal}
  {The Journal of chemical physics}\ }\textbf {\bibinfo {volume} {113}},\
  \bibinfo {pages} {9901} (\bibinfo {year} {2000})}\BibitemShut {NoStop}%
\bibitem [{\citenamefont {Komsa}\ \emph {et~al.}(2013)\citenamefont {Komsa},
  \citenamefont {Kurasch}, \citenamefont {Lehtinen}, \citenamefont {Kaiser},\
  and\ \citenamefont {Krasheninnikov}}]{komsa2013point}%
  \BibitemOpen
  \bibfield  {author} {\bibinfo {author} {\bibfnamefont {H.-P.}\ \bibnamefont
  {Komsa}}, \bibinfo {author} {\bibfnamefont {S.}~\bibnamefont {Kurasch}},
  \bibinfo {author} {\bibfnamefont {O.}~\bibnamefont {Lehtinen}}, \bibinfo
  {author} {\bibfnamefont {U.}~\bibnamefont {Kaiser}},\ and\ \bibinfo {author}
  {\bibfnamefont {A.~V.}\ \bibnamefont {Krasheninnikov}},\ }\href@noop {}
  {\bibfield  {journal} {\bibinfo  {journal} {Physical Review B}\ }\textbf
  {\bibinfo {volume} {88}},\ \bibinfo {pages} {035301} (\bibinfo {year}
  {2013})}\BibitemShut {NoStop}%
\bibitem [{\citenamefont {Krasheninnikov}\ and\ \citenamefont
  {Nordlund}(2010)}]{krasheninnikov2010ion}%
  \BibitemOpen
  \bibfield  {author} {\bibinfo {author} {\bibfnamefont {A.}~\bibnamefont
  {Krasheninnikov}}\ and\ \bibinfo {author} {\bibfnamefont {K.}~\bibnamefont
  {Nordlund}},\ }\href@noop {} {\bibfield  {journal} {\bibinfo  {journal}
  {Journal of applied physics}\ }\textbf {\bibinfo {volume} {107}},\ \bibinfo
  {pages} {3} (\bibinfo {year} {2010})}\BibitemShut {NoStop}%
\bibitem [{\citenamefont {Pizzochero}\ and\ \citenamefont
  {Yazyev}(2018)}]{pizzochero2018single}%
  \BibitemOpen
  \bibfield  {author} {\bibinfo {author} {\bibfnamefont {M.}~\bibnamefont
  {Pizzochero}}\ and\ \bibinfo {author} {\bibfnamefont {O.~V.}\ \bibnamefont
  {Yazyev}},\ }\href@noop {} {\bibfield  {journal} {\bibinfo  {journal} {2D
  Materials}\ }\textbf {\bibinfo {volume} {5}},\ \bibinfo {pages} {025022}
  (\bibinfo {year} {2018})}\BibitemShut {NoStop}%
\bibitem [{\citenamefont {Kotakoski}\ \emph {et~al.}(2010)\citenamefont
  {Kotakoski}, \citenamefont {Jin}, \citenamefont {Lehtinen}, \citenamefont
  {Suenaga},\ and\ \citenamefont {Krasheninnikov}}]{kotakoski2010electron}%
  \BibitemOpen
  \bibfield  {author} {\bibinfo {author} {\bibfnamefont {J.}~\bibnamefont
  {Kotakoski}}, \bibinfo {author} {\bibfnamefont {C.~H.}\ \bibnamefont {Jin}},
  \bibinfo {author} {\bibfnamefont {O.}~\bibnamefont {Lehtinen}}, \bibinfo
  {author} {\bibfnamefont {K.}~\bibnamefont {Suenaga}},\ and\ \bibinfo {author}
  {\bibfnamefont {A.~V.}\ \bibnamefont {Krasheninnikov}},\ }\href@noop {}
  {\bibfield  {journal} {\bibinfo  {journal} {Physical Review B}\ }\textbf
  {\bibinfo {volume} {82}},\ \bibinfo {pages} {113404} (\bibinfo {year}
  {2010})}\BibitemShut {NoStop}%
\bibitem [{\citenamefont {Zobelli}\ \emph {et~al.}(2007)\citenamefont
  {Zobelli}, \citenamefont {Gloter}, \citenamefont {Ewels}, \citenamefont
  {Seifert},\ and\ \citenamefont {Colliex}}]{zobelli2007electron}%
  \BibitemOpen
  \bibfield  {author} {\bibinfo {author} {\bibfnamefont {A.}~\bibnamefont
  {Zobelli}}, \bibinfo {author} {\bibfnamefont {A.}~\bibnamefont {Gloter}},
  \bibinfo {author} {\bibfnamefont {C.}~\bibnamefont {Ewels}}, \bibinfo
  {author} {\bibfnamefont {G.}~\bibnamefont {Seifert}},\ and\ \bibinfo {author}
  {\bibfnamefont {C.}~\bibnamefont {Colliex}},\ }\href@noop {} {\bibfield
  {journal} {\bibinfo  {journal} {Physical Review B}\ }\textbf {\bibinfo
  {volume} {75}},\ \bibinfo {pages} {245402} (\bibinfo {year}
  {2007})}\BibitemShut {NoStop}%
\bibitem [{\citenamefont {Garcia}\ \emph {et~al.}(2014)\citenamefont {Garcia},
  \citenamefont {Raya}, \citenamefont {Mariscal}, \citenamefont {Esparza},
  \citenamefont {Herrera}, \citenamefont {Molina}, \citenamefont {Scavello},
  \citenamefont {Galindo}, \citenamefont {Jose-Yacaman},\ and\ \citenamefont
  {Ponce}}]{garcia2014analysis}%
  \BibitemOpen
  \bibfield  {author} {\bibinfo {author} {\bibfnamefont {A.}~\bibnamefont
  {Garcia}}, \bibinfo {author} {\bibfnamefont {A.~M.}\ \bibnamefont {Raya}},
  \bibinfo {author} {\bibfnamefont {M.~M.}\ \bibnamefont {Mariscal}}, \bibinfo
  {author} {\bibfnamefont {R.}~\bibnamefont {Esparza}}, \bibinfo {author}
  {\bibfnamefont {M.}~\bibnamefont {Herrera}}, \bibinfo {author} {\bibfnamefont
  {S.~I.}\ \bibnamefont {Molina}}, \bibinfo {author} {\bibfnamefont
  {G.}~\bibnamefont {Scavello}}, \bibinfo {author} {\bibfnamefont {P.~L.}\
  \bibnamefont {Galindo}}, \bibinfo {author} {\bibfnamefont {M.}~\bibnamefont
  {Jose-Yacaman}},\ and\ \bibinfo {author} {\bibfnamefont {A.}~\bibnamefont
  {Ponce}},\ }\href@noop {} {\bibfield  {journal} {\bibinfo  {journal}
  {Ultramicroscopy}\ }\textbf {\bibinfo {volume} {146}},\ \bibinfo {pages} {33}
  (\bibinfo {year} {2014})}\BibitemShut {NoStop}%
\bibitem [{\citenamefont {Yoshimura}\ \emph {et~al.}(2018)\citenamefont
  {Yoshimura}, \citenamefont {Lamparski}, \citenamefont {Kharche},\ and\
  \citenamefont {Meunier}}]{yoshimura2018first}%
  \BibitemOpen
  \bibfield  {author} {\bibinfo {author} {\bibfnamefont {A.}~\bibnamefont
  {Yoshimura}}, \bibinfo {author} {\bibfnamefont {M.}~\bibnamefont
  {Lamparski}}, \bibinfo {author} {\bibfnamefont {N.}~\bibnamefont {Kharche}},\
  and\ \bibinfo {author} {\bibfnamefont {V.}~\bibnamefont {Meunier}},\
  }\href@noop {} {\bibfield  {journal} {\bibinfo  {journal} {Nanoscale}\
  }\textbf {\bibinfo {volume} {10}},\ \bibinfo {pages} {2388} (\bibinfo {year}
  {2018})}\BibitemShut {NoStop}%
\bibitem [{\citenamefont {Kretschmer}\ \emph {et~al.}(2020)\citenamefont
  {Kretschmer}, \citenamefont {Lehnert}, \citenamefont {Kaiser},\ and\
  \citenamefont {Krasheninnikov}}]{kretschmer2020formation}%
  \BibitemOpen
  \bibfield  {author} {\bibinfo {author} {\bibfnamefont {S.}~\bibnamefont
  {Kretschmer}}, \bibinfo {author} {\bibfnamefont {T.}~\bibnamefont {Lehnert}},
  \bibinfo {author} {\bibfnamefont {U.}~\bibnamefont {Kaiser}},\ and\ \bibinfo
  {author} {\bibfnamefont {A.~V.}\ \bibnamefont {Krasheninnikov}},\ }\href@noop
  {} {\bibfield  {journal} {\bibinfo  {journal} {Nano letters}\ }\textbf
  {\bibinfo {volume} {20}},\ \bibinfo {pages} {2865} (\bibinfo {year}
  {2020})}\BibitemShut {NoStop}%
\end{thebibliography}%

\end{document}

\renewcommand{\thetable}{S\arabic{table}}
\renewcommand{\thefigure}{S\arabic{figure}}
\renewcommand{\thepage}{S\arabic{page}}

\title{Supplemental Material: Defect engineering in two-dimensional pentagonal PdTe$_{2}$: Tuning electronic, optical, and magnetic properties}
\author{Poonam Sharma}
%\email{pspoonamsharma44@gmail.com}
\author{ Vaishali Roondhe}
%\email{@gmail.com}
\author{Alok Shukla}
\email{shukla@iitb.ac.in}
\affiliation{Department of Physics, Indian Institute of Technology Bombay, Mumbai
400076, India}
\maketitle

\begin{center}
 { Supplemental Material}   
\end{center}

\vspace{2mm}
This supporting document contains the following entries: the third and sixth are tables, while the rest are figures.
\vspace{3mm}

\raggedright

1. Optimized \emph{p}-PdTe$_2$ monolayer structure on Pd(100) substrate .\\ 
\vspace{1.3mm}
2. Partial density of states (PDOS) of the pristine~\emph{p}-PdTe$_2$ monolayer. \\ 
\vspace{1.3mm}
3. Calculated band gaps using GGA+SOC and HSE06+SOC methods.\\ 
\vspace{1.3mm}
4. Shifts in the VBM/CBM were obtained using the HSE06+SOC approach, as compared to PBE+SOC results.\\ 
\vspace{1.3mm}
5. Total density of states (TDOS) for different considered vacancies in the~\emph{p}-PdTe$_2$ monolayer without SOC.\\ 
\vspace{1.3mm}
6. Calculated excess electronic polarizability for V$_{Pd+Te}$, V$_{Te+Te}$, and V$_{Pd+4Te}$ defective configurations.\\ 
\vspace{1.3mm}
7. Formation energies of charged defects mono-vacancies as a function of inverse cell dimension.\\ 
\vspace{1.3mm}
8. AIMD simulated structures of the~\emph{p}-PdTe$_2$ monolayer for a time period of 5ps at 400 K corresponding to V$_{Te}$. \\ 
\vspace{1.3mm}
9. The (a,c,e) real and (b,d,f) imaginary parts of the total electronic polarizability corresponding to V$_{Pd+Te}$, V$_{Te+Te}$ and V$_{Pd+4Te}$, respectively.\\
\vspace{1.3mm}
10. The (a,c,e) real and (b,d,f) imaginary parts of the excess electronic polarizability corresponding to V$_{Pd+Te}$, V$_{Te+Te}$ and V$_{Pd+4Te}$, respectively.s

\newpage

\begin{figure*}[ht]
\centering\includegraphics[width=0.25\linewidth]{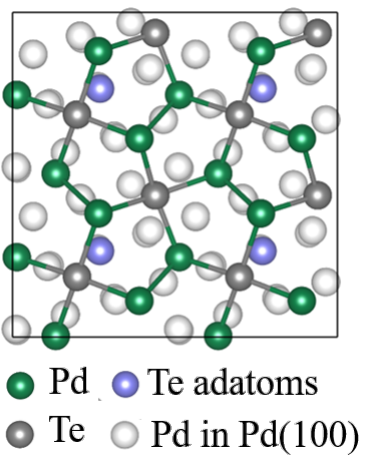}\caption{Optimized~\emph{p}-PdTe$_2$ monolayer structure on the top of Te adatoms layer on Pd(100) substrate.}
\label{fig:figS2} 
\end{figure*}

\begin{figure*}[ht]
\centering\includegraphics[width=0.6\linewidth]{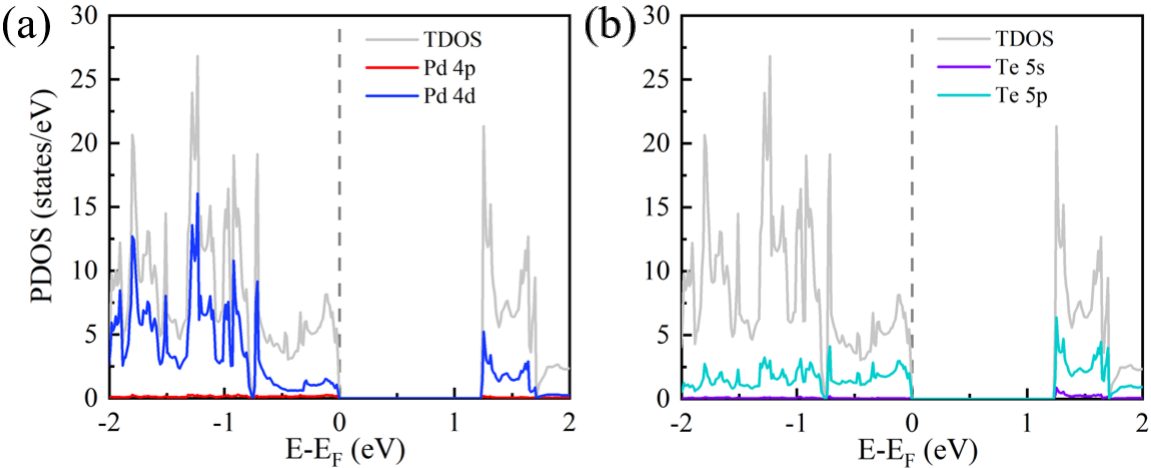}\caption{Partial density of states of (a) Pd (4p and 4d orbitals) and (b) Te (5s and 5p orbitals) atoms. The 5$p$ orbitals of Te atoms mainly contribute to the PDOS as compared to the 4$d$ orbitals of Pd atoms. The SOC is incorporated here.}
\label{fig:figS1} 
\end{figure*}

\begin{figure*}[ht]
\includegraphics[width=0.5\linewidth]{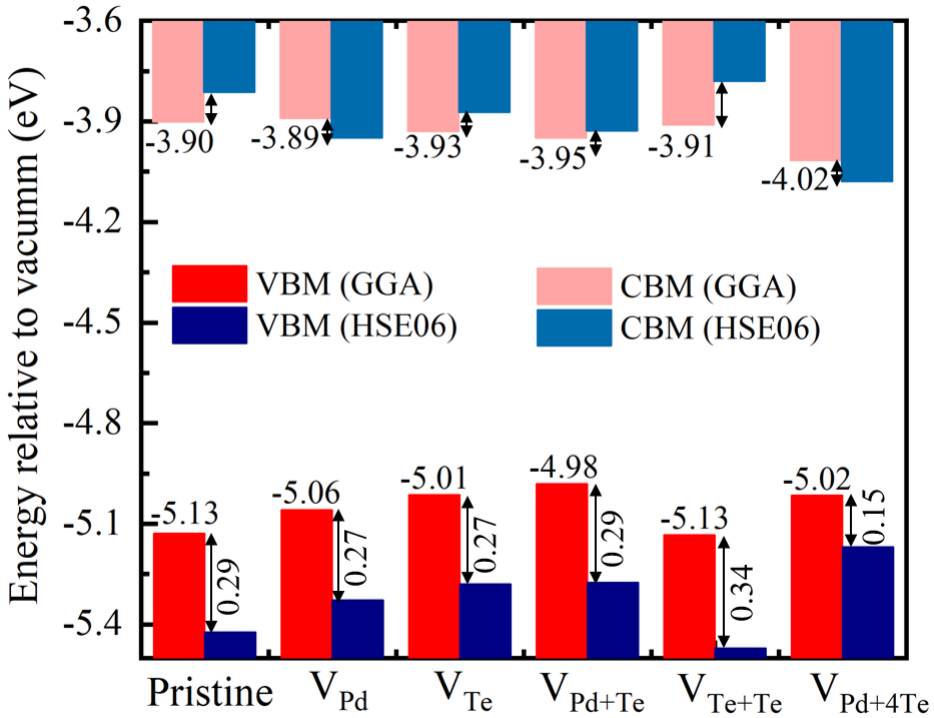}\caption{ Shifts in the VBM/CBM obtained using HSE06+SOC approach, as compared to PBE+SOC results. Additionally, the absolute values of the band edges are also indicated for PBE+SOC cases. Because the shifts are very small for CBMs as compared to the VBMs, their magnitudes are not indicated in the figure. However, they are (from left to right): 0.09 eV, 0.06 eV, 0.06 eV, 0.02 eV, 0.13 eV, 0.06 eV.}
\label{fig:vpd-vte-form-energy} 
\end{figure*}

\begin{figure*}[ht]
\centering\includegraphics[width=0.7\linewidth]{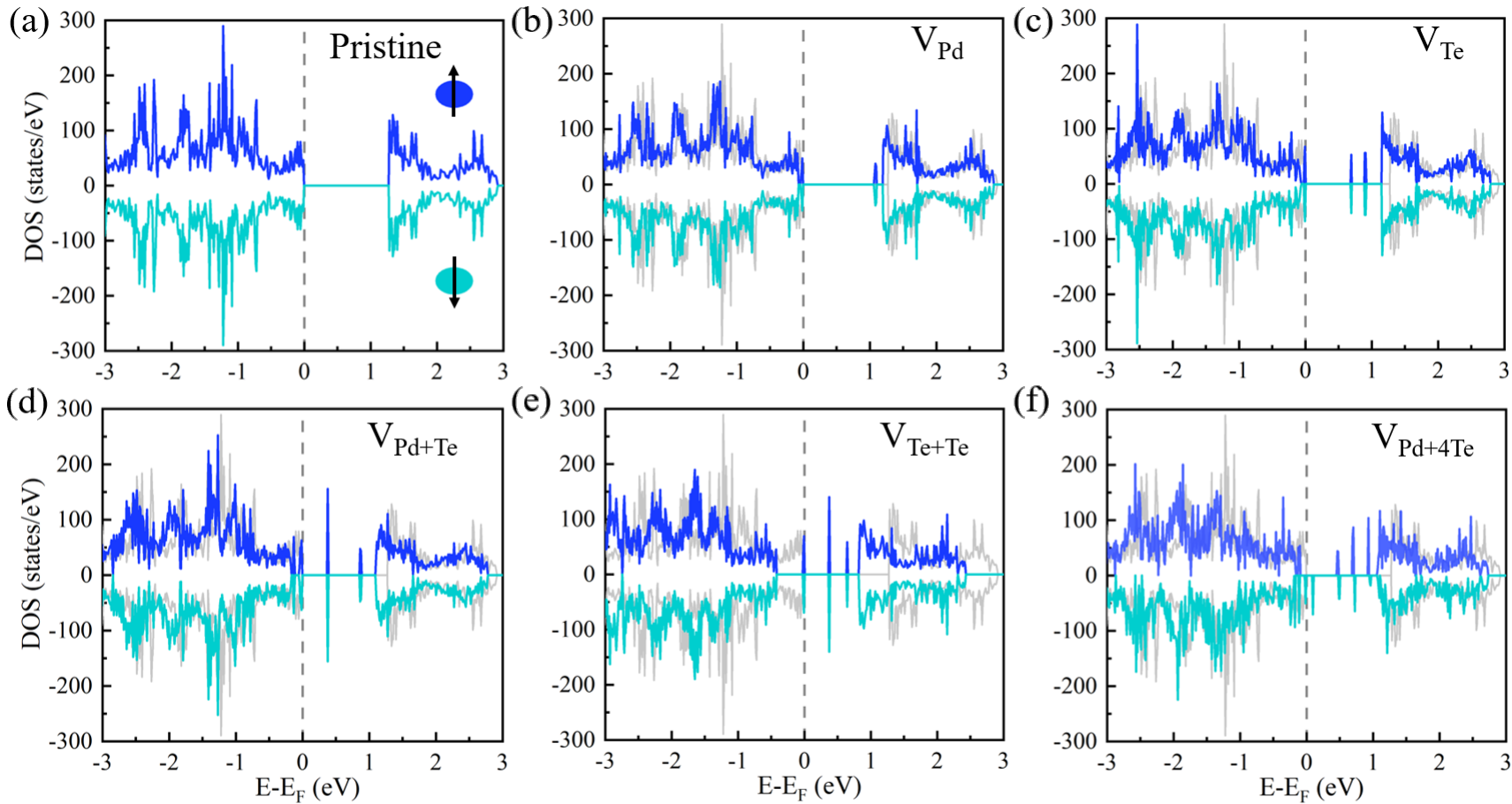}\caption{Spin-polarized total density of states corresponding to (a) pristine, (b) V$_{Pd}$, (c) V$_{Te}$, (d) V$_{Pd+Te}$, (e) V$_{Te+Te}$, and (f) V$_{Pd+4Te}$ vacancies in the~\emph{p}-PdTe$_2$ monolayer computed using GGA. The gray color in the background represents the TDOS corresponding to the pristine 4$\times$4$\times$1 supercell of the~\emph{p}-PdTe$_{2}$ monolayer.}
\label{fig:figS2} 
\end{figure*}

\begin{figure*}[ht]
\centering\includegraphics[width=1\linewidth]{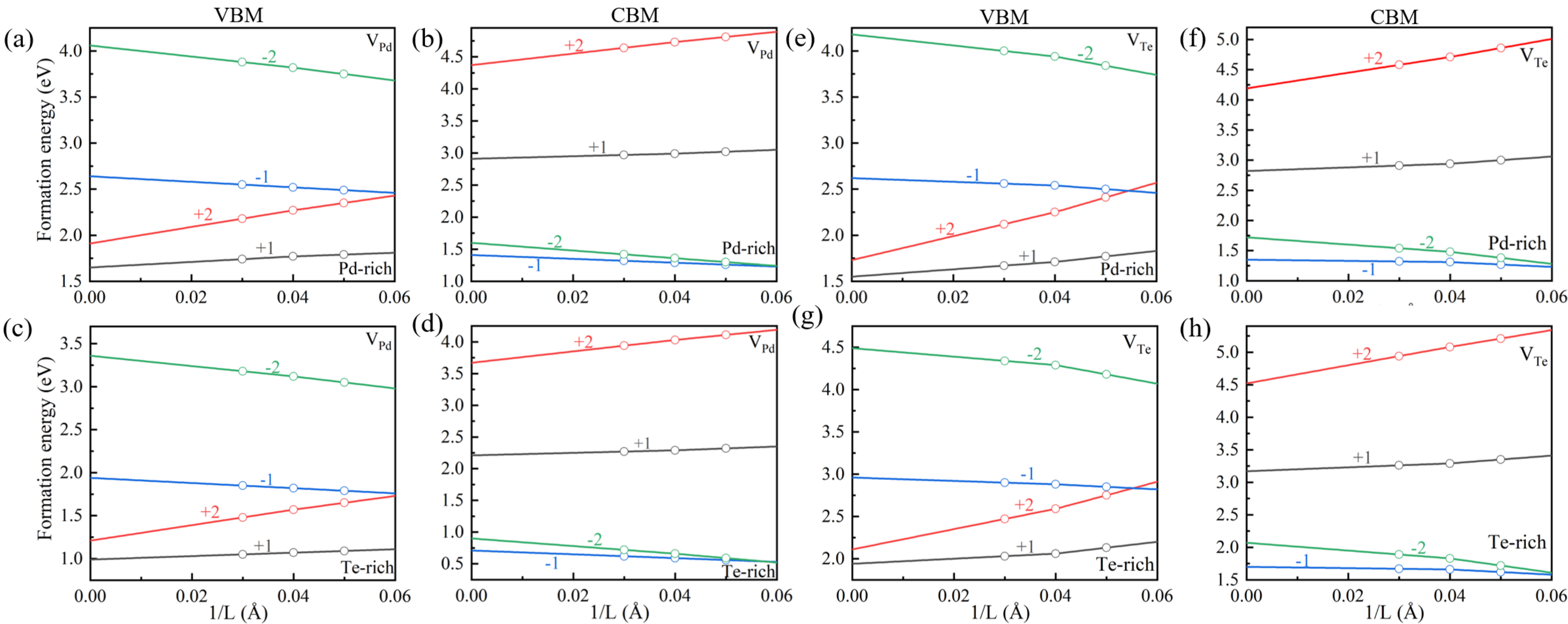}\caption{Formation energies of charged defects corresponding to (a-d) V$_{Pd}$ and (e-f) V$_{Te}$ in \emph{p}-PdTe$_2$ monolayer as a function of inverse cell dimension (L$_a$$\approx$L$_b$$\approx$L$_c$) at VBM and CBM in Pd-rich and Te-rich conditions.}
\label{fig:form} 
\end{figure*}

\begin{figure*}[ht]
\centering\includegraphics[width=0.5\linewidth]{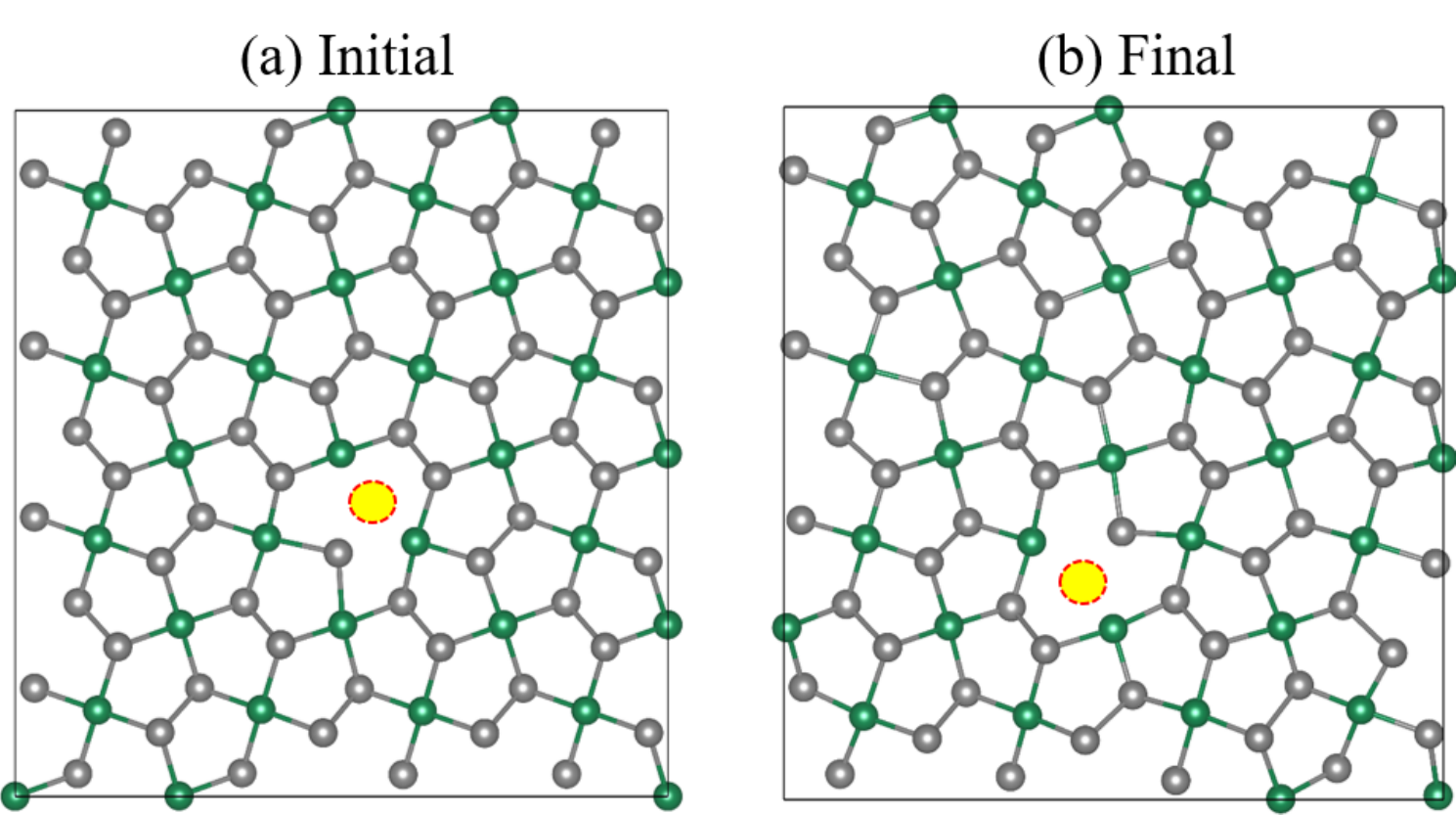}\caption{AIMD simulated structures of~\emph{p}-PdTe$_2$ monolayer with the (a) initial and (b) final positions of V$_{Te}$ after a time period of 5ps at 400 K. The highlighted circle indicates the V$_{Te}$ position.}
\label{fig:aimd} 
\end{figure*}

\begin{figure*}[ht]
\centering\includegraphics[width=0.7\linewidth]{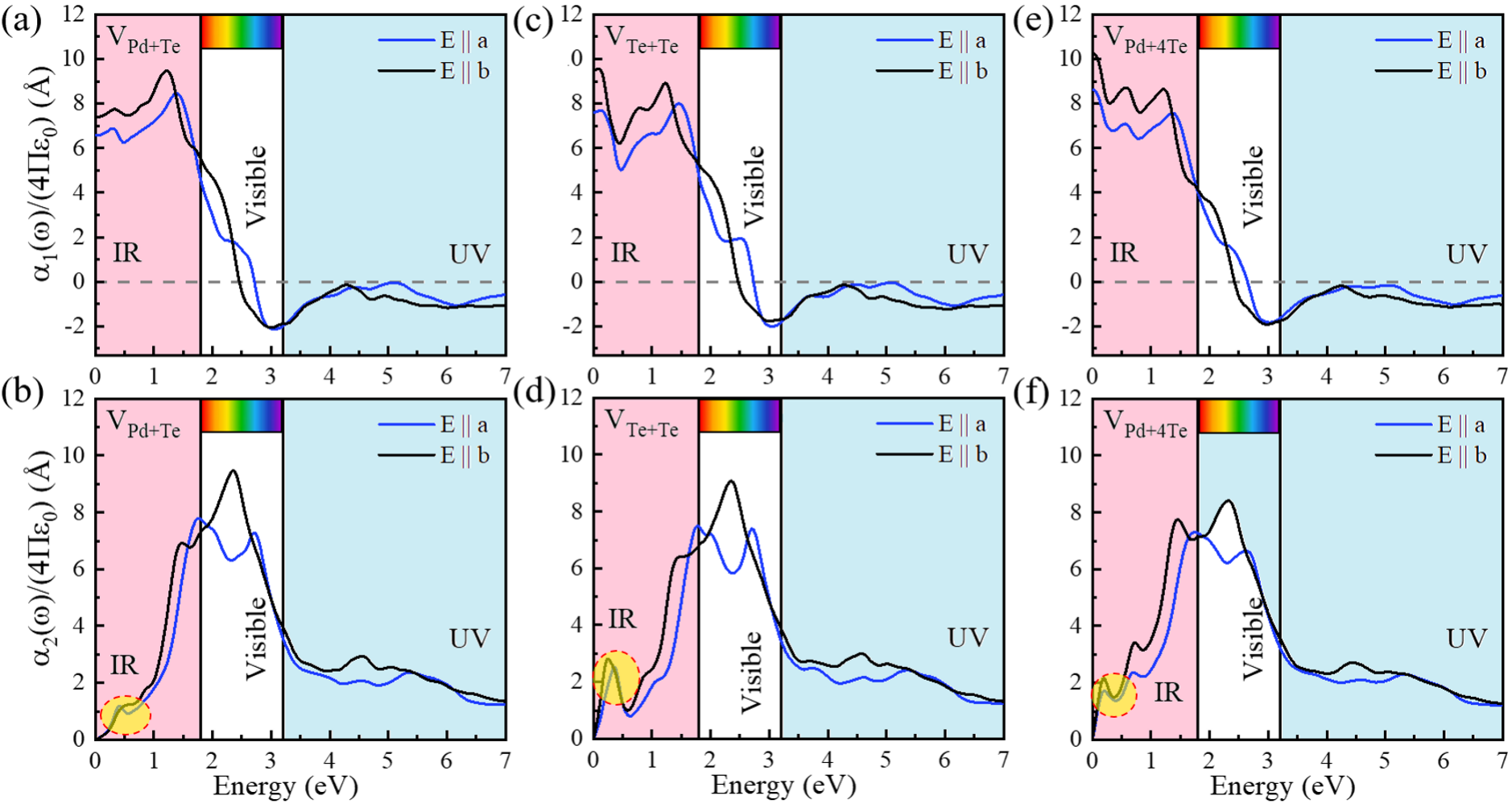}\caption{The (a,c,e) real and (b,d,f) imaginary parts of the electronic polarizability per unit area corresponding to V$_{Pd+Te}$, V$_{Te+Te}$ and V$_{Pd+4Te}$, respectively. The SOC is included in all the calculations.}
\label{fig:opti} 
\end{figure*}

\begin{figure*}[ht]
\centering\includegraphics[width=0.7\linewidth]{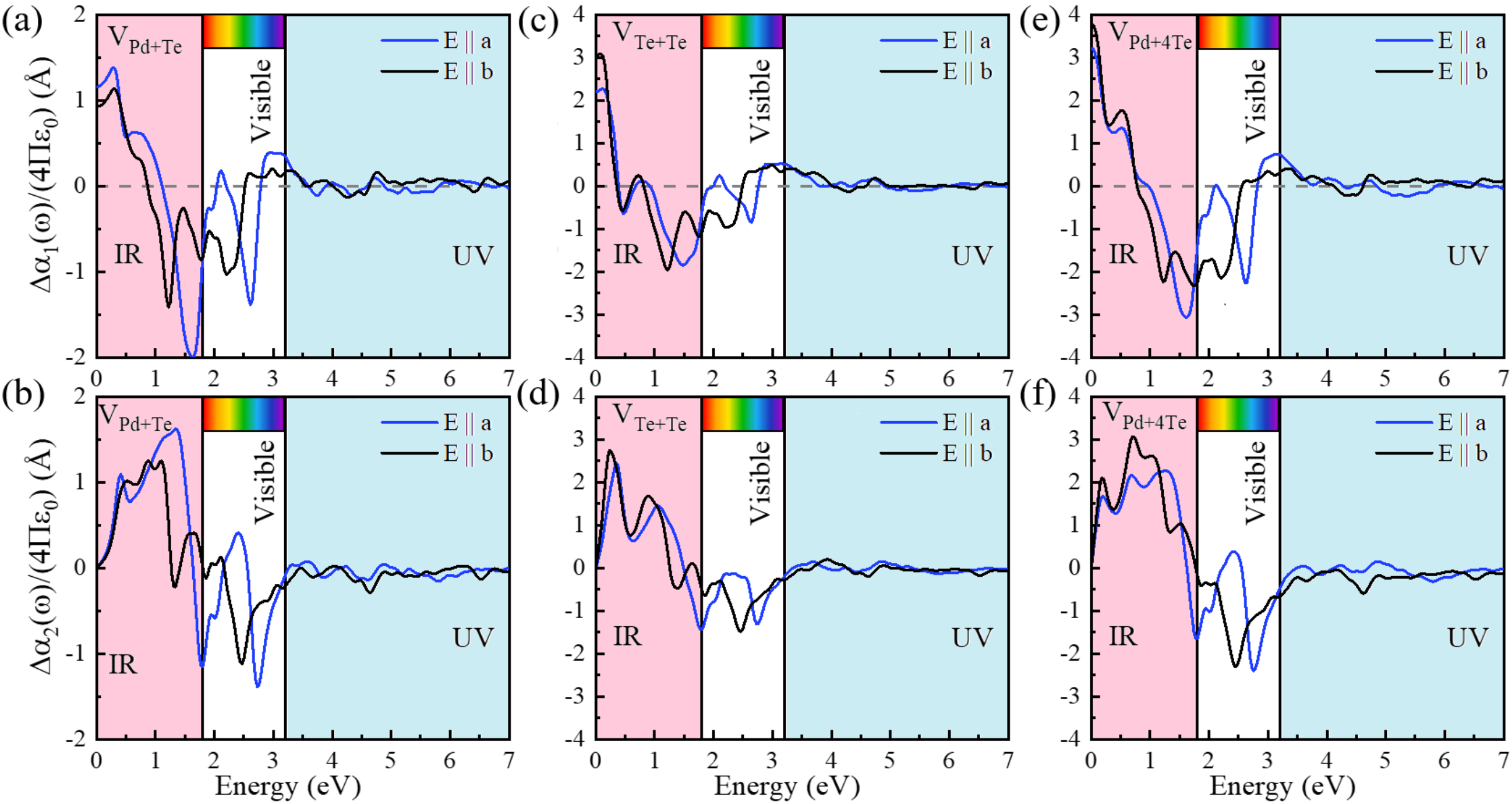}\caption{Calculated (a,c,e) real ($\Delta$$\alpha$$_1(\omega)$) and (b,d,f) imaginary ($\Delta$$\alpha$$_2(\omega)$) parts of excess electronic polarizability per unit area corresponding to V$_{Pd+Te}$, V$_{Te+Te}$ and V$_{Pd+4Te}$, respectively. The SOC is included in all the calculations.}
\label{fig:opti_excess} 
\end{figure*}

\begin{table}[ht]
\centering %\setlength{\tabcolsep}{2pt}
\caption{\label{tab:table1} Calculated band gaps of pristine \emph{p}-PdTe\(_2\) monolayer and various vacancy configurations in a 4$\times$4$\times$1 supercell using GGA+SOC and HSE06+SOC methods.}

\begin{ruledtabular}
\begin{tabular}{ccccccc}
Configurations$\rightarrow$ & Pristine & V$_{Pd}$  & V$_{Te}$  & V$_{Pd+Te}$  & V$_{Te+Te}$  & V$_{Pd+4Te}$\tabularnewline
\hline 
GGA+SOC  & 1.24  & 1.17  & 1.09  & 1.03  & 1.23 & 1.00\tabularnewline
HSE06+SOC  & 1.68  & 1.37  & 1.41  & 1.35  & 1.67& 1.09\tabularnewline
\end{tabular}
\end{ruledtabular}

\label{Table:table1} 
\end{table}

\begin{table}
\caption{Calculated excess static electronic polarizability per unit area ($\Delta\alpha_{1}(0)$), the locations of most intense (MI) peaks (eV) of the real ($\Delta\alpha_{1}(\omega)$) and imaginary ($\Delta\alpha_{2}(\omega)$) parts of the excess electronic polarizability per unit area  for V$_{Pd+Te}$, V$_{Te+Te}$ and V$_{Pd+4Te}$.}

\begin{ruledtabular}
\begin{tabular}{c|c|c|c|c|c|ccc}
 &  \multicolumn{3}{c|}{$E\parallel a$ }  & \multicolumn{3}{c}{$E\parallel b$ } &  & \tabularnewline
\hline 
Vacancy types$\rightarrow$ & V$_{Pd+Te}$  & V$_{Te+Te}$  & V$_{Pd+4Te}$ & V$_{Pd+Te}$  & V$_{Te+Te}$  & V$_{Pd+4Te}$  &  & \tabularnewline
\hline 
$\Delta\alpha_{1}(0)$  & 1.15 & 2.20  & 3.21  & 0.92  & 3.04  & 3.78        \tabularnewline
$\Delta\alpha_{1}(\omega)$ (MI, eV)  & 0.29 & 0.11  & 0.52  & 0.29 & 0.09 & 0.52      \tabularnewline
$\Delta\alpha_{2}(\omega)$ (MI, eV) & 1.33 & 0.34  & 1.24  & 0.87  & 0.23  & 0.72    \tabularnewline

\end{tabular}\label{tab:defS} 
\end{ruledtabular}

\end{table}

\bibliographystyle{apsrev4-2}
\bibliography{pdte2}